\documentclass{article}
\usepackage{arxiv}
\usepackage[superscript,biblabel]{cite}
\usepackage{authblk}
\usepackage{amsmath}
\usepackage{amsfonts}
\usepackage{amssymb}
\usepackage{gensymb}
\usepackage{latexsym}
\usepackage{booktabs,array,multirow}
\usepackage{graphicx}
\usepackage{physics}
\usepackage{subcaption}
\usepackage{siunitx}
\usepackage{float}
\usepackage{xcolor}
\usepackage{textcomp}
\usepackage{tabulary}
\usepackage[utf8]{inputenc}
\usepackage{xr}
\graphicspath{{Pictures/}}

\makeatletter
\newcommand*{\addFileDependency}[1]{
\typeout{(#1)}
\@addtofilelist{#1}
\IfFileExists{#1}{}{\typeout{No file #1.}}
}\makeatother

\author[1]{
{Rémi~Blinder\thanks{To whom correspondence should be addressed: \texttt{remi.blinder@uni-ulm.de}}}%
}
\author[1]{Yuliya~Mindarava}%
\author[2]{Martin Korzeczek}
\author[3]{Alastair Marshall\thanks{Present address: Qruise GmbH, Am Hölzersbach 7, 66113 Saarbr\"{u}cken, Germany}}
\author[4]{Felix Glöckler\thanks{Present address: Sensific GmbH, 89081 Ulm, Germany}}
\author[4]{Steffen Nothelfer\thanks{Present address: Hensoldt, Wörthstraße 85, 89077 Ulm, Germany}}
\author[4]{Alwin Kienle}
\author[5]{Christian Laube}
\author[5]{Wolfgang Knolle}
\author[6]{Christian Jentgens}
\author[2,7]{Martin B. Plenio}
\author[1,7]{Fedor Jelezko}

\affil[1]{Institute for Quantum Optics, Albert-Einstein Allee 11, Ulm University, 89081 Ulm, Germany }
\affil[2]{Institute of Theoretical Physics, Albert-Einstein Allee 11, Ulm University, 89081 Ulm, Germany }
\affil[3]{NVision Imaging Technologies GmbH, 89081 Ulm, Germany}

\affil[4]{Institute for Laser Technologies in Medicine and Metrology at the University of Ulm (ILM), Helmholtzstr. 12, 89081, Ulm, Germany}

\affil[5]{Leibniz Institute of Surface Engineering, Leipzig 04318, Germany}

\affil[6]{Pureon AG, Kreuzlingerstrasse 1, 8574 Lengwil, Switzerland}

\affil[7]{Centre for Integrated Quantum Science and Technology (IQST), Ulm 89081, Germany}

\renewcommand{\headeright}{Preprint}
\renewcommand{\undertitle}{Preprint}

\title{$^{13}$C Hyperpolarization with Nitrogen-Vacancy Centers in Micro- and Nanodiamonds for Sensitive Magnetic Resonance Applications}

\date{}                     

\begin{document}

\maketitle

\begin{abstract}

Nuclear hyperpolarization is a known method to enhance the signal in nuclear magnetic resonance (NMR) by orders of magnitude. The present work  addresses the   $^{13}$C hyperpolarization in diamond micro- and nanoparticles,  using the optically pumped nitrogen-vacancy center (NV) to polarize $^{13}$C spins at room temperature. 
Consequences of the small particle size are mitigated by using a combination  of surface treatment improving  the $^{13}$C relaxation ($T_1$) time, as well as that of NV, and applying a technique for NV  illumination based on a microphotonic structure.  Monitoring the light-induced redistribution of the NV spin state populations with electron paramagnetic resonance, a strong  polarization enhancement for the NV spin state is observed in a narrow spectral region corresponding to  about 4\%  of these defect centers.  
By combining adjustments to the `PulsePol' sequence  and slow sample rotation, the  NV-$^{13}$C polarization transfer rate is improved further. The hyperpolarized $^{13}$C NMR signal is observed in particles of \SI{2}{\micro\meter} and \SI{100}{\nano\meter} median sizes, with enhancements over the thermal signal  (at $\SI{0.29}{\tesla}$ magnetic field),  of 1500 and 940,  respectively. The present demonstration of room-temperature hyperpolarization anticipates the development of agents based on nanoparticles for sensitive magnetic resonance  applications.

\end{abstract}

\section*{Introduction}

As a technique improving the sensitivity of Nuclear Magnetic Resonance (NMR) and Magnetic Resonance Imaging (MRI), hyperpolarization is widely explored for enhancing  
the acquisition of the nuclear precession signal.\cite{eills2023} Besides its applications in medical imaging,\cite{nikolaou2014,wang2019} it holds the promise of benefiting multiple research fields including metabolomics,\cite{ribay2023} drug screening and monitoring,\cite{kim2019,bertarello2022} the study of proteins,\cite{sadet2019} nano/mesoporous systems \cite{rankin2019} or battery materials.\cite{haber2022} Signal enhancements by above 4 orders of magnitude were demonstrated in the work from Ardenkjær-Larsen \textit{et al.} in 2003,\cite{larsen2003}  employing the method, nowadays known as dissolution Dynamical Nuclear Polarization (dissolution DNP), which enabled \textit{in-vivo} imaging of metabolic processes.\cite{nelson2013} Despite these accomplishments, alternative concepts are being investigated, with the prospect of reducing hardware cost and complexity (that are both elevated in dissolution DNP, due to the use of cryogens) or, to bring magnetic resonance signal boosts to new systems.
To circumvent the need for cryogenic temperatures, yet achieve high nuclear signal enhancements, a proposed approach is the use of DNP protocols with electrons that are initialized out-of-equilibrium in a spin state with above-thermal polarization. Early on, in 1990, this was demonstrated by Henstra \emph{et al.},\cite{henstra1990} who used the photoexcited state of pentacene molecules in naphtalene to produce high nuclear spin polarization at room temperature. 
In recent developments, the same method was successfully applied to polarize other matrices,\cite{negoro2018, hamachi2021} and its combination with Nuclear Overhauser Effect enhancement in the liquid state was demonstrated, allowing for polarization of external substances.\cite{eichhorn2022,matsumoto2022}

An interesting prospect in terms of novel applications of DNP is the generation of hyperpolarization within nanoparticles ($<$\SI{200}{\nano\meter}). This stems from the following reasons:  1) 
such nanoparticles can diffuse efficiently by circulation in the blood flow,\cite{alexis2008} making them candidate for the application as MRI tracers and 2)  owing to the high surface-to-volume ratio, they could play the role of renewable source of polarization for   external substances. 
The possibility of performing MRI acquisitions enhanced by hyperpolarization was demonstrated with $^{13}$C in diamond~\cite{kwiatkowski2018,waddington2019} or $^{29}$Si in silicon~\cite{kwiatkowski2017} nanoparticles, both polarized from  thermally polarized endogenous spin defects in a cryogenic environment, at liquid Helium temperatures. Although this scheme performs  well under the   cryogenic setting, it shows limited efficiency under ambient conditions.

Nanodiamonds attract interest due to several factors, that is, the long  $^{13}$C spin polarization lifetime ($T_1$) in diamond,\cite{reynhardt2001} the good biocompatibility,\cite{yang2019} and the possibility of producing single-crystalline nanoparticles with lattice properties similar to bulk.\cite{mochalin2011}
Diamond can also host a wide variety of paramagnetic defects\cite{loubser1978} that can be used for DNP. One example is  the substitutional nitrogen or P1 center  ($S=1/2$),\cite{reynhardt1998_dnp_1, bretschneider2016, shimon2022} which is incorporated during growth.  Specifically in nanodiamonds, near-surface spin  defects,\cite{yavkin2015,peng2019} 
have also been employed.\cite{kwiatkowski2018,waddington2019,yoon2019,boele2020} 
Finally, hyperpolarization was demonstrated using the nitrogen-vacancy  center (NV), either in bulk crystal \cite{london2013,king2015,scheuer2016,parker2019}
 or diamond powder ($>$\SI{1}{\micro\meter} particles).\cite{ajoy2018,miyanishi2020,gierth2020} DNP techniques relying on the  NV spin, $S=1$,  draw  on the essential capability that it can be  initialized to a high degree (up to 92\%) into the $m_s=0$ state,\cite{waldherr2011} upon excitation with short pulses of green light, at arbitrary magnetic field and  ambient conditions.
By combining optical pumping with dynamical nuclear polarization, it becomes possible to transfer the spin polarization generated on the NV  to surrounding nuclei. 
 Harnessing  NV as a polarization source holds the potential to push the boundaries of conventional DNP methods. This approach could potentially achieve  high levels of nuclear polarization, without the requirement of cryogenic temperatures.

However, material and DNP protocols must be optimized in order to fulfill requirements of using color centers in nanoparticles, such as  NV in nanodiamonds.
Previous attempts to hyperpolarize $^{13}$C spins from NV in  diamond nanocrystals ($<$\SI{200}{\nano\meter})  resulted in only a very low degree of nuclear polarization.\cite{miyanishi2020,gierth2020} In this work, we address the primary challenges affecting the efficiency of nuclear hyperpolarization in nanodiamonds. First, we mitigate the impact of surface magnetic noise on $^{13}$C relaxation time  by appropriate surface treatment. Second, we enhance NV initialization by improving sample illumination. Furthermore, we increase the number of NV that are involved simultaneously in polarization transfer to nuclear spins by adjusting parameters in a DNP sequence (PulsePol) \cite{schwartz2018} to enhance its  robustness to  the orientation-induced broadening of the NV spectrum in powder samples. Last, we implement slow sample rotation ($\leq$25\degree/s) to bring different subsets of NV into the  excitation region of the adapted DNP  sequence and cycle through all of them. Thanks to these improvements, we achieve efficient polarization transfer from NV to $^{13}$C spins within diamond particles of \SI{2}{\micro\meter} and \SI{100}{\nano\meter} typical sizes, resulting in NMR signal enhancements over the thermal signal  of 1500 and 940, respectively, at the polarization field of about $\SI{0.29}{\tesla}$.

\section*{Results and discussion}

\begin{figure}[htp]
\centering
\includegraphics[width=0.8\linewidth]{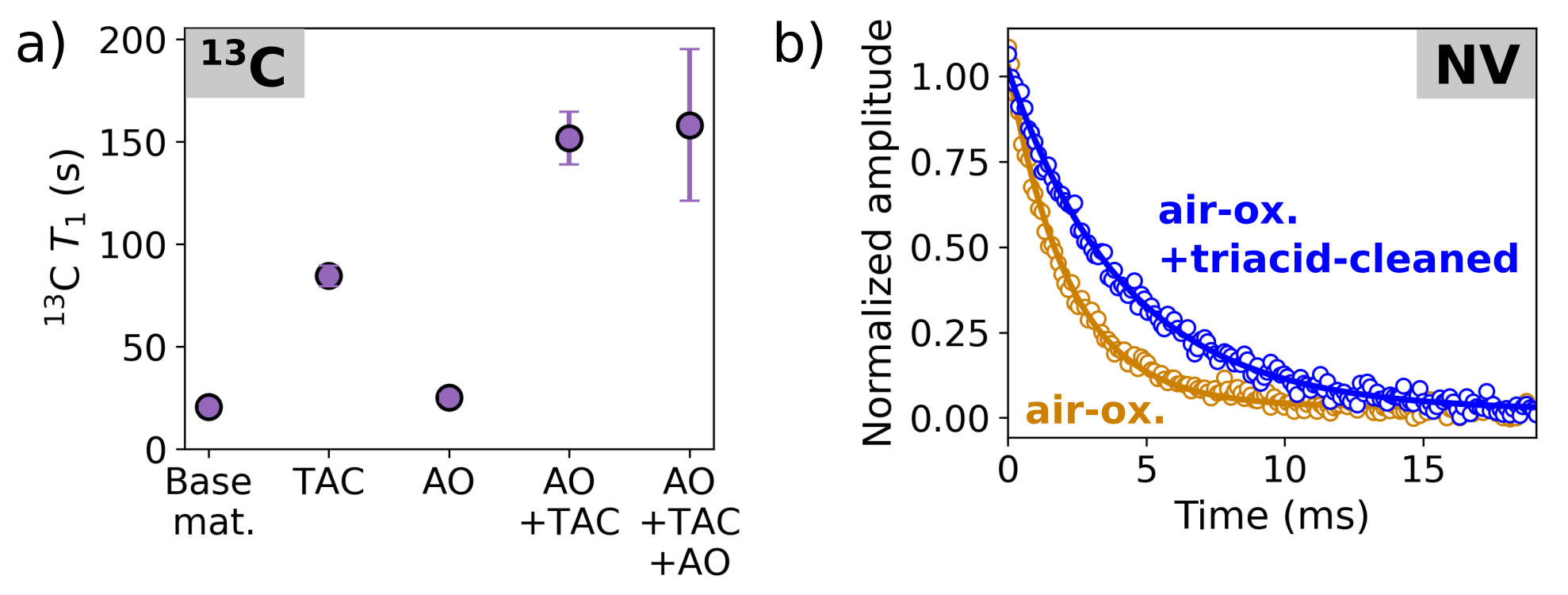}
\caption{Improving the spin  longitudinal relaxation times of $^{13}$C, NV   
 by  surface treatment of nanodiamonds  (\SI{100}{\nano\meter}) reducing contributions to  broadband magnetic noise  a) Evolution of the $^{^{13}\mathrm{C}}T_1$ measured in  NMR ($B=\SI{7.05}{\tesla}$) following treatments of the base commercial material:  air-oxidation (AO), triacid cleaning (TAC), or combinations (AO+TAC and  AO+TAC+AO).  Both combinations lead to an about seven-fold improvement over the relaxation time in the original material b) For the electron-irradiated and annealed sample, containing NV, comparison of the  $^{\mathrm{NV}}T_1$  taken on the $m_S=0$ to +1 transition at $B=\SI{0.24}{\tesla}$, following application of a laser pulse. Solid lines are exponential decay fits, providing  $^{\mathrm{NV}}T_1=\SI{2.20\pm0.03}{\milli\second}$  and   
$\SI{4.23\pm0.07}{\milli\second}$  following the air-oxidation (AO) and the combined  (AO+TAC) treatment, respectively.   }
\label{fig:13C_NV_T1}
\end{figure}

In this work, we use  diamond powders from Pureon AG (Lengwil, Switzerland) with median sizes of about \SI{2}{\micro\meter} and \SI{100}{\nano\meter} (see Supporting Information (SI), section \ref{SI:Material} for size distribution parameters). Substitutional nitrogens  (P1 centers) are converted into NV by high-energy electron irradiation and annealing, as described in a previous work.\cite{mindarava2020_conv} In this process, oxidation at high temperature (\SI{620}{\degreeCelsius}) in air is performed to remove graphitic residues.\cite{laube2017}  Besides the resulting  NV concentrations, given in Table~\ref{table:Diam_powder_properties}, we also consider the  $^{13}$C spin-lattice relaxation time  ($^{^{13}\mathrm{C}}T_1$), which is a crucial  parameter for  hyperpolarization experiments. The $^{^{13}\mathrm{C}}T_1$ determines not only the attainable level of nuclear hyperpolarization but also the duration allowed between hyperpolarization and signal detection in the investigated sample.
While microdiamonds  are expected to exhibit a  $^{^{13}\mathrm{C}}T_1$ as in bulk crystals of similar defect composition,  the spin-lattice relaxation in nanodiamonds ($<$\SI{200}{\nano\meter})  typically  shows contribution  related to  surface spins (dangling bonds etc.).~\cite{panich2015}
In such small particles,  in order to prolong the $^{^{13}\mathrm{C}}T_1$, it is thus important to work  towards suppression of the surface magnetic noise. 
Besides the air oxidation, our \SI{100}{\nano\meter} particles were as well treated  with a three-acid mixture, as detailed in  Experimental section.
To validate this approach, we investigated the NMR $^{^{13}\mathrm{C}}T_1$ at different  stages of treatment of the commercial nanodiamonds. The data in Figure~\ref{fig:13C_NV_T1}a was acquired at $B=\SI{7.05}{\tesla}$ magnetic field, remarkably providing  evidence that removal of relaxation-inducing sources  is better achieved through the combination of air-oxidation (AO) and triacid cleaning (TAC), rather than by  performing each treatment separately. The combination (AO+TAC) indeed  leads to $^{^{13}\mathrm{C}}T_1 = \SI{152 \pm 12}{\second}$ at \SI{7.05}{\tesla}, which constitutes an improvement by over a factor of seven with respect to the base commercial material ($^{^{13}\mathrm{C}}T_1 = \SI{21 \pm 1}{\second}$). That improvement is dominantly accounted for by the triacid cleaning (leading  alone to an about four-fold increase). We explain these results as follows: air-oxidation achieves a first part of the required surface cleaning, by efficiently removing the surface graphite and the related dangling bonds. The triacid cleaning achieves an important second part, by removing paramagnetic metallic residues. The iron element could be detected by X-ray photoemission spectroscopy (XPS) on smaller nanodiamonds from the same manufacturer, thus constitutes the most likely origin of the magnetic noise (see SI, section~\ref{SI:XPS}). 
To investigate further the impact of  the  acid treatment step, we  measure the longitudinal relaxation time of NV   ($^{\mathrm{NV}}T_1$) by  pulsed electron paramagnetic resonance (EPR), on the air-oxidized sample, before and after acid cleaning. From the results shown in Figure~\ref{fig:13C_NV_T1}b, it can be seen that the combined  treatment prolongs $^{\mathrm{NV}}T_1$  by more than a factor of two, compared to air oxidation  alone (Table~\ref{table:Diam_powder_properties}). We correlate this observation with the disappearance of a broad ($\sim\SI{3000}{G}$) signal in X-band EPR, that we attribute to the removal of the iron   impurities (SI, section~\ref{SI:surface_epr}). 
Here, observing the short   $^{\mathrm{NV}}T_1$ before the treatment with acid as in Figure~\ref{fig:13C_NV_T1}b is a telltale sign of magnetic noise extending in the high-frequency ($\sim$ GHz) range, thereby impacting the  relaxation of spins having comparable or lower resonant frequencies.\cite{rosskopf2014}  Last, although air-oxidation and triacid cleaning might provide a different balance of  functional groups on the  surface, this (possible) difference does not seem to affect the longitudinal relaxation, as visible from Figure~\ref{fig:13C_NV_T1}a. Indeed, after applying again air oxidation (AO+TAC+AO), no detectable change in $^{^{13}\mathrm{C}}T_1$ is observed.

\begin{figure}[ht]
\centering
\includegraphics[width=0.75\linewidth]{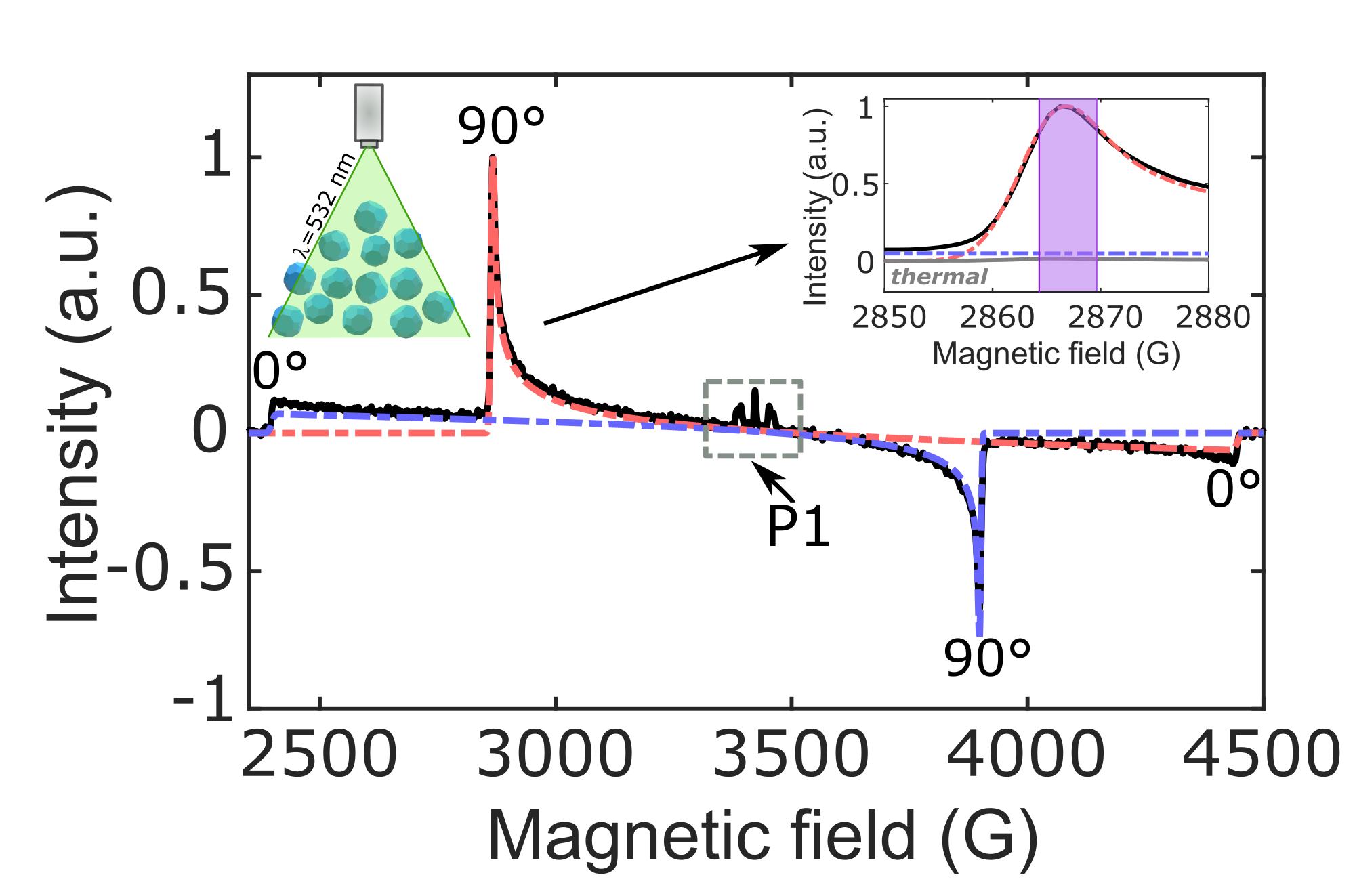}
\caption{Pulsed X-band EPR spectrum  under illumination, for the sample of \SI{2}{\micro\meter} particles. The experimental data (solid black line) was acquired  by echo detection following a \SI{700}{\micro\second} laser pulse,  using the microwave frequency $\nu=\SI{9.59}{\giga\hertz}$.  Dashed red and blue lines are  simulated spectra for the two separate spin transitions of NV fitting the experimental spectrum (see SI, sect.~\ref{SI:illumination:spectrum} for the description of the simulation). The resonance positions for NV having their axis either aligned  (0\textdegree) or perpendicular  (90\textdegree) to the magnetic field, locate at the spectrum extremities and on the intense peaks, respectively (as indicated). Besides NV, the P1 spectrum appears  in the central part. The right  inset shows a zoom on the low-field peak of the optically polarized spectrum and its comparison to the acquired thermal signal (grey), showing here a 85-fold  enhancement with illumination. The shaded blue area in the right inset  highlights a region corresponding to a range of $\Delta=(2\pi) \SI{15}{\mega\hertz}$ width in the frequency domain, in which  transitions for about 4\% of all NV are addressed.}
\label{fig:optic_polar}
\end{figure}

We turn our attention to the process of NV initialization by optical pumping, which is yet another prerequisite to performing (efficient)  hyperpolarization. To monitor this process, we use again EPR. In addition to being  a valuable tool for determining the  concentration, $T_1$ and $T_2$ times of NV, EPR allows precise tracking of the redistribution of the NV spin state populations under illumination.\cite{drake2015}  In fact, although  EPR spectra of optically polarized NV ensembles in powder samples have  been reported,\cite{mindarava2020_13C, miyanishi2020} performing homogeneous  illumination of NV in macroscopic  quantities of  micro- and nanodiamond  remains a challenge, owing to the strong light scattering created by the difference in refractive index between the diamond ($n=2.4$, at \SI{532}{\nano\meter} wavelength) and its surroundings. 
To overcome this issue, we couple light into a structure of closely packed waveguides forming an array, in which the diamond powder is embedded together with a solution of ethylcinnamat acting as buffer between the structure and the diamond.  The details of the illumination protocol are presented in SI, section \ref{SI:illumination}.

In Figure~\ref{fig:optic_polar} is shown the NV spectrum measured with  pulsed EPR for the \SI{2}{\micro\meter} particles, using the microstructured waveguide array for illumination with green light (\SI{532}{\nano\meter}).  The spectrum  shows a broad structure,  spreading equally on each side of the  $g\sim2$ electron resonance field  (on which the signal from P1 centers, seen in Figure~\ref{fig:optic_polar}, is centered as well). Considering the large value of the zero-field splitting for NV ($D=\SI{2.87}{\giga\hertz}$), 
the important  inhomogeneous broadening arises  in powder samples  as the axis of NV within randomly   oriented  diamond particles has  uniform distribution across the complete solid angle. In the dark, the spectrum has a characteristic structure, known as the Pake doublet,\cite{pake1948} showing two intense positive peaks, corresponding to NV being oriented perpendicular to the magnetic field.\cite{mindarava2020_13C, yavkin2019}  
Under illumination (Figure~\ref{fig:optic_polar}), the NV spectrum exhibits positive enhancement in the low-field region and negative enhancement in the high-field part (thus   a zero-crossing near the $g=2$ resonance).
These changes result from the redistribution of populations in the NV ground spin state, here induced by optical pumping. Importantly, in EPR, the signal scales  with the difference in population $(\Delta \rho)_{\mathrm{NV}}$ between the two states involved in the  spin transition. Thus, by measuring  in a chosen spectral region in the dark, and comparing the signal to that observed under illumination, the  population difference  ratio $(\Delta \rho)_{\mathrm{NV}} / (\Delta \rho)_{\mathrm{NV,dark}}$  can be estimated (SI, section \ref{SI:illumination:pulsed}). 
In the spectral region in Figure~\ref{fig:optic_polar} in which the highest spectral intensity is attained, that is at the low field peak ($B \SI{ \approx 0.287}{\tesla}$), we determine  $(\Delta \rho)_{\mathrm{NV}} / (\Delta \rho)_{\mathrm{NV,dark}}=46$ and 61, following a \SI{400}{\micro\second} laser pulse (used later on for NV initialization),   for the samples containing \SI{2}{\micro\meter} and \SI{100}{\nano\meter} particles, respectively (Figure~\ref{fig:SI_illumination_pulse} in SI).  The \SI{100}{\nano\meter} particles being more efficient  light scatterers, the mass to be  exposed to the laser beam was decreased two-fold in order to keep similar  illumination conditions. Considering that the signal intensity in the dark is governed by the Boltzmann statistics (details in SI, section \ref{SI:illumination:pulsed}), the \emph{absolute} population difference  is estimated as   $(\Delta \rho)_{\mathrm{NV}}\approx 3\%$ (2.4 and 3.2\%, respectively).

\begin{figure}[ht]
\centering
\includegraphics[width=0.5\linewidth]{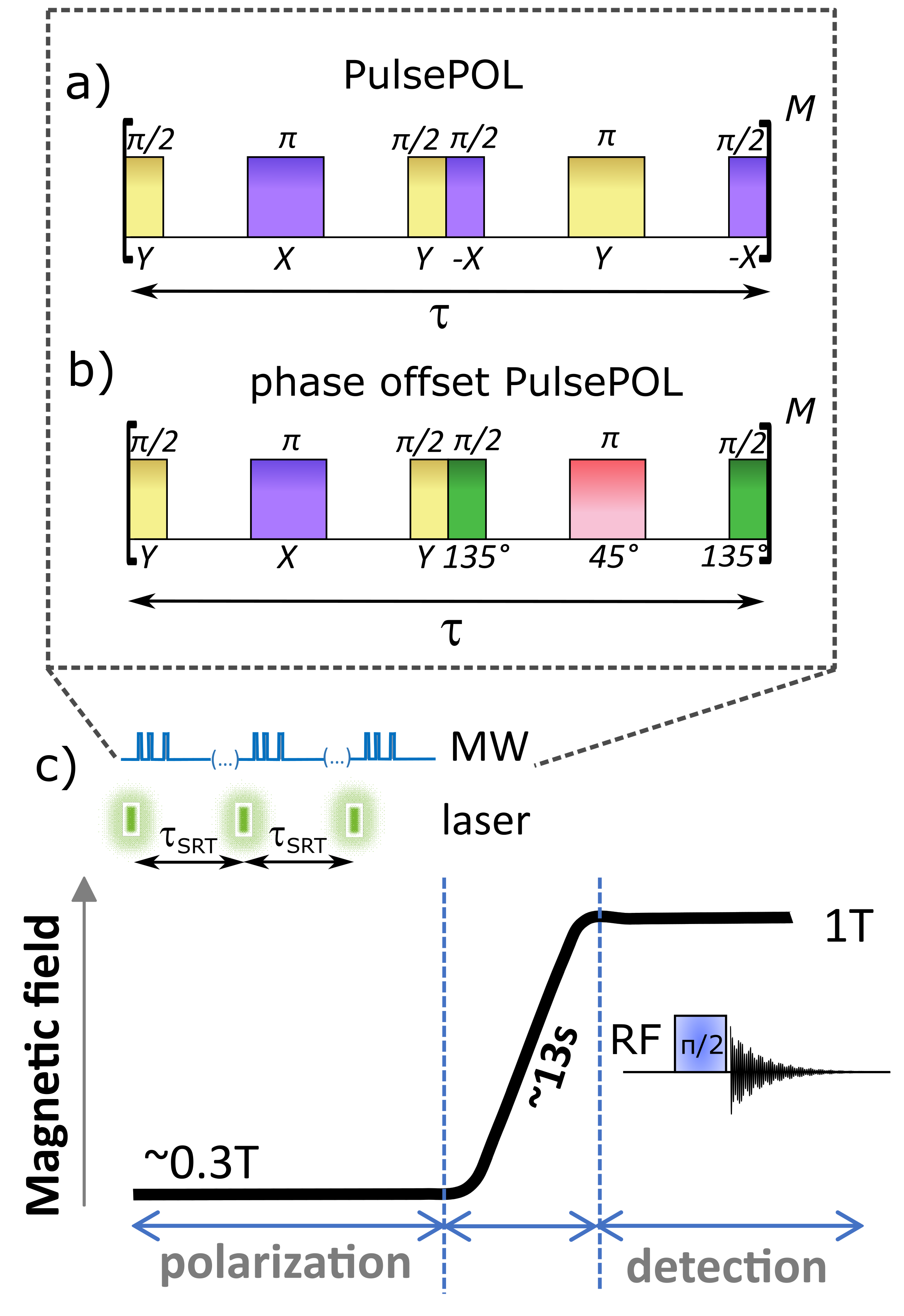}
\caption{ Sequences and timings. a) Conventional PulsePol sequence and b) phase-offset variant.
Both the sequences in a) and b) include equal delays between the $\pi$ and $\pi/2$ pulses. By adjusting these waiting times, one varies the  interval  $\tau$ (length of the 6-pulse block), which determines the resonant  condition for polarization transfer. The represented pattern  is repeated $M$ times, so that $M\times\tau$ is the duration of the  interaction with the nuclear spins. c) Illustration of the hyperpolarization protocol. One polarization cycle consists in  performing  NV initialization  by a laser pulse and polarization transfer using one of the sequences in  (a,b). The cycle   is repeated at regular  intervals (with a period $\tau_{\mathrm{SRT}}$) in order to build up nuclear polarization.  For NMR detection, the magnetic field is ramped up to \SI{1}{\tesla}, which takes about \SI{13}{\second}, and the $^{13}$C signal is acquired \emph{in-situ}}. 
\label{fig:hp_protocol}
\end{figure}

We now focus on the DNP sequence used for NV-$^{13}$C polarization transfer. Considering that the NV spectrum spans over an extended spectral range (Figure~\ref{fig:optic_polar}), corresponding in the frequency domain to over several GHz, it appears essential 
to carefully select the protocol for NV-$^{13}$C polarization transfer, and of utmost importance to look for DNP sequences offering a high degree of robustness to existing parameter inhomogeneities.  First, the polarization transfer sequence should reduce the impact of the strong spectral broadening, by working over a wide range of frequency detunings for NV. Second, efficient operation in the presence of Rabi drive imperfections is also desired  - to alleviate inhomogeneities in the MW field or in the NV transition dipole. In this work, we consider the PulsePol sequence, represented in Figure~\ref{fig:hp_protocol}a~\cite{schwartz2018} and its variant, shown in Figure~\ref{fig:hp_protocol}b.\cite{tratzmiller2021} We  highlight a notable difference to  the integrated solid effect (ISE) sequence, which was first introduced by Henstra \emph{et al.},\cite{henstra1990} later used in microdiamonds \cite{ajoy2018, miyanishi2020} and compared to PulsePol.\cite{schwartz2018} PulsePol (or its variant) uses discrete pulses, while the ISE exploits the dynamics of the  electron-nuclear transitions occuring during a long, continuous  pulse.\cite{chen2015,scheuer2017} 
With PulsePol, the peak microwave power determines the NV spectral range ($\Delta_{\mathrm{pol}}$) in which  NV-$^{13}$C polarization transfer is effective, that is  $\Delta_{\mathrm{pol}} \sim \Omega_1$ where  $\Omega_1$ is the Rabi frequency for NV.  This property  brings some benefits with respect to the  speed of polarization: in the  range covered by $\Delta_{\mathrm{pol}}$, the NV  are addressed in parallel,  while sequences such as the ISE manipulate them sequentially,  by action of a frequency (or field) sweep. As $\Omega_1$ has no theoretically imposed limit, it is sensible to operate the instrumentation at the highest output power. Working with a \SI{40}{\watt} amplifier guarantees  in our system that  the Rabi frequency of NV is $\Omega_1 = (2\pi)\SI{10.5}{\mega\hertz}$. 
 We use this microwave source to drive NV that are oriented at \SI{90}{\degree} to the magnetic field. This allows enhancing the fraction of centers  that are participating to the polarization dynamics, due to the larger solid angle that is covered along the equator of a sphere. For these NV, we choose to drive the transition that shows the highest NV polarization enhancement under illumination, that is the low field peak, visible at  $B=\SI{0.287}{T}$  in  Figure~\ref{fig:optic_polar}, as the latter  shows stronger absolute enhancement than the high-field peak ($B=\SI{0.390}{T}$).  
The protocol for hyperpolarization is represented in Figure~\ref{fig:hp_protocol}c. It consists in repeating the cycle of consecutive laser pulse initialization and microwave driving, in order to  gradually build up polarization for the $^{13}$C spins.

For improving the robustness properties of PulsePol, a possibility is to use the variant in Figure~\ref{fig:hp_protocol}b from Tratzmiller \emph{et al.},\cite{tratzmiller2021} which relies on alternative pulse phase  definitions  compared to the original sequence in Figure~\ref{fig:hp_protocol}a (hence the label `phase-offset PulsePol').  Besides  the increased robustness against frequency detuning, which could be verified by performing $^{13}$C hyperpolarization in single crystal diamond (SI, sect.~\ref{SI:pulsepol_phaseoffset}), the phase-offset variant is also expected to mitigate   inhomogeneities in the Rabi drive to a degree comparable to what is achieved in the standard PulsePol. 
The resonance condition for polarization transfer can be expressed similarly in both sequences as $\tau = \frac{n\pi}{\omega_I}$ where $\tau$ is the length of the 6-pulse block (Figure~\ref{fig:hp_protocol}) and  $\omega_I$ is the Larmor frequency of the target nuclear spins.  However,   the phase-offset PulsePol  has half-integer resonances (see SI, sect.~\ref{SI:pulsepol_4p5vs3p5}), contrasting with the odd-integer resonances $n=1,3,5,7,$ etc. as in the original PulsePol.
In the present scenario, in which the microwave excitation is resonant with NV which are at perpendicular orientation to the magnetic field, the choice of the resonance index $n$ revealed to be non-trivial.
Similarly to the case of the original PulsePol in Figure~\ref{fig:hp_protocol}a, in which  the $n=3$ resonance is predicted to be optimal for polarization transfer,\cite{schwartz2018} the $n=3.5$ condition was expected to offer better performance using the phase-offset PulsePol. However, our results suggest an improvement in the polarization transfer rate by using rather  $n=4.5$. This can be explained considering the dynamics involving the $^{14}$N nuclear spin of the NV, whose state can evolve  during the application of the PulsePol sequence, owing to the presence of   transitions that affect  simultaneously the NV and $^{14}$N spin (which show finite probability as the transverse magnetic field breaks the defect $C_{3v}$ symmetry). Simulations demonstrate that choosing the $n=4.5$ resonance is in fact  quasi-optimal  for refocusing the NV-$^{14}$N interaction and thereby avoiding  inhibition of the polarization transfer to  $^{13}$C by this dynamic (see SI, sect.~\ref{SI:pulsepol_phaseoffset} for comparison of the different resonances  and the description of NV-$^{14}$N interaction). 

Although using the more robust    sequence variant  in Figure~\ref{fig:hp_protocol}b provides a gain in the number of NV participating in the polarization transfer process,  it is  beneficial to implement strategies for increasing further the number of contributing NV. We now discuss such additional methods, optimize them by detecting the hyperpolarized signal in the \SI{2}{\micro\meter} sample, before applying them as well to the \SI{100}{\nano\meter} nanodiamonds.

\begin{figure}[bt]
\centering
\includegraphics[width=\linewidth]{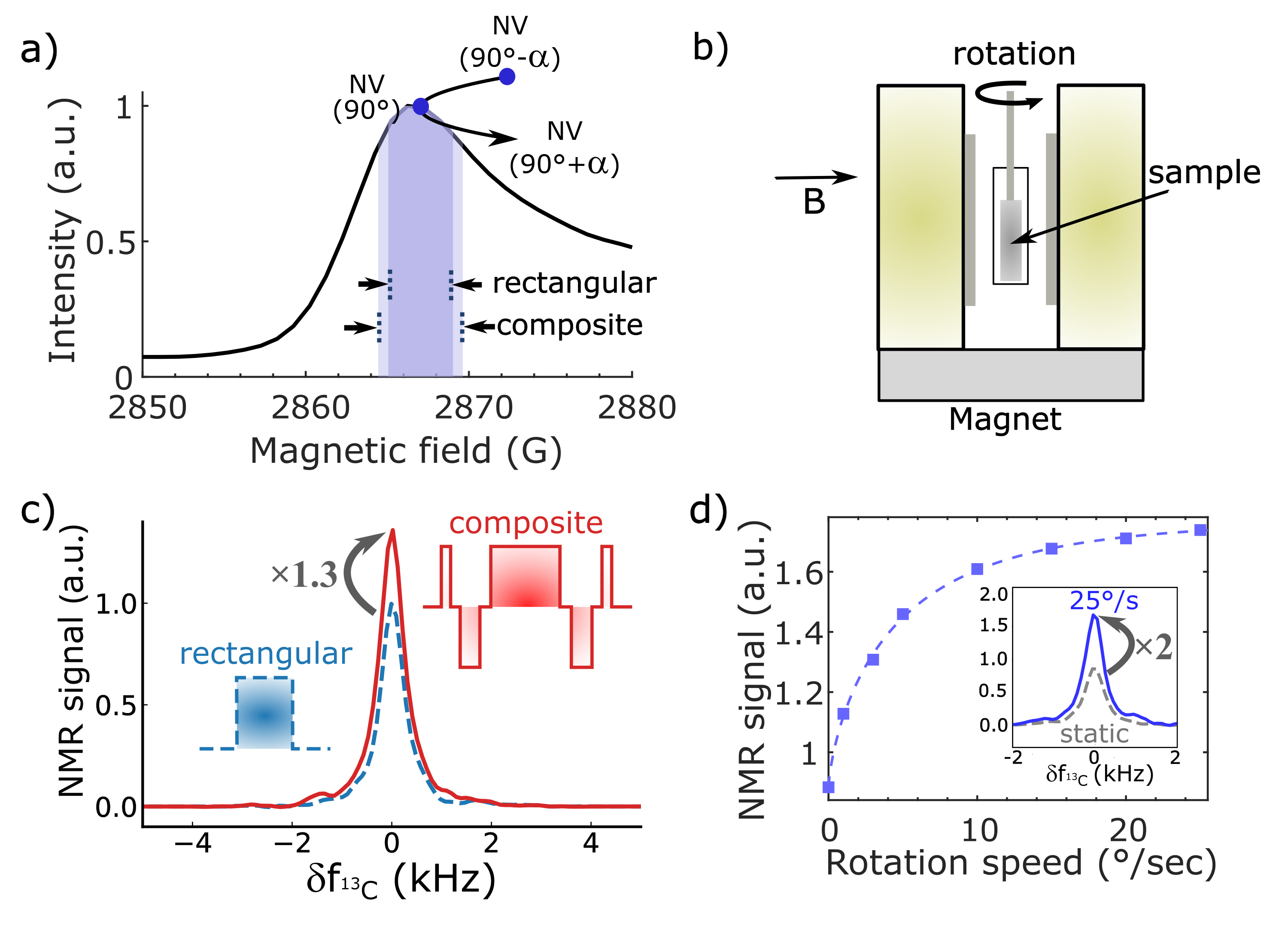}
\caption{a) NV EPR spectral line with an illustration of strategies to increase the number of NV participating in polarization transfer. The microwave excitation bandwidth is extended by using composite pulses. Sample rotation enables a larger number of NV with various orientations to the magnetic field to pass through the microwave excitation region. b) Illustration showing the orientation of the rotation axis in the experimental setup c) Comparison of the hyperpolarized $^{13}$C NMR signals obtained using rectangular and composite pulses. d) Hyperpolarized $^{13}$C NMR signals obtained with sample rotation at different speeds. The dashed line is a stretched exponential fit as described in the text, yielding parameters $\omega_{\mathrm{r},0}=\SI{4.9(5)}{\degree/s}$  and  $\beta = 0.74(6)$ as the stretch exponent.  Acquisitions represented on this figure were taken on the \SI{2}{\micro\meter} sample. }
\label{fig:NVproportion}
\end{figure}

The first method consists in replacing the individual rectangular pulses  by ``composite'' versions to provide further robustness to frequency detuning. We consider composite pulses described by Shaka~\emph{et al.}~\cite{shaka1987}, which are built by adding `sidebands' of alternating phases around a central rectangular pulse. In this work, the pulses by Shaka~\emph{et al.} are  refined by the introduction of waiting times, and then numerically optimized, in order to improve their robustness.  
In our conditions $\Omega_1 = (2\pi)\SI{10.5}{\mega\hertz}$ we determined that the simplest (two-sideband) version of the pulses  leads  to efficient operation of PulsePol in the range $\Delta_{\mathrm{pol}}=  (2\pi) \SI{15}{\mega\hertz}$ (see SI, sect.~\ref{SI:pulsepol_composite} for description of the pulses optimization and  sequence robustness properties).  
Owing to the asymmetric shape of the peak, the magnetic field was offset by a few gauss to maximize the number of addressed NV, as shown in inset of Figure~\ref{fig:optic_polar}, where we have represented a spectral region corresponding to the $\SI{15}{\mega\hertz}$ of excitation width. We estimate that this spectral region  covers approximately a  fraction of 4\% of all centers, corresponding to NV oriented with a tilt ($\theta$) from the magnetic field such that $|\theta-
90\degree|\lesssim 2.5\degree$  (see SI, sect. \ref{SI:sec_hamiltonian_diag_90deg} and \ref{SI:sec_fraction_delta_pol}).     In Figure~\ref{fig:NVproportion}c, the  hyperpolarized $^{13}$C signals obtained using the standard and  the optimized two-sideband pulse  are compared, showing an improvement by about 30\%  compared to the latter (setting   parameters such as the repetition number $M=68$, and the sequence repetition time  $\tau_{\mathrm{SRT}}=\SI{1.5}{\milli\second}$, which appeared to be optimal, from SI, sect.~\ref{SI:pulsepol_otherparams}). We note that the duration of the pulses is constrained by the  chosen PulsePol resonance condition, as it is visible from Figure~\ref{fig:hp_protocol}, the length of the $\pi$ pulse should not exceed $\tau/4$. This restricted us from using `higher-order' versions (i.e. with more than two sidebands) of the pulses, even though these are predicted to be more efficient. These would be within reach using a more powerful amplifier.  

As an additional approach for enhancing the signal, we implemented slow sample rotation, around an axis perpendicular to the magnetic field (Figure~\ref{fig:NVproportion}a-b). This allows cycling through different NV subsets, as those defined by the excitation region of PulsePol  will be progressively replaced, upon rotation, by NV of different initial orientation. The slow sample rotation implemented as in Figure~\ref{fig:NVproportion}b will bring, in fact, the axis of any NV to a direction orthogonal to the magnetic field, ensuring its spin transitions meet the \SI{90}{\degree} resonances. We rotate at speeds up  to 25 \degree/s -the maximum allowed by our   rotation stage- and obtain an increase in the $^{13}$C signal  that is remarkably two-fold at the maximum velocity (Figure~\ref{fig:NVproportion}d). This suggests that  switching regularly the subset of NV  that  is addressed by the  polarization transfer sequence is favorable, compared to using continuously the same subset.

 We can understand this observation, from the dynamics of the $^{13}$C polarization buildup. Without applying rotation, the increase of signal with the polarization time follows a so-called stretched exponential function $s(t)=A \left (1-\exp{-(\frac{t}{T_{\mathrm{pol}}})^\beta} \right )$, with $T_{\mathrm{pol}}=\SI{29(4)}{\second}$ and $\beta=0.70(8)$ (SI, Figure~\ref{fig:SI_buildup_2um_norotation}). Consequently, efficient polarization occurs at short times,  $s(t)\sim (\frac{t}{T_{\mathrm{pol}}})^{\beta}$, as  $\beta < 1$.  
It was previously observed that:~\cite{beatrez2023} 1) initially, the buildup occurs due to direct polarization of $^{13}$C spins from NV or spin diffusion on short length scales, while at longer times, it acquires (slower) dynamics impacted by the less efficient spin diffusion at longer scales, and 2) 
the NV  acts  in its very close surroundings,  as a relaxation sink for nuclear spins shortening their $T_1$. 
 Here, we exploit this dynamic to benefit  from the efficient polarization creation regime at short times (by rotating the sample) but still address  distant $^{13}$C spins, with a long nuclear $T_1$. 
From Figure~\ref{fig:NVproportion}d, increasing the rotation speed ($\omega_{\mathrm{r}}$) first enhances steeply the signal $s(\omega_{\mathrm{r}})$, which reaches an apparent saturation at high speeds. This behavior  can be fitted as  
$s(\omega_{\mathrm{r}})=s_0 + a \left (1-\exp{-(\omega_{\mathrm{r}}/\omega_{\mathrm{r},0})^\beta}  \right)$, that is a stretched exponential, from which  $\omega_{\mathrm{r},0}=\SI{4.9(5)}{\degree/s}$ is extracted and $\beta=0.74(6)$.
The stretched exponential behavior  reminds of that of the  polarization buildup. In fact, it is instructive to 
consider the overall rotation angle during the nuclear polarization buildup time $\omega_{\mathrm{r}} T_{\mathrm{pol}}$. At a speed $\omega_{\mathrm{r}}\approx \omega_{\mathrm{r},0}$ one has   $\omega_{\mathrm{r}} T_{\mathrm{pol}} =   \omega_{\mathrm{r},0} T_{\mathrm{pol}} = \SI{142(20)}{\degree}$, which is close to the condition when  \emph{half a turn} ($\SI{180}{\degree}$) is performed  during $T_{\mathrm{pol}}$.  Thus, the significant signal increase provided by rotation in the range of speed $\omega_{\mathrm{r}}<\omega_{\mathrm{r},0}$ occurs as more of the different NV subsets are progressively being  involved in the polarization transfer, until all NV participate (at  $\omega_{\mathrm{r}} \approx  \omega_{\mathrm{r},0}$).  At the maximum speed (25 \degree/s), we measure a polarization buildup time $T_{\mathrm{pol}}=\SI{34(2)}{\second}$ comparable to that in the static sample.  As the buildup dynamic is ultimately determined by the relaxation properties ($T_1$) of the spins that are being polarized, our observation  indicates 
that $^{13}$C populations with comparable $T_1$ are being addressed, in the experiments with and without rotation. Since the signal is maximally increased by  rotation (25 \degree/s), we use this setting in the rest of our experiments.

\begin{figure}[ht]
\centering
\includegraphics[width=\linewidth]{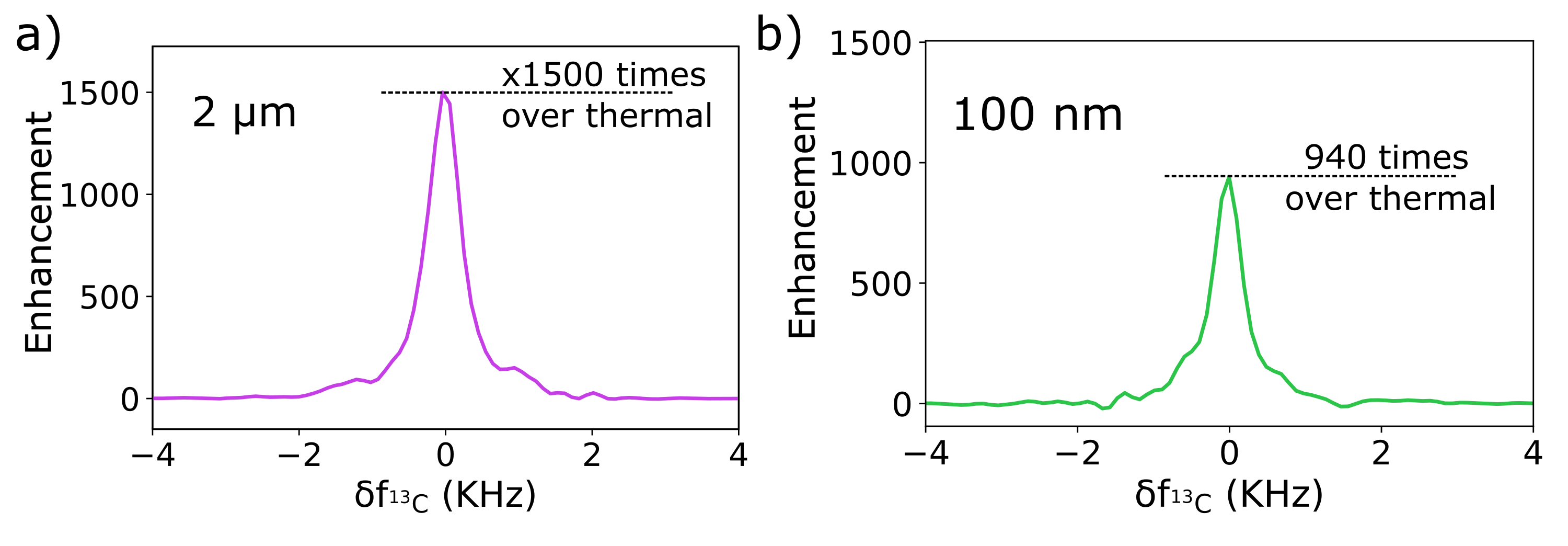}
\caption{$^{13}$C NMR signal obtained after 5 min of hyperpolarization for a) $\SI{2}{\micro\meter}$ and b) $\SI{100}{\nano\meter}$
diamond particles. The spectra intensities are normalized by the $^{13}$C thermal signal at the polarization field (\SI{0.287}{\tesla}), showing enhancements of 1500
and 940 times in a) and b), respectively.}
\label{fig:Hyper_signal}
\end{figure}

\begin{table}[h!]
	\centering
	\caption[]{Summary of the sample properties. P1, NV concentrations and NV relaxation times are determined by X-band EPR.  The $T_2$ time of NV  is measured by  Hahn echo ($\pi/2-\pi$) in pulsed EPR and the $T_1$ time is obtained by fitting the decay of the echo-detected NV signal following a laser pulse. The $^{13}$C hyperpolarization is characterized by the buildup with the time $T_{\mathrm{pol}}$, and the enhancement $\epsilon_{\mathrm{13C}}$ (at the polarization field, $B=\SI{0.287}{\tesla}$), after combining all approaches for signal improvement  described in the text. }
		\begin{tabular}{ c  c  c  c  c  c  c }
		
        Particle size  & [P1]  & [NV] &  $^{\mathrm{NV}}T_2$  &  $^{\mathrm{NV}}T_1$   &  $T_{\mathrm{pol}}$  & enhanc. $\epsilon_{\mathrm{13C}}$ \\ \hline \hline
		 \SI{2}{\micro\meter}  &  $\SI{39(2)}{ppm}$  &  $\SI{8.2\pm 0.5}{ppm}$ & $\SI{2.6\pm0.1}{\micro\second}$ &  $\SI{3.82\pm0.04}{\milli\second}$ 
   &  \SI{34(2)}{\second}   & 1500 \\
		
		  \SI{100}{\nano\meter}  & $\SI{18(1)}{ppm}$ & $\SI{3.4\pm 0.2}{ppm}$   & $\SI{2.7\pm0.1}{\micro\second}$  &  $\SI{4.23\pm0.07}{\milli\second}$ &  \SI{32(5)}{\second}  & 940\\
		
	\end{tabular}
	\label{table:Diam_powder_properties}
\end{table}

The $^{13}$C hyperpolarized signal that results from combining the above-mentioned strategies is represented in Figure~\ref{fig:Hyper_signal}. The corresponding enhancements over the  $^{13}$C thermal signal ($\epsilon_{13\mathrm{C}}$) at the  polarization field   are $\epsilon_{13\mathrm{C}}^{\SI{2}{\micro\meter}}=1500$  
and $\epsilon_{13\mathrm{C}}^{\SI{100}{\nano\meter}}=940$. To our knowledge, this is the first demonstration of hyperpolarization using an optically pumped color center  at ambient conditions  in diamond  nanoparticles ($<\SI{200}{\nano\meter}$). It is worthwhile to remark the speed at which the signal in Figure~\ref{fig:Hyper_signal} is achieved, i.e.   running the DNP protocol for \SI{5}{\minute}, which is sufficient to reach the maximum of the buildup for both samples.  We measured $\SI{34(2)}{\second}$  and $\SI{32(5)}{\second}$ as buildup times ($T_{\mathrm{pol}}$) in the $\SI{2}{\micro\meter}$ and $\SI{100}{\nano\meter}$ particles, respectively. Relaxation times of $\SI{142\pm16}{\second}$ and $\SI{116\pm16}{\second}$ govern the decay of the hyperpolarized signal at \SI{1}{\tesla} magnetic field   (SI, sect.~\ref{SI:pol_buildup}). 
A significantly stronger   enhancement is observed in the \SI{2}{\micro\meter} sample. 
In the simplest description, for sample of similar polarization buildup times and neglecting effects of NV-NV interaction, we expect the enhancement factor to scale with the efficiency of NV initialization  and  their density, so that $\epsilon_{\mathrm{13C}} \propto (\Delta\rho)_{\mathrm{NV}}   \cdot  \mathrm{[NV]} $. Thus, from the NV concentrations (Table~\ref{table:Diam_powder_properties}),  and from the determined $(\Delta\rho)_{\mathrm{NV}}$, 
one would expect $\epsilon_{13\mathrm{C}}^{\SI{2}{\micro\meter}}/\epsilon_{13\mathrm{C}}^{\SI{100}{\nano\meter}} = \frac{0.024}{0.032}\cdot\frac{\SI{8.2}{ppm}}{\SI{3.4}{ppm}}= 1.8$, which  is close to what we observe, i.e.,   $\frac{1500}{940}=1.6$. To explain the small  discrepancy between $1.8$ and $1.6$, we would like to point to a possible role of NV-NV interactions, which should result in faster NV decoherence in the case of dense NV ensembles during the application of PulsePol.
 Although this is outside of the scope of the present work,  it might be beneficial, when working with extremely dense NV ensembles,  to apply sequence modifications for  refocussing the dipole-dipole interaction between the different NV.~\cite{tratzmiller2022thesis}

Our findings allow discussing the spatial distribution of the  polarized $^{13}$C spins,  
which is of fundamental interest. Considering the natural isotopic content (1.1\%)  of $^{13}$C in our samples,  the spread of polarization  by spin diffusion  is predicted to follow the coefficient   $D=\SI{6.7e-15}{\centi\meter\square\per\second}$~\cite{parker2019} . Taking into account  our buildup times,  $T_{\mathrm{pol}}\sim\SI{30}{\second}$, we obtain as    diffusion length   $l  \sim  \sqrt{D T_{\mathrm{pol}}} = \SI{4}{\nano\meter}$.  For randomly distributed defects, the nearest-neighbour distance is $r_{\mathrm{NN}} = 0.55 n^{-1/3}$, where $n$ is defect density. We calculate that the NV concentrations in Table~\ref{table:Diam_powder_properties} correspond to  
comparable distances, $r_{\mathrm{NN}} = \SI{4.9}{\nano\meter}$ and $r_{\mathrm{NN}} = \SI{6.5}{\nano\meter}$, for the  \SI{2}{\micro\meter} and   \SI{100}{\nano\meter} samples, respectively.  In our samples, spin diffusion therefore leads the polarization to reach most parts   
of the diamond lattice between the different NV, which is in line with conclusions  drawn previously in the context of single crystal experiments.~\cite{ajoy2019}

Although our efforts were focused on NV in the diamond material, our approach can be extended to centers with optical pumping capability in different host lattices. Spectral broadening induced by the anisotropy, that we observe for  NV, will occur similarly for other known defects, such as  pentacene in napthtalene crystals, or the divacancy in silicon carbide ($S=1$). Our results highlight the interest of using the PulsePol sequence, that addresses  all the centers in a given spectral range in parallel, while performing sample rotation  to progressively bring into resonance centers that are initially unaddressed.  Using PulsePol is also favorable for optimizing sequence timings, as illustrated by the small fraction of time it takes in the full polarization cycle. To obtain the signal in Figure~\ref{fig:Hyper_signal}, the sequence in Figure~\ref{fig:hp_protocol}b was used with a duration $M\tau\approx\SI{50}{\micro\second}$  (for $M=68$), which is much shorter than the laser initialization pulse ($\SI{400}{\micro\second}$), the repetition interval ($\tau_{\mathrm{SRT}}=\SI{1.5}{\milli\second}$) and the decay of NV polarization ($^{\mathrm{NV}}{T_1}>$ ms). 
There exists intervals, both between the pulses  and after, where the PulsePol sequence could be applied  at different carrier frequencies. In future developments, this could be considered for improving our bandwidth ($\Delta_{\mathrm{pol}}$)  and thus  enhancing further the hyperpolarized signal (SI, sect.~\ref{SI:increase_deltapol}). Frequency sweeps, in contrast,  typically address the centers slowly and sequentially, which can make the optimization of sequence timings more difficult and constrained.

Let us consider from the enhancement factors in Figure~\ref{fig:Hyper_signal} the absolute nuclear polarization, defined as $p_{13\mathrm{C}}=\frac{{N_{\uparrow}-N_{\downarrow}}}{{N_{\uparrow}+N_{\downarrow}}}$ 
where $N_{\uparrow}, N_{\downarrow}$ are the populations of the two  Zeeman spin states of  $^{13}$C. The value of $p_{13\mathrm{C}}$ constitutes a figure of merit that is independent on the detection field. 
Applying the enhancement values observed in Figure~\ref{fig:Hyper_signal} to the Boltzmann population, $p_{13\mathrm{C},\mathrm{thermal}}\approx \frac{\hbar\omega_I}{2 k_B T} = \SI{2.52e-7}{}$ (room temperature, \SI{0.287}{\tesla}), we obtain $0.038\%$  and $0.024\%$ as $p_{13\mathrm{C}}$  for the \SI{2}{\micro\meter} and \SI{100}{\nano\meter} powder samples, respectively. 
We discuss this outcome in the light of the work from King \emph{et al.}~\cite{king2015}, who employed single crystal diamond with  similar bulk material properties as our powders (electron-irradiated type Ib diamond). In that work, after performing DNP at  \SI{0.42}{\tesla} magnetic field, the value of 6\% of absolute $^{13}$C  polarization was observed, which exceeds by a factor of 160   our best result (\SI{2}{\micro\meter} particles). We  attribute this discrepancy to the combination of several factors as follows. First, King \emph{et al.} performed NV-$^{13}$C  polarization transfer using NV aligned to the magnetic field in single crystal diamond, which represents the ideal conditions for generating high electron spin polarization by optical pumping. Although the precise enhancement of NV signal  by illumination was not reported, it is common to reach, under such conditions, 
a population difference between the optically pumped ($m_S=0$) and any of the optically depleted ($m_S=\pm1$) states that obeys $(\Delta\rho)_{\mathrm{NV}}>30\%$,\cite{drake2015} i.e., improved more than $10$-fold compared to our powders. We note that, besides the absence of light scattering, the suppression of spin state mixing for this NV alignment also  likely plays a role in the improvement on $(\Delta\rho)_{\mathrm{NV}}$ (which accounts for a factor of about two, according to the  model fitting our powder spectrum, described in SI, sect.~\ref{SI:illumination:spectrum}).   Second,  NV in single crystal belong to four  homogeneously  oriented subsets. This enables working conveniently with a fourth of all NV ($25\%$), which yields in comparison to our powder another factor  $25/4$ ($\approx 6$). Thus, it is not surprising to observe  $>60$ increases in the enhancement factor in single crystal, compared to powder in the present conditions of our work.

From these considerations, one can suggest possible ways to increase the $^{13}$C enhancement further by adjusting some  experimental parameters. We expect that  combining the the application of the sequence at different frequencies (multiplexing) and pulse shaping   would enable increasing the width of the spectral range $\Delta_{\mathrm{pol}}$ to  \SI{100}{\mega\hertz} , which corresponds to about 16\% of NV (SI, sect.  \ref{SI:increase_deltapol}). 
Better initialization of NV could be achieved through the use of higher laser intensities, however, 
in our experimental configuration, the laser intensity that can be applied was  limited by sample heating. Temperature elevation in our micro- and nanodiamonds  was estimated from resonance shift in the NV EPR spectrum,  to be  \SI{51}{\kelvin} for the \SI{2}{\micro\meter}  particles and \SI{24}{\kelvin} for the \SI{100}{\nano\meter} particles (SI, section~\ref{SI:illumination:heating}).  
We employed, for both samples, a configuration in which the diamond particles are densely packed, which favors localized heating.  Better heat dissipation would be attained using an illumination geometry  that allows also for the sample to be   sparsely dispersed,
in a gel or a liquid (possibly still using  microstructured waveguides  to counter light scattering).
Although the realisation might present some challenges, this could be combined with some designs  already employed  for efficient cooling.~\cite{ajoy2022}

Dispersion in a fluid might promote particle reorientation by thermal motion, which  in fact  has been a proposed  alternative to mechanically rotating the  sample during the application of the polarization sequence.~\cite{chen2015}   However,  the fluid viscosity, which determines the tumbling rate for nanodiamonds of a given size, might need to be carefully controlled as the reorientation dynamics should not make the nanodiamonds depart strongly from their orientation while NV  are being driven by microwave. 
For a particle with  $r=\SI{80}{\nano\meter}$ as hydrodynamic radius, using the Stokes-Einstein-Debye relation for the rotational diffusion coefficient, $D_{\mathrm{r}}=k_\mathrm{B} T/8\pi\eta r^3$ in glycerol of dynamic viscosity $\eta = \SI{1.4}{\pascal\second}$ at room temperature ($T=\SI{293}{\kelvin}$), one has $D_{\mathrm{r}}=\SI{0.22}{\per\second}$.   In the conditions of our experiment, the region of efficient polarization transfer  lies within a $90 \pm \SI{2.5}{\degree}$ angle range from the magnetic field direction. Considering the RMS diffusion angle  $\delta = \sqrt{2 D_{\mathrm{r}} t}$, thermal motion will drive the nanodiamond out of this orientation range after a time $t\approx \SI{4.3}{\milli\second}$.  The `residence time' is longer by two orders of magnitude than the duration of application of PulsePol in our protocol ($M\tau \approx \SI{50}{\micro\second}$), thus microwave driving with PulsePol  will remain effective. The full polarization transfer cycle including initialization, microwave driving and waiting time, takes \SI{1.5}{\milli\second} and therefore could be applied 3 times, on average, before the nanodiamond moves to a less favorable orientation. This number of repetitions lies   below what we established by mechanical rotation  ($>  130$ polarization cycles, corresponding to $>\SI{200}{\milli\second}$), however, the viscosity of the fluid could be varied, e.g. by moderate cooling, to make the number of repetitions comparable. Our findings suggest indeed  that \emph{slow}  reorientation, that ensures  most nanodiamonds enter the bandwidth during the nuclear $T_1$ time, is already sufficient to provide a significant improvement in the hyperpolarized signal.


Turning to the polarization decay time ($^{^{13}\mathrm{C}}T_1$), it can be seen from the above-reported values that  the decay in our samples occurs must faster than what has been reported for ultrapure single crystals,~\cite{reynhardt2001} which we attribute to the high concentration of paramagnetic species (P1, NV and others), providing  relaxation channels via flip-flops, or via their own $T_1$ process.~\cite{ajoy2019} We now discuss consequences in terms of applications and possible routes for material development.

Despite being low by the standards of ultrapure diamond, the  $^{13}$C $T_1$ times we observe are expected to be sufficient for some applications.
A possible follow-up experiment to our work would consist in using ``core-shell'' particles, in which it is possible to obtain comparable $T_1$ times for a highly ($\sim100\%$ $^{13}$C) isotopically enriched shell, overgrown on naturally abundant nanodiamonds containing NV,  similar to the material  used in the current work.~\cite{mindarava2021} As originally reported, this would enable mediating spin polarization transferred from NV  in the ``core'', towards the surface, using spin diffusion in the enriched shell.  This opens exciting perspectives of applications in which external substances could be hyperpolarized. 
This could apply to solids or very  viscous liquids, as polarization transfer could be achieved by  cross-polarization sequences. In conventional liquids, it could occur via the nuclear Overhauser effect, although the optimization of  polarization transfer might require  considering specific dynamics of the fluid at the surface of the nanoparticles.~\cite{matsumoto2022} Besides, provided NV initialization and microwave driving can repeatedly  be  applied, such nanodiamonds would constitute a  replenishable source of polarization. In particular, this could be of direct interest for  NMR detected \emph{in-situ}, as there, it is sufficient to have nuclear $T_1$ times of the order of one minute (or even slightly lower)  both in the diamond nanoparticle and in the external  substance.~\cite{matsumoto2022}

For another envisioned application, using nanodiamonds in MRI as tracers, it must be kept in mind that the $^{13}$C polarization must survive while the nanodiamonds  diffuse to the location of interest (e.g.,  tumor) which can be long.~\cite{winter2022}
This points to a possible  limitation of the material employed in the present study, for which the $^{^{13}\mathrm{C}}T_1$ restricts the time before  
detection to a few minutes. 
An approach for improving the nuclear $T_1$ consists in performing electron-irradiation on a sample with low nitrogen content, with the rationale of keeping the overall content of paramagnetic spins low.  This was demonstrated to provide hour-long lifetimes in single crystals while still allowing efficient  $^{13}$C hyperpolarization. \cite{scheuer2016} 
Recently, long coherence for NV  under Hahn echo and dynamical decoupling  were observed in milled chemical vapor deposition diamond, suggesting the possibility of producing $\sim$\SI{200}{\nano\meter} particles with low magnetic noise levels, comparable to that in high-quality bulk material \cite{wood2022}. 
In smaller particles (\SI{100}{\nano\meter} and below), however, one might need to consider as well dangling bonds, or impurities in the diamond which are thought to be stabilized as a co-product of surface irregularities.~\cite{chou2023,peng2019} The desired reduction in the number of the surface-related defects  in particles of \SI{100}{\nano\meter} and below might require the combination of material engineering for producing  particles of near-spherical shape with a smooth surface,  and approaches for reduction or passivation of possibly remaining defects.

\section*{Conclusion}

Combining improvements on the material and  in the dynamical nuclear polarization protocol, we demonstrated the hyperpolarization of $^{13}$C in diamond particles down to \SI{100}{\nano\meter} size, using the optically pumped NV center as polarization source. In the nanoparticles, the impact of the magnetic noise  from surface sources is reduced  by combining  treatments of air-oxidation and triacid cleaning, ensuring  a long  $T_1$ for the $^{13}$C spins. For improved NV initialization, the light scattering by the nanoparticles is mitigated by performing illumination using microstructured waveguides.  To reduce the impact of the spectral broadening of NV in powder samples, the PulsePol sequence is optimized, and slow sample rotation ($25$ \degree/s) is performed. In contrast with protocols previously used for nanodiamonds ($<\SI{200}{\nano\meter}$), our approach does not require cryogenic temperatures. The current results   establish NV-containing nanodiamonds as a platform for hyperpolarization based on $^{13}$C spins at ambient conditions, promising future  applications in the fields of sensitive nuclear magnetic resonance  or MRI.

\clearpage
\pagebreak
\section*{Experimental Section}

Diamond powders with  approximately  $\SI{2}{\micro\meter}$  and $\SI{100}{\nano\meter}$ median particle size (Pureon AG, type Ib, HPHT, MSY1.5-2.5 and MSY 0-0.2) were used as starting material. To form NV, the samples were electron-irradiated with a dose of $3 \times 10^{18}$ cm$^{-2}$ in a 10~MeV electron accelerator operating under permanent argon flow. The temperature during the irradiation was maintained at 800 \textdegree C by controlling the dose per pulse and repetition frequency of the accelerator.  To clean the surface of diamond particles  from graphitic residues, the air oxidation (AO) treatment consists in annealing the samples at 620 \textdegree C for 5 hours in air.\cite{mindarava2020_conv}
For removing metallic residues,  the tri-acid cleaning (TAC) treatment consists in exposition to a high-concentration three-acid mixture of   HNO$_3$(40\%), HClO$_4$(70\%), H$_2$SO$_4$(90\%) in 1:1:1 proportions, under conditions of  $\SI{200}{\degreeCelsius}$ temperature and  \SI{6}{bar} pressure, for approximately 2 hours. To prepare samples for the hyperpolarization experiment, a mass of   \SI{12}{\milli\gram} was used for the $\SI{2}{\micro\meter}$ powder. For the $\SI{100}{\nano\meter}$ sample,  \SI{6}{\milli\gram} was used.

The longitudinal relaxation times of $^{13}$C nuclear spins shown in Figure~\ref{fig:13C_NV_T1}a  were obtained using a 300MHz WB NMR system at room temperature and  $B=\SI{7.05}{\tesla}$ magnetic field, equipped with an Avance III console (Bruker Biospin GmbH).

 Characterization of paramagnetic defect concentrations and their spin properties, as well as  the implementation of polarization transfer from NV to $^{13}$C nuclear spins, were made with the X-band  Bruker Elexsys E580 spectrometer (9-10 GHz). The  ER-4122-SHQE resonator was used for spin-counting using continous wave (CW) EPR. The pulsed FlexLine resonator ER-4118X-MD5 with the traveling wave tube (TWT) amplifier from Applied System Engineering, Inc. (Model no. 117) served to characterize the NV $T_1$ and $T_2$ times. 
Nuclear hyperpolarization  was performed using the pulsed-ENDOR EN-4118X-MD4 resonator. To address the duty cycle limitations of the TWT amplifier, a 40 Watt CW solid-state microwave amplifier from AR Deutschland GmbH (40S6G18B 6-18 GHz) was employed, replacing the TWT. The sequences shown in Figure~\ref{fig:hp_protocol} were implemented with the Bruker SpinJet AWG.
To perform \emph{in situ} detection of the hyperpolarized $^{13}$C NMR signal, we used the RF coil of the ENDOR resonator. A tank circuit was added outside the resonator for tuning and matching, allowing for detection of NMR signal around the resonance frequency of $^{13}$C spins at \SI{1}{\tesla} ($\nu_{\mathrm{13C}}\sim\SI{10.7}{\mega\hertz}$). The NMR signal is detected using a KEA$^2$ console (Magritek GmbH) 
using a 500W RF amplifier from (BT00500-ALPHA-SA from Tomco Technologies) for pulse amplification. 
 To obtain the data represented in  Figure~\ref{fig:NVproportion} and~\ref{fig:Hyper_signal}, the hyperpolarized signal was accumulated  5 times. Acquisition of the $^1$H NMR signal from a water sample of known volume was used as an  intensity  reference in order to determine the $^{13}$C signal enhancement values in Figure~\ref{fig:Hyper_signal} (see SI, sect.~\ref{SI:enhancement_thermal}, for details on the enhancement determination). 
 

For illumination, a \SI{3}{\watt} \SI{532}{ \nano\meter} Diode-Pumped Solid State (DPSS) laser from Laserglow Technologies  was used and pulsed using TTL modulation. A \SI{1.5}{\milli\meter}  multimode fiber delivers light onto the waveguide structure as shown in SI, Figure~\ref{fig:SI_brush}.
Due to losses, the power at the output of fiber is \SI{1.6}{\watt}. The laser pulse length was set to \SI{700}{\micro\second} when measuring spectra as in Figure~\ref{fig:optic_polar},  and to \SI{400}{\micro\second} in the hyperpolarization experiment . Considering the repetition time of \SI{1.5}{\milli\second}, the  \SI{400}{\micro\second} correspond to an average laser power of \SI{420}{\milli\watt}. At this power, the illumination induces some heating of the samples, which was evaluated to be  \SI{51}{\kelvin} and \SI{24}{\kelvin} for the \SI{2}{\micro\meter} and \SI{100}{\nano\meter} respectively (SI, section~\ref{SI:illumination:heating}). 

The X-ray photoelectron spectroscopy (XPS) acquisitions described  in SI were performed using a Physical Electronics instrument (PHI 5800). The equipment was equipped with a hemispherical electron analyzer, a monochromatic Al K$\alpha$ X-ray source (\SI{1486.6}{\electronvolt}), and a flood gun to prevent sample charging. The  spectra were recorded with a pass energy of \SI{93.9}{\electronvolt} and  \SI{29.35}{\electronvolt} for the survey and the high-resolution spectrum, respectively. The XPS spectra were calibrated using the C1s peak as a reference, which corresponds to a binding energy of  \SI{284.7}{\electronvolt}.


\section*{Acknowledgements}

We acknowledge Junichi Isoya for providing the single crystal diamond used for pre-testing the  polarization sequences and Joachim Bansmann for assistance with XPS acquisition. 

F. J. and M. B. P. acknowledge support via the ERC Synergy Grant HyperQ (Grant no. 856432) and the EU  via the  project QuMicro (Grant No. 01046911). F. J. acknowledges support from the EU via the project FLORIN (Grant no. 101086142), QUANTERA via the project “Microfluidics Quantum Diamond Sensor”, Carl Zeiss Foundation via IQST,  the projects QPHOTON  and Ultrasens-Vir, and the DFG. M. B. P. acknowledges support via  the EU project SPINUS (Grant no. 101135699), the EU project C-QuENS (Grant no. 101135359) and the German Federal Ministry of Education and Research (BMBF) under the funding program quantum technologies - from basic research to market via the project QuE-MRT (FKZ: 13N16447) and via the project QSens: Quantensensoren für die biomedizinische Diagnostik (QMED) (Grant no. 03ZU1110FF).

\section*{Conflict of Interest}

The authors have no conflict of interest to declare.

\section*{Author Contributions}

R. B. and Y. M. conducted the experiments and analyzed the data. M. K. developed the composite pulses and performed numerical simulations of the robustness of PulsePol. R.  B. conducted complementary sequence  simulations  (at specific NV orientations).  
A. M. performed the automation of the complete DNP experiment. F. G., S. N. and A. K. contributed to initial simulations of light scattering, and performed the design of the waveguide array structure, F. G. optimized the 3d-printing process.  W. K. and C. L. performed the electron-irradiation and annealing and air-oxidation. C. J. provided the base material and information on the size distribution. R. B., Y. M. and M. K wrote the manuscript with input from all others.  M. B. P., A. K. and F. J. conceived the study. 

\section*{Data Availability}

The data that support the findings of this study are available from the corresponding authors upon reasonable request.




\bibliography{Hyper_article} 

\end{document}


\maketitle
 
\tableofcontents

\section{Material}
\label{SI:Material}

\begin{figure}[htp]
\centering
\includegraphics[width=1\linewidth]{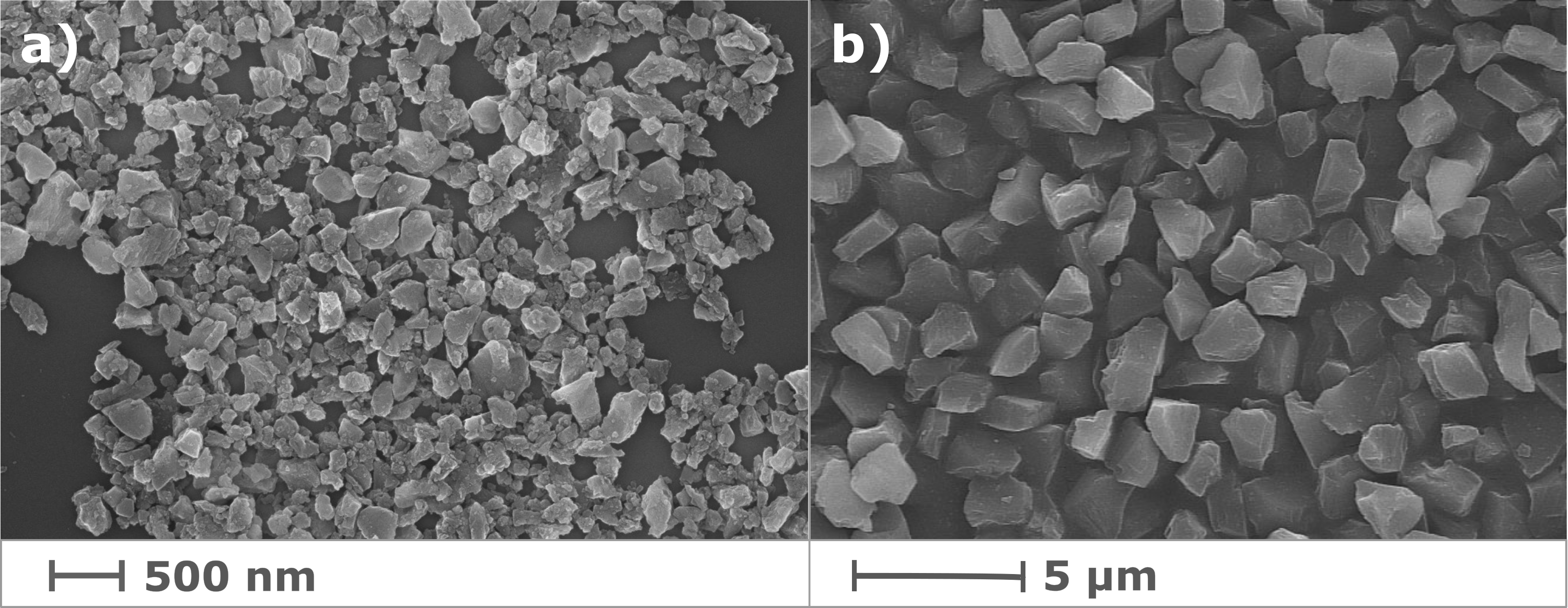}
\caption{Scanning electron microscopy images of a) \SI{100}{\nano\meter} and b) \SI{2}{\micro\meter} diamond powders, respectively Pureon AG, MSY0-0.2 and  MSY1.5-2.5.}
\label{fig:SI_SEM}
\end{figure}

The commercial samples  are Pureon AG, type Ib, HPHT,  MSY1.5-2.5 and  Pureon AG, type Ib, HPHT,  MSY0-0.2 having  about \SI{2}{\micro\meter} and \SI{100}{\nano\meter} median size, respectively. Scanning electron microscopy images  are shown in Fig.~\ref{fig:SI_SEM}. The size distribution parameters for the batches employed in this work are given in  Table~\ref{table:SI_particles_median_sigma}. \vspace{20pt}

 \begin{table}[h!]
	\centering
    \begin{tabular}{ccc}
   
    \rowcolor[HTML]{FFFFFF} 
    sample                         & median size (\SI{}{\micro\meter})                                                               & standard deviation  (\SI{}{\micro\meter})                                                       \\ \hline \hline
    Pureon AG, MSY1.5-2.5, type Ib & 1.95 & 0.31  \\ 
    Pureon AG, MSY0-0.2, type Ib   & 0.115 & 0.028 \\
    \end{tabular}
    \caption{Size distribution properties for the batches of diamond powder samples used in the present work.}	\label{table:SI_particles_median_sigma}
\end{table}

\clearpage
\pagebreak

\section{Surface treatment    characterization}

\subsection{Monitoring surface spin defects by EPR}
\label{SI:surface_epr}

\begin{figure}[htp]
\centering
\includegraphics[width=0.33\linewidth]{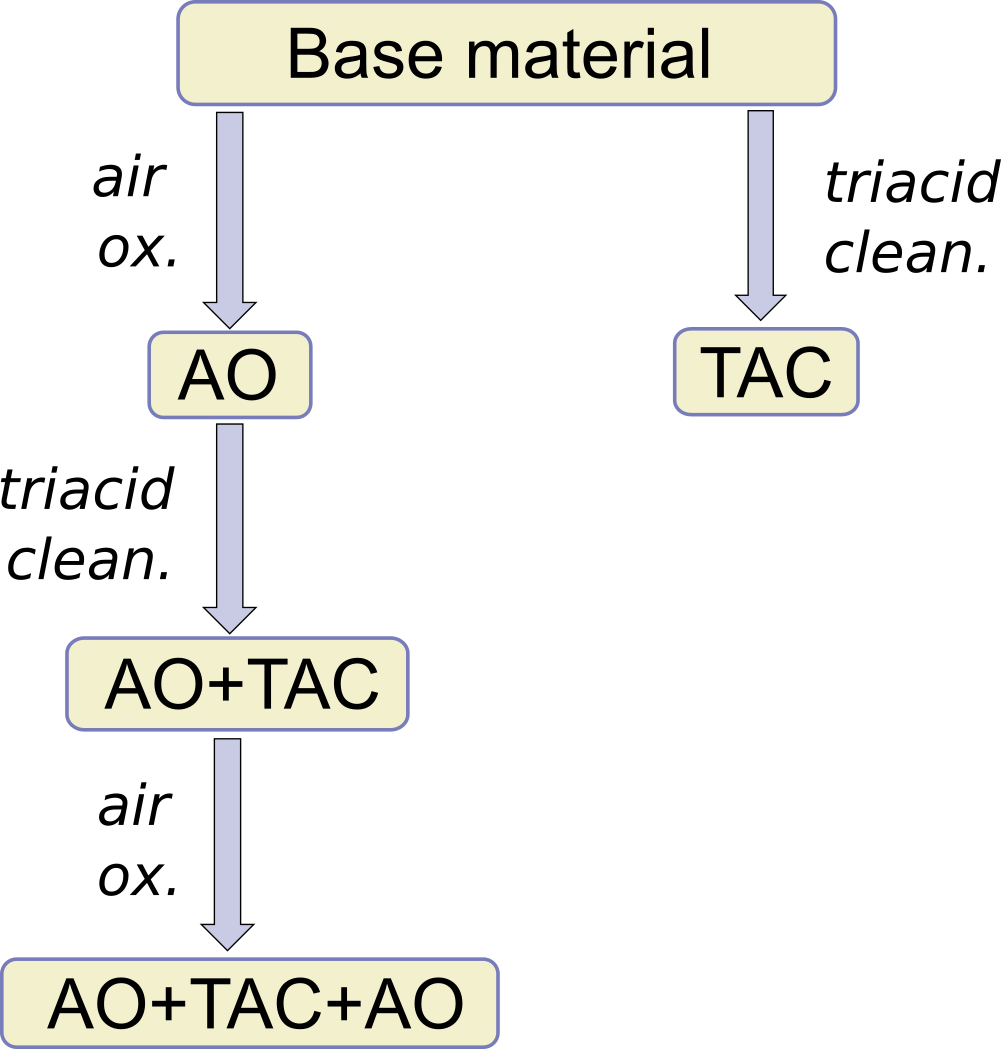}
\caption{Description of the surface treatment steps for the \SI{100}{\nano\meter} sample. }
\label{fig:SI_ao_tac_desc}
\end{figure}

EPR spectra were measured on the commercial  \SI{100}{\nano\meter} sample (Pureon AG MSY0-0.2  Type Ib, labelled `base material') and after the several surface treatment steps  summarized in Fig.~\ref{fig:SI_ao_tac_desc}.

\begin{figure}[htp]
\centering
\includegraphics[width=1\linewidth]{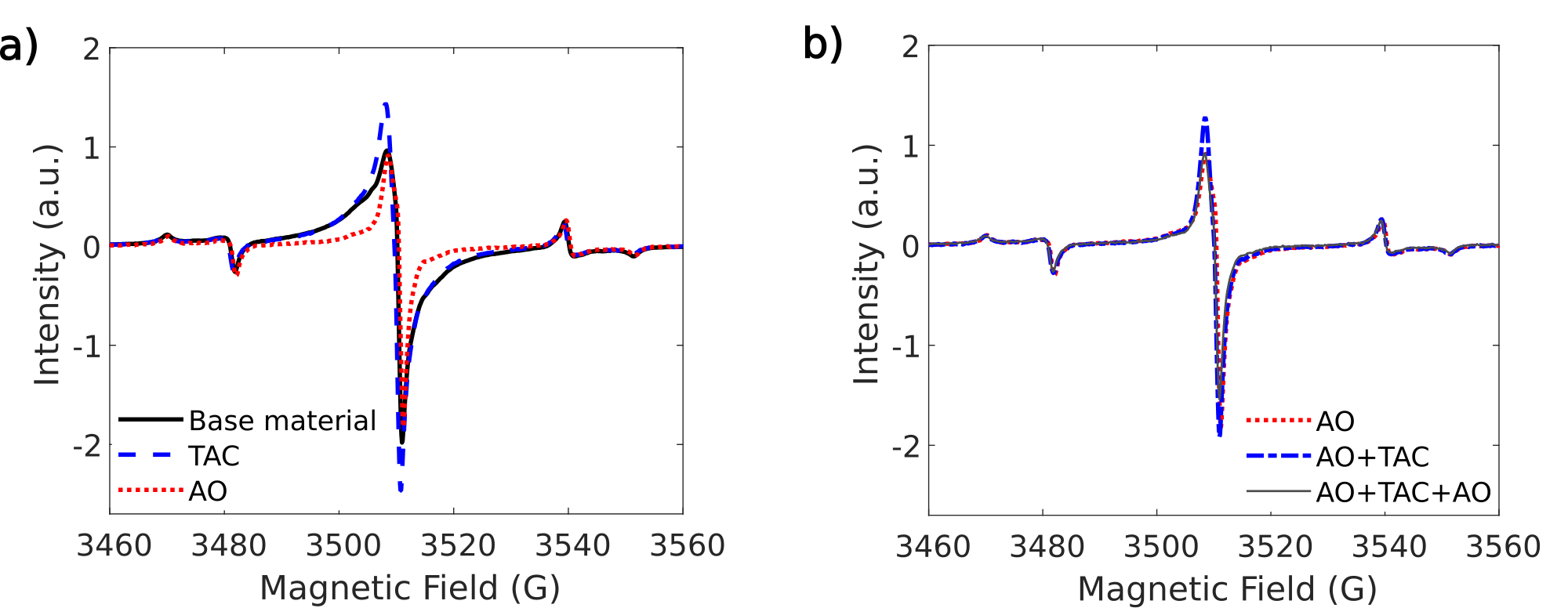}
\caption{CW EPR spectra in a narrow spectral range (\SI{100}{G}) around the $g=2$ resonance, at low microwave power ($P_{\mathrm{MW}}=\SI{0.150}{\micro\watt}$). The acquisition frequency is $\nu=\SI{9.84}{\giga\hertz}$. The spectra are normalized to the sample mass and $Q$-factor. a) Spectra showing the effect of the first treatment steps (AO or TAC). 
The satellites at \SI{3480}{G} and \SI{3540}{G}  correspond to the hyperfine lines of P1 centers. These satellites are unaffected by surface treatments, as expected for bulk impurities. The signal in the center comprises both the P1 central line, and $g\sim2$ paramagnetic species. Although a  minor  change is observed with the triacid (TAC), a strong modification is observed following air-oxidation (AO), as the latter leads to the disappearance of one central component that we attribute to  the efficient removal of graphite and  associated dangling bonds. b) Effect of the successive  treatment combinations. A small change in the intensity of the central line after the AO+TAC step, however, the spectrum remains essentially unaffected.
}
\label{fig:SI_EPR_low_power_central}
\end{figure}

\begin{figure}[htp]
\centering
\includegraphics[width=0.7\linewidth]{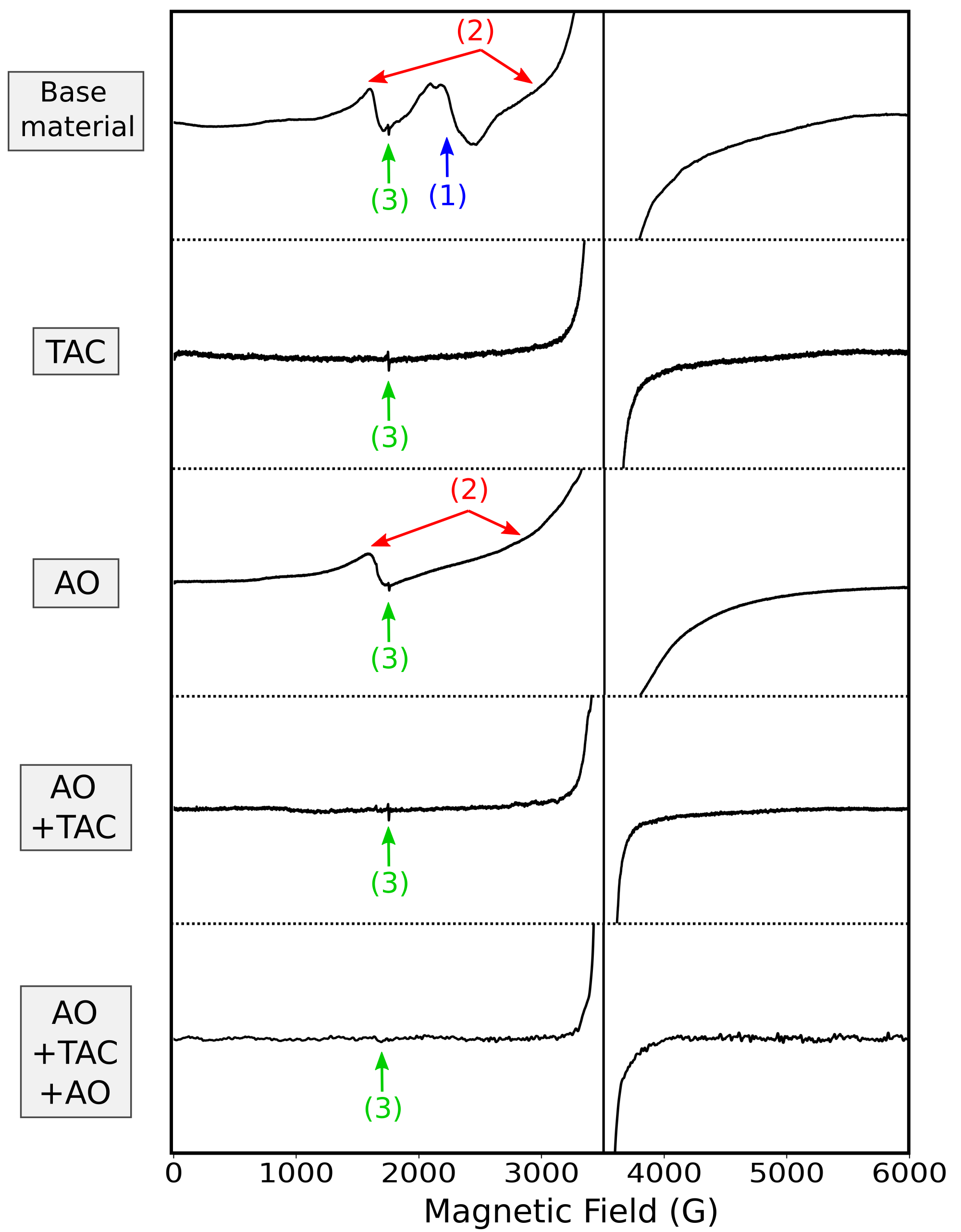}
\caption{CW EPR spectra in a broad spectral range (0-\SI{6000}{G}), at high microwave power ($P_{\mathrm{MW}}=\SI{23.7}{\milli\watt}$). The acquisition frequency is $\nu=\SI{9.84}{\giga\hertz}$ and the data is normalized to the sample mass and $Q$-factor. The spectra are also trimmed vertically in order to better visualize, specifically, the broad contributions. The first spectrum shows an impurity contribution (1) removed both by the AO and TAC treatment. Another contribution, labelled (2),  is removed by the TAC step only, and is associated to  paramagnetic species of iron (see text). The contribution (3), which appears in the half-field region (at $g\sim 4$),  might be associated with impurities of $S>1/2$ multiplicity within the nanodiamonds. After   the last treatment step  (AO+TAC+AO), the contribution (3) is likely still present, but not observed due to longer spectrum averaging and possible magnetic field or frequency drifts. }
\label{fig:SI_EPR_widerange}
\end{figure}

In this study, the starting material is the commercial sample, with no NV centers formed by irradiation and annealing. To characterize the surface treatments, it is useful to perform two  types of EPR measurements:

\begin{itemize}
\item A narrow scan around $g=2$ (in a $\sim \SI{100}{G}$ range) at low microwave power. This allows resolving bulk and subsurface impurities~\cite{yavkin2015}, as well as  dangling bonds with long relaxation times (e.g. in graphite). The corresponding acquisitions are shown in Fig.~\ref{fig:SI_EPR_low_power_central}.
\item An acquisition in a wide spectral range ($\SI{6000}{G}$),  at higher microwave power, in order  to resolve impurities with a broad spectra and short lifetimes.  The spectra are shown in Fig.~\ref{fig:SI_EPR_widerange}.
\end{itemize}

It is possible to observe  the disappearance of a $g\sim2$ contribution following the air-oxidation (AO) treatment (Fig.~\ref{fig:SI_EPR_low_power_central}a). We attribute this effect to the removal of dangling bonds in graphite covering the nanodiamonds. This is associated with a strong change in color of the nanoparticles, initially black, which turn bright. 
in the color of the nanoparticles, which are initially black and become bright after the treatment.

The wide-range, high-power spectrum in Fig.~\ref{fig:SI_EPR_widerange}. reveals also the presence of several broad contributions in the base material (covering a $\sim \SI{3000}{G}$ range). Although one contribution (labelled (1) in Fig.~\ref{fig:SI_EPR_widerange}) is removed both by the AO and TAC treatments, another important contribution (labelled (2) in Fig.~\ref{fig:SI_EPR_widerange}) persists after air-oxidation, but is removed  by the  triacid (TAC) treatment. Iron constitutes a likely contaminant, as it is present in the beads used in the manufacturing process, for milling the material down to the desired particle  size. In smaller nanodiamonds from the same manufacturer that underwent the AO treatment, the presence of iron is confirmed by XPS (sect.~\ref{SI:XPS}).  Consequently: 1) we attribute the broad EPR contribution that is removed by the triacid cleaning,  to iron-related spin impurities 2) we hypothesize that the fast fluctuations of these paramagnetic impurities  explain the   shortening effect on the $T_1$ of $^{13}$C and NV (detailed in sections \ref{SI:sec_13C_T1}  and \ref{SI:sec_NV_T1}, respectively).

\clearpage

\subsection{$^{13}$C $T_1$ relaxation}
\label{SI:sec_13C_T1}

\begin{figure}[htp]
\centering
\includegraphics[width=0.6\linewidth]{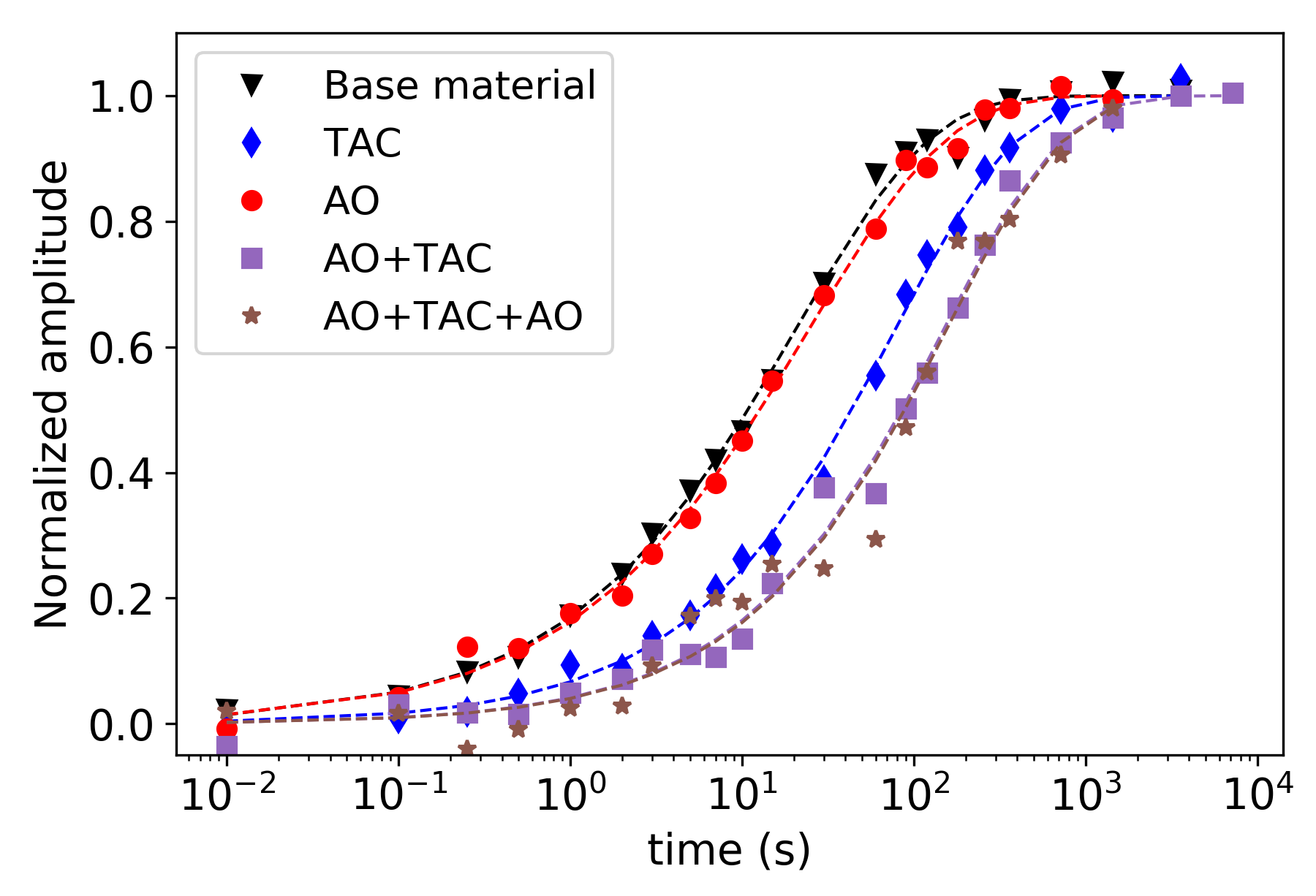}
\caption{$^{13}$C relaxation for the \SI{100}{\nano\meter} particles, at different stages of surface treatment (AO: air-oxidation, TAC: triacid-cleaning), acquired by saturation-recovery with conventional NMR at $B=\SI{7.05}{\tesla}$. Dashed lines are stretched exponential recovery fits.}
\label{fig:SI_100nm_T1}
\end{figure}

In Fig.~\ref{fig:SI_100nm_T1} are represented  $^{^{13}\mathrm{C}}T_1$ acquisitions performed with conventional NMR at $B=\SI{7.05}{\tesla}$, using the same batch of samples as in the EPR investigation in section~\ref{SI:surface_epr}.
Acquisition was made with  the NMR saturation-recovery sequence, which consists in observing the return to thermal polarization after a train of saturating pulses (here corresponding to 20 $\pi/2$ pulses spaced  by \SI{5}{\milli\second}). The experimental data points in Figure~\ref{fig:SI_100nm_T1} consist of the spectrum integral at each value of the time interval between saturation and readout. 
Stretched exponential fits $s(t)=A \left (1-\exp{-(\frac{t}{T_{\mathrm{1}}})^\beta} \right )$ are shown as dashed lines in Fig.~\ref{fig:SI_100nm_T1} and the fitted parameters are  
listed in Table~\ref{table:SI_100nm_T1}.

 \begin{table}[h!]
	\centering

	\begin{tabular}{|rcc|}
    \hline
    \rowcolor[HTML]{e6e6e6} 
    \SI{100}{\nano\meter} sample        & $^{^{13}\mathrm{C}}T_1$ (s) & stretch exponent $\beta$ \\ \hline 
    Base material & 21(1)                       & 0.55(3)                  \\ 
    TAC           & 80(4)                       & 0.61(3)                  \\ 
    AO            & 25(2)                       & 0.54(3)                  \\ 
    AO+TAC        & 152(13)                     & 0.63(4)                  \\ 
    AO+TAC+AO     & 158(37)                     & 0.63(9)                   \\  \hline           
    \end{tabular}
 	\caption{ Fitted parameters for the $^{13}$C $T_1$ relaxation of the  \SI{100}{\nano\meter} sample at different stages of surface treatment (samples are  labelled as in Fig.~\ref{fig:SI_100nm_T1}). }
	\label{table:SI_100nm_T1}
\end{table}

\subsection{NV $T_1$ relaxation}
\label{SI:sec_NV_T1}

The NV longitudinal relaxation times were measured on the  \SI{100}{\nano\meter} samples, that were as well electron-irradiated and annealed.  Here, we compare two treatments (AO, AO+TAC). The air-oxidation (AO) was done both before and after the electron irradiation and simultaneous annealing. Triacid cleaning (for the AO+TAC sample) was performed only after the second air-oxidation. Single exponential fits of the experimental data as shown in main text, Fig.~\ref{fig:13C_NV_T1}b provide the relaxation times  given in Table~\ref{table:SI_100nm_T1_NV}. 

In contrast to our other measurements, the magnetic field was set to the lowest at which intensity in the EPR spectrum could be detected considering the field region in main text, Fig.~\ref{fig:optic_polar}, that is, using $\nu = \SI{9.59}{\giga\hertz}$, $B=\SI{2403}{G}$. This allows being selective to NV centers that are aligned to the magnetic field, and avoid state-mixing induced by the misalignement.  The $^{\mathrm{NV}}T_1$ measured with this method reflects exclusively the rate of the single quantum transitions ($m_S=0$ to $\pm1$), and thus can be compared to the typical $^{\mathrm{NV}}T_1$ measurements reported in the literature. However, our determination is performed at higher magnetic field than in most cases of literature reports (using, e.g., zero-field). The increased magnetic field can lead to further decoupling of the NVs from certain noise sources, and thus favor a longer $T_1$.

 \begin{table}[h!]
	\centering

	\begin{tabular}{|rc|}
    \hline
    \rowcolor[HTML]{e6e6e6} 
    \SI{100}{\nano\meter} sample        & $^{\mathrm{NV}}T_1$ \\ \hline 
    AO  &  $\SI{2.20\pm0.03}{\milli\second}$                  \\ 
    AO + TAC    &  $\SI{4.23\pm0.07}{\milli\second}$              \\ 
    \hline           
    \end{tabular}
 	\caption{$^{\mathrm{NV}}T_1$ corresponding to the fits  shown in main text, Fig.~\ref{fig:13C_NV_T1}b.   }
	\label{table:SI_100nm_T1_NV}
\end{table}


\subsection{XPS on smaller nanoparticles ($\sim \SI{25}{\nano\meter}$)}
\label{SI:XPS}

\begin{figure}[htp]
\centering
\includegraphics[width=0.8\linewidth]{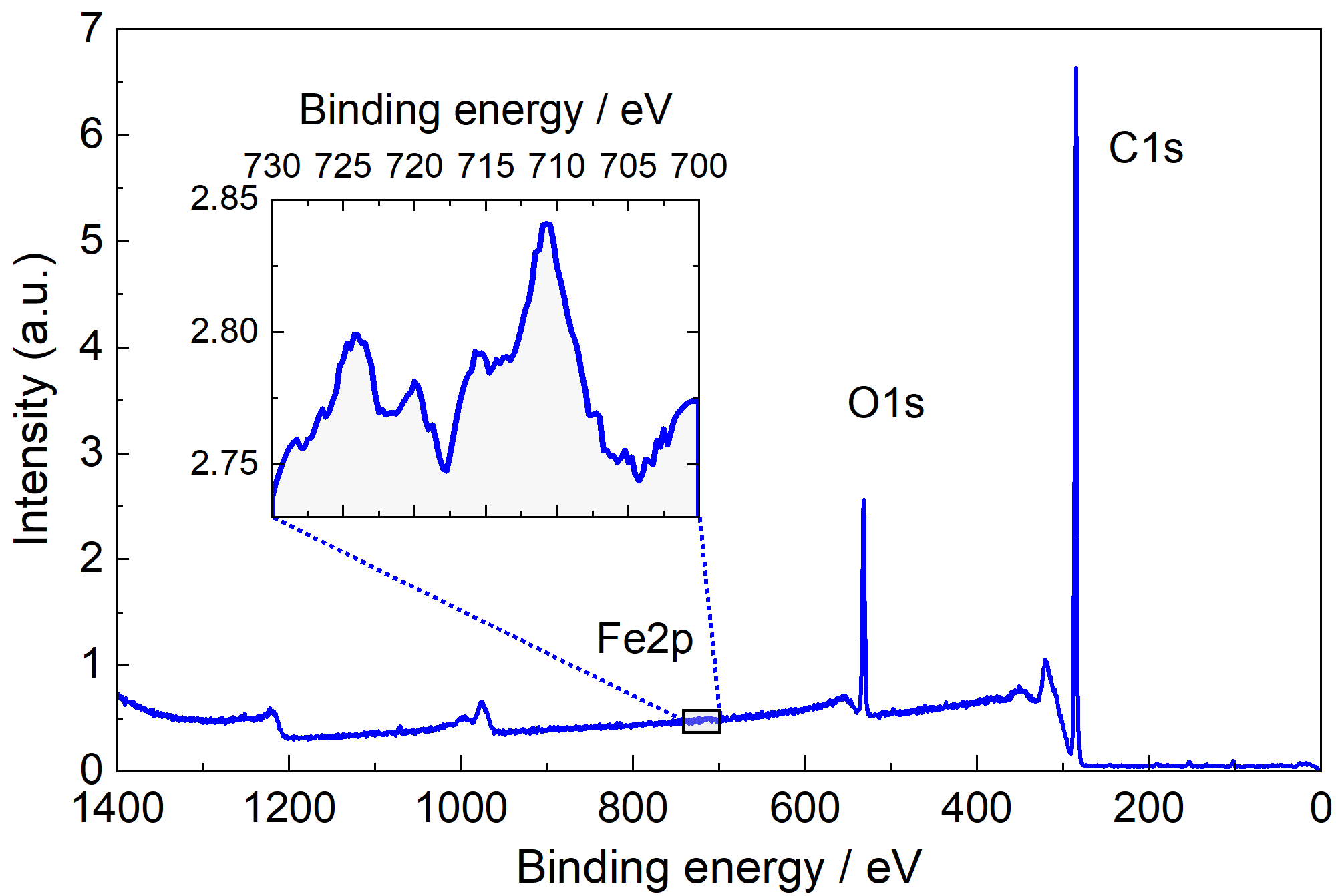}
\caption{X-ray photoemission spectroscopy spectrum of the sample Pureon AG, MSY0-0.5, type Ib, of \SI{25}{\nano\meter} median size, showing the carbon (C1s) and oxygen (O1s) peaks. Inset: acquisition in the [700,730] eV range, showing the signal characteristic of iron (Fe2p).  }
\label{fig:SI_XPS_25nm}
\end{figure}

X-ray photoemission spectrum on a sample of  \SI{25}{\nano\meter} median size is shown in Fig.~\ref{fig:SI_XPS_25nm}. The sample is from the same manufacturer  as the  \SI{100}{\nano\meter} \SI{2}{\micro\meter} samples  (Pureon AG), however contamination induced by the fabrication process (milling) is more likely to present due to the smaller size. 
The Fe2p peak of iron could be detected, as shown in inset of Fig.~\ref{fig:SI_XPS_25nm}. 
In the \SI{100}{\nano\meter} sample (AO treatment), no iron could be detected by XPS, which we attribute to the lower concentration at the surface of bigger particles.

\pagebreak

\pagebreak
\section{Sample illumination}
\label{SI:illumination}

\subsection{Illumination protocol}
\label{SI:illumination:protocol}

\begin{figure}[ht]
\centering
\includegraphics[width=0.7\linewidth]{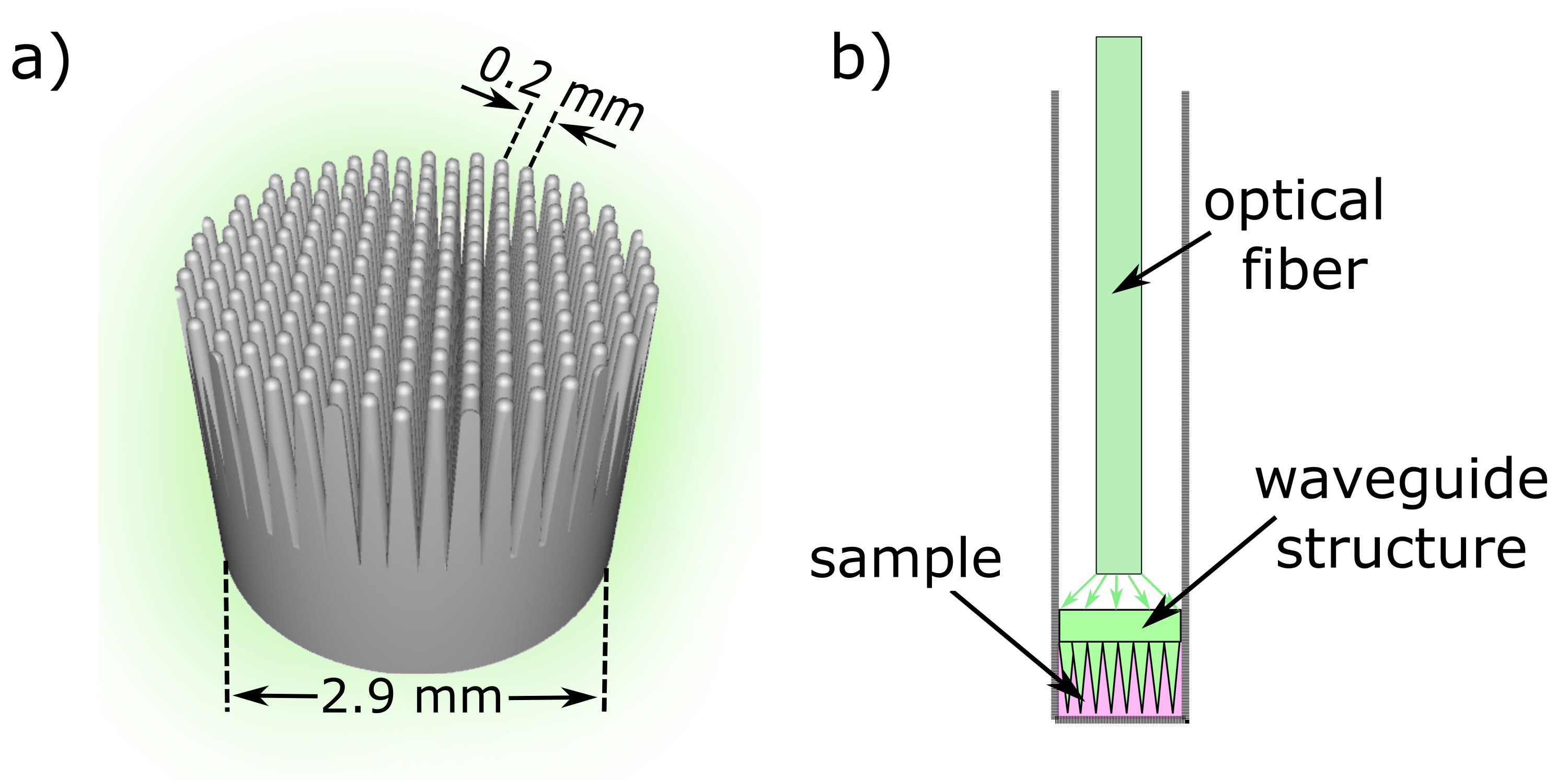}
\caption{a) 3D printed micro-photonic structure consisting of more than a hundred tips. b) Schematic illustration of the 3D structure inside the EPR tube, which provides optical access to the contained diamond powder.}
\label{fig:SI_brush}
\end{figure}

In contrast to single diamond crystals, the propagation of light inside of diamond powder of micro- and nanosize is hindered 
by strong light scattering, due to the high refractive index of diamond ($n=2.4$ at the wavelength of \SI{532}{\nano\meter}). In the context of hyperpolarization, 
there is a requirement to perform homogeneous optical pumping of large 
amounts of sample. A microstructured waveguide array was developed to ensure an efficient illumination of such quantities of 
diamond powder and circumvent the issue caused by light scattering. 
 
The waveguide arrays make it possible to directly access a larger volume of micro- or nanodiamonds with light, leading to a better polarization of NV centers upon illumination. 

The micro-pillar array was manufactured using a commercially available direct laser writing system, employing the 2-photon-polymerization technique (Photonic Professional GT+, Nanoscribe GmbH, Eggenstein-Leopoldshafen, Germany). We used the resin IP-S (Nanoscribe GmbH, Eggenstein-Leopoldshafen, Germany) due to its excellent optical properties in the VIS and 10x lens with NA=0.3 for fast prototyping progression. The structures were printed on a silicon wafer using the dip-in approach, where the resin acts as its own immersion medium. The slicing and hatching distances were chosen as 5~\textmu m and 1~\textmu m, respectively, which, although showing visible steps between layers, still produced good quality results. Uncured resin was dissolved and removed by submerging the structure in PGMEA and cleaning was done by a subsequent bath and rinsing with isopropyl alcohol during which UV curing was applied to ensure the complete polymerization of all components and to reduce shrinkage. The structure itself comprises a base with a diameter of 2.9~mm on which a hexagonal pattern of approximately 200 pillars, each around 1~mm in height, is placed. The pillars are truncated cones which are topped by a hemispherical cap (see Figure \ref{fig:SI_brush}). The distance between the revolution axes of neighboring pillars is \SI{200}{\micro\meter}.

In the sample preparation, the diamond powder was mixed with  ethylcinnamat (C$_{11}$H$_{12}$O$_{2}$), the latter, with a refractive index ($\lambda= \SI{532}{\nano\meter}$) $n=1.57$, has the following use for index matching  1) it reduces losses by internal reflection inside the structured array and 2) it mitigates further the effects of light scattering by the diamond particles.  The mixture of diamond powder and ethylcinnamat  (with a mass equal or lower to that of the diamond powder) provides a viscous fluid, which was placed in a sapphire tube of (\SI{4}{\milli\meter} OD, Situs Technical  GmbH, Wuppertal, Germany) and the waveguide structure placed  on top. As shown on Fig.~\ref{fig:SI_brush}b, an optical fiber (\SI{1.5}{\milli\meter} diameter) delivers the illumination.  In separate trials, glycerol was used as the index-matching solution. As an alternative, water could be used, if working below the melting point. While water is in general a convenient solvent for room temperature operation, it has the issue of its strong microwave absorption above the melting point. A specific resonator design or sample geometry would be required for working with water-containing samples at room temperature.


\clearpage
\subsection{Phenomenological description of the NV    spectrum in X-band EPR under optical pumping} 
\label{SI:illumination:spectrum}

\subsubsection{Redistribution of state populations} 

The EPR spectrum measured under illumination (as in main text, Fig.~\ref{fig:optic_polar}) shows the contribution  from NV centers in randomly oriented crystallites. Therefore, all possible NV center orientations must be considered in the spectrum description. 

Let us consider the case of a NV center aligned to the magnetic field. For this NV, the eigenstates, defined by the quantum number $m_s$, can be written as $\ket{-1},\ket{0},\ket{1}$. Illumination depletes equally the $\ket{\pm1}$ states, by population transfer to $\ket{0}$, so that the occupation probability of each state ($\rho_{m_S}$) differs from that in thermal equilibrium. It obeys

\begin{equation}
	\rho_{m_S} =  (1-p_{\mathrm{NV}}) \rho_{m_s, \mathrm{thermal}} + p_{\mathrm{NV}}\lvert\bra{m_s}\ket{0}\rvert^2,
	\label{eq:NVpolar_aligned}
\end{equation}

\noindent  where $p_{\mathrm{NV}}$ is the degree of initialization of the NV center.  The $\lvert\bra{m_s}\ket{0}\rvert^2$ term equals to 1 for $m_s=0$, and 0 otherwise. For $p_{\mathrm{NV}}=1$, the  probability of being into the $m_s = 0$ state is maximum (100\%).

When NV centers are misaligned with respect to the magnetic field,  $m_s$ no longer remains a good quantum number, due to state mixing. In this case, one can define new eigenstates $\ket{s_{i,\theta}}$, where $i = 1, 2, 3$ and $\theta$ is the angle between the NV axis and the orientation of the magnetic field. The misalignment-induced state mixing leads to different occupation probabilities compared to the aligned case. In our present phenomenological description, it is taken into account by performing the following replacement in Eq.~\ref{eq:NVpolar_aligned}:  

\begin{equation}
    \lvert\bra{m_s}\ket{0}\rvert^2 \rightarrow \lvert\bra{s_{i,\theta}}\ket{0}\rvert^2,
	\label{eq:NVoccupation_theta}
\end{equation}

\noindent so that instead of Eq.~\ref{eq:NVpolar_aligned}, the occupation probability ($\rho_{i,\theta}$) of NV center to $\ket{s_{i,\theta}}$ state can be expressed as

\begin{equation}
	\rho_{i,\theta} =  (1-p_{\mathrm{NV}}) \rho_{i,\theta, \mathrm{thermal}} + p_{\mathrm{NV}}\lvert\bra{s_{i,\theta}}\ket{0}\rvert^2.
	\label{eq:NVpolar}
\end{equation}

In EPR, the signal is summed over each transition and scales with the population difference between the two states involved in the transition. It follows from Eq.~\ref{eq:NVpolar} that the signal ($S_{\mathrm{EPR}}$), e.g. in the field-swept spectrum, can be predicted as

\begin{equation}
	S_{\mathrm{EPR}} =  (1-p_{\mathrm{NV}}) S_{\mathrm{EPR,thermal}}+ p_{\mathrm{NV}}S_{\mathrm{EPR,pol.}},
	\label{eq:NVpolar_spectrum_EPR}
\end{equation} 

\noindent  where $S_{\mathrm{EPR,thermal}}$ and  $S_{\mathrm{EPR, pol.}}$ correspond respectively to thermally polarized NVs, and  NVs with occupation probabilities given by  $\lvert\bra{s_{i,\theta}}\ket{0}\rvert^2$ (Eq.~\ref{eq:NVoccupation_theta}).

\begin{figure}[htp!]
\centering
\includegraphics[width=1\linewidth]{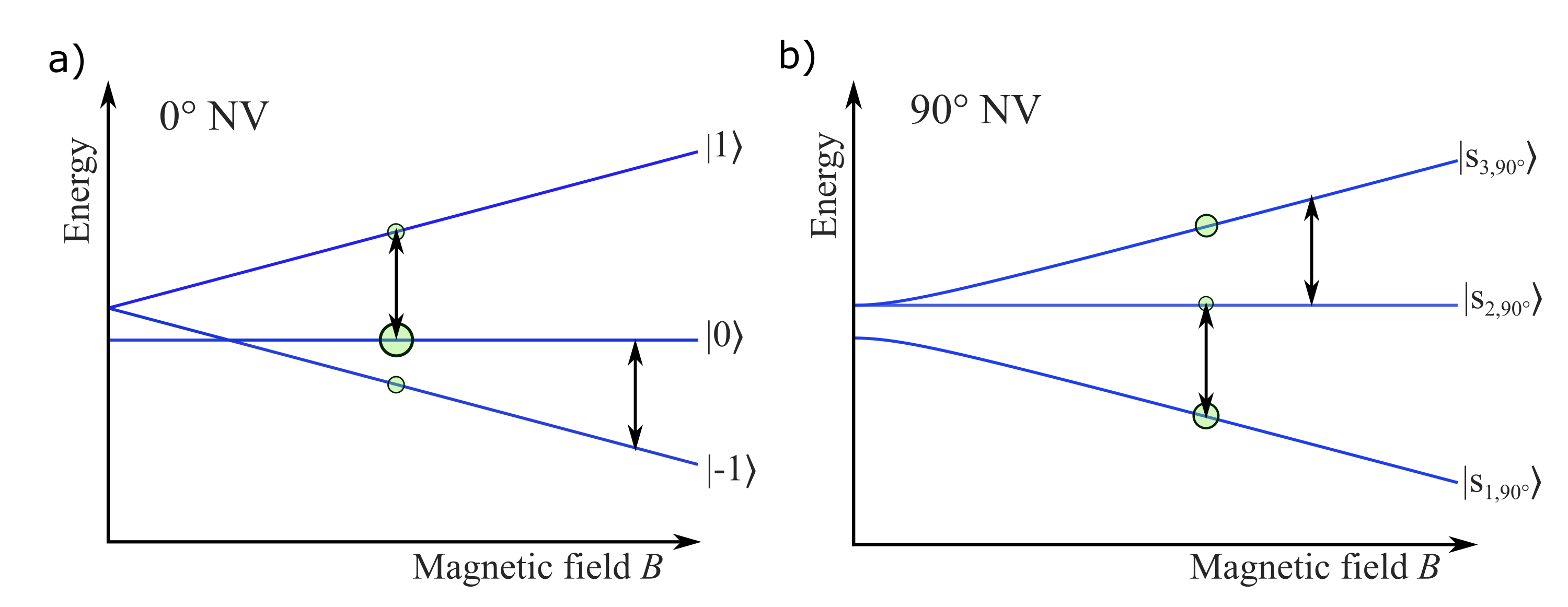}
\caption{Energy levels of an NV center with 
a) 0° and b) 90° orientation to the magnetic field. The lines represented in  a) result from direct diagonalization of the Hamiltonian. The ones in b) are obtained after the derivation in section  \ref{SI:sec_hamiltonian_diag_90deg} (Eq.~\ref{eq:E_NV90}). The arrows indicate the position of the transitions in a field-swept experiment (the microwave frequency, $\nu \approx \SI{9.6}{\giga\hertz}$). The diameter of the circles on the energy levels depicts the corresponding occupation probabilities under optical pumping, according to the description in section~\ref{SI:illumination:spectrum}, with the parameter describing the optical initialization of the NV center in Eq.~\ref{eq:NVpolar}, $p_{\mathrm{NV}}=0.3$.}
\label{SI_energy_levels}
\end{figure}

Remarkably, the occupation probabilities predicted by this equation are strongly orientation-dependent  \cite{drake2015}.
The energy level plots in Fig.~\ref{SI_energy_levels} illustrate the probabilities in the 0° and 90° case, calculated for $p_{\mathrm{NV}}=0.3$.

Although this description is of convenient use, we insist on its phenomenological character. 
 In reality, the mechanism responsible for the change in occupation probabilities under illumination involves the different rates in the intersystem crossing (ISC) from the $^3$E excited state~\cite{goldman2015}. A complete description of the EPR spectrum under illumination should thus likely consider the change in polarization dynamics induced by state mixing within the $^3$E excited state~\cite{tetienne2012}. Nevertheless, the good fit of our model to the experimental data, shows that the  phenomenological description already captures the main features of the spectral changes induced by illumination.

\subsubsection{Spectral fits}

\begin{figure}[ht]
\centering
\includegraphics[width=0.75\linewidth]{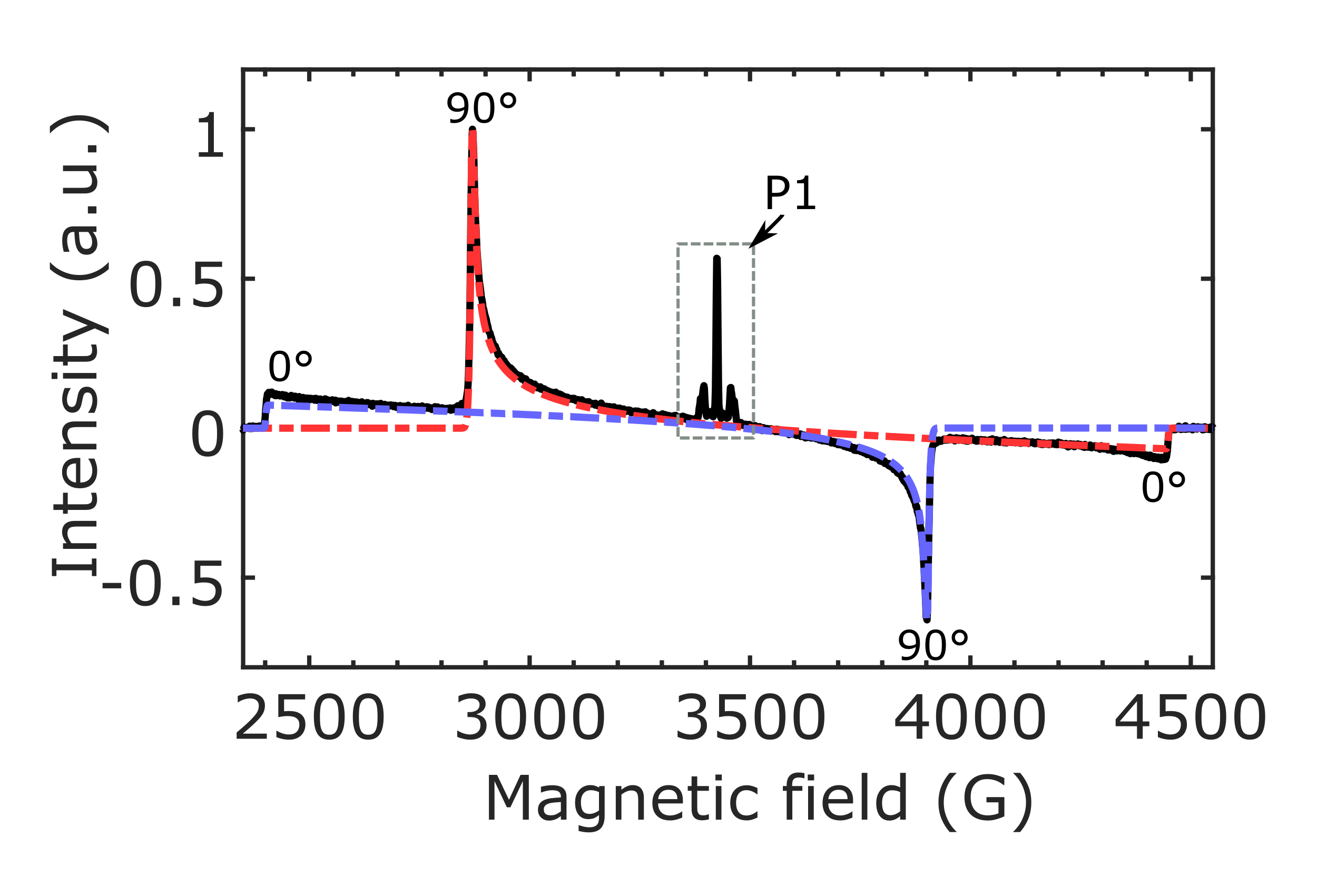}
\caption{NV EPR spectrum obtained after illumination with \SI{700}{\micro\second} laser pulses for \SI{100}{\nano\meter}-sized nanodiamonds with a sample mass of 12 mg (black solid). The mass was subsequently decreased for hyperpolarization experiments, to ensure better illuminations conditions. Contributions from the two separate transitions fitting the experimental spectrum  are represented in red and blue dashed curves, respectively. }
\label{fig_spectrum_MSY100_illum}
\end{figure}

Simulations of the NV spectra (main text, Fig.~1 and  Fig.~\ref{fig_spectrum_MSY100_illum})  were implemented using  the \texttt{pepper} in the EasySpin package in Matlab  \cite{stoll2006}.  The spin system is built considering the Hamiltonian in  Eq.~\ref{eq:Hamil_general}.
To obtain $S_{\mathrm{EPR,thermal}}$ in Eq.~\ref{eq:NVpolar_spectrum_EPR}, one can set the experiment setting \texttt{Temperature} to the wanted temperature in Kelvins (for ambient conditions,  \texttt{Temperature = 298}). To obtain $S_{\mathrm{EPR,pol}}$, the \texttt{Temperature} setting is set to \texttt{[1 0 0]} (Matlab list).
When the signal is  significantly enhanced above thermal ($S_{\mathrm{EPR}}/S_{\mathrm{EPR,thermal}} \gg 1$), as an approximation, one can keep only the second term in Eq.~\ref{eq:NVpolar_spectrum_EPR}, which then yields $S_{\mathrm{EPR}} \approx  p_{\mathrm{NV}}S_{\mathrm{EPR,pol.}}$.

Besides the orientation-induced broadening, secondary contributions to the inhomogeneous broadening of the NV center are the  magnetic interaction to the surrounding spin bath (consisting of P1, other NVs, and additional paramagnetic defects), strain, and electric fields induced by local charges~\cite{mittiga2018}. 
The interaction to the paramagnetic spin bath is considered by defining a lorentzian broadening, using the linewidth parameter \texttt{lw}. To take into account strain and  electric fields induced by local charges -both generating center-to-center variation of $e_x,e_y$ terms in Eq.~\ref{eq:Hamil_general}-  one can in fact define $e_y=0$, and add a distribution of $e_x$, characterized by its full width at half maximum FWHM($e_x$). Indeed, since powder averaging is performed in the simulation, the $e_y$ term need not to be considered. We approximate the distribution of the term $e_x$ as a gaussian, which is  done through the parameter \texttt{DStrain} (\texttt{DStrain}=[0,FWHM($e_x$)]).  We find:

\begin{itemize}
\item For the \SI{2}{\micro\meter} particles (main text, Fig. 1): $D=\SI{2869}{\mega\hertz}$ , $lw=0.4$ mT, FWHM($e_x$)= \SI{21}{\mega\hertz},
\item For the \SI{100}{\nano\meter} particles (Fig.~\ref{fig_spectrum_MSY100_illum}): $D=\SI{2869}{\mega\hertz}$, $lw=\SI{0.4}{\milli\tesla}$, FWHM($e_x$)=\SI{27}{\mega\hertz},
\end{itemize}

\begin{figure}[p]
\centering
\includegraphics[width=0.8\linewidth]{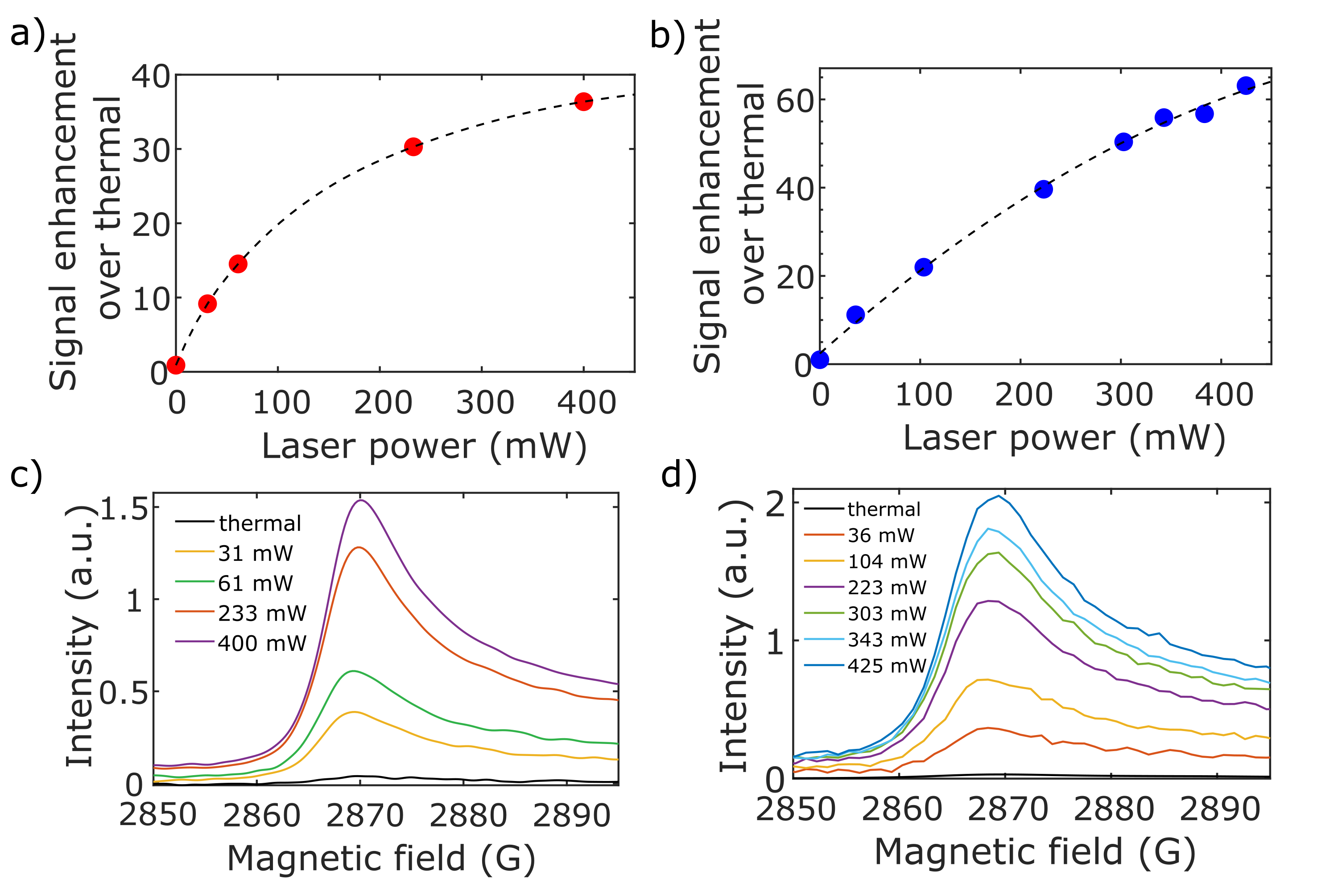}
\caption{Enhancement of the NV EPR signal under continuous  illumination at different laser powers for the samples used for hyperpolarization (a. \SI{2}{\micro\meter},  b. \SI{100}{\nano\meter}). Dashed black lines are a guide to the eye. The enhancement corresponds to the intensity of the low field EPR peak corresponding to NV centers with 90$\degree$ to the magnetic field,  normalized to its intensity measured in the dark.  Corresponding spectra are shown in c. and d. respectively. }
\label{fig:SI_signal_enhance_optic_polar}
\end{figure}

\begin{figure}[p]
\centering
\includegraphics[width=0.45\linewidth]{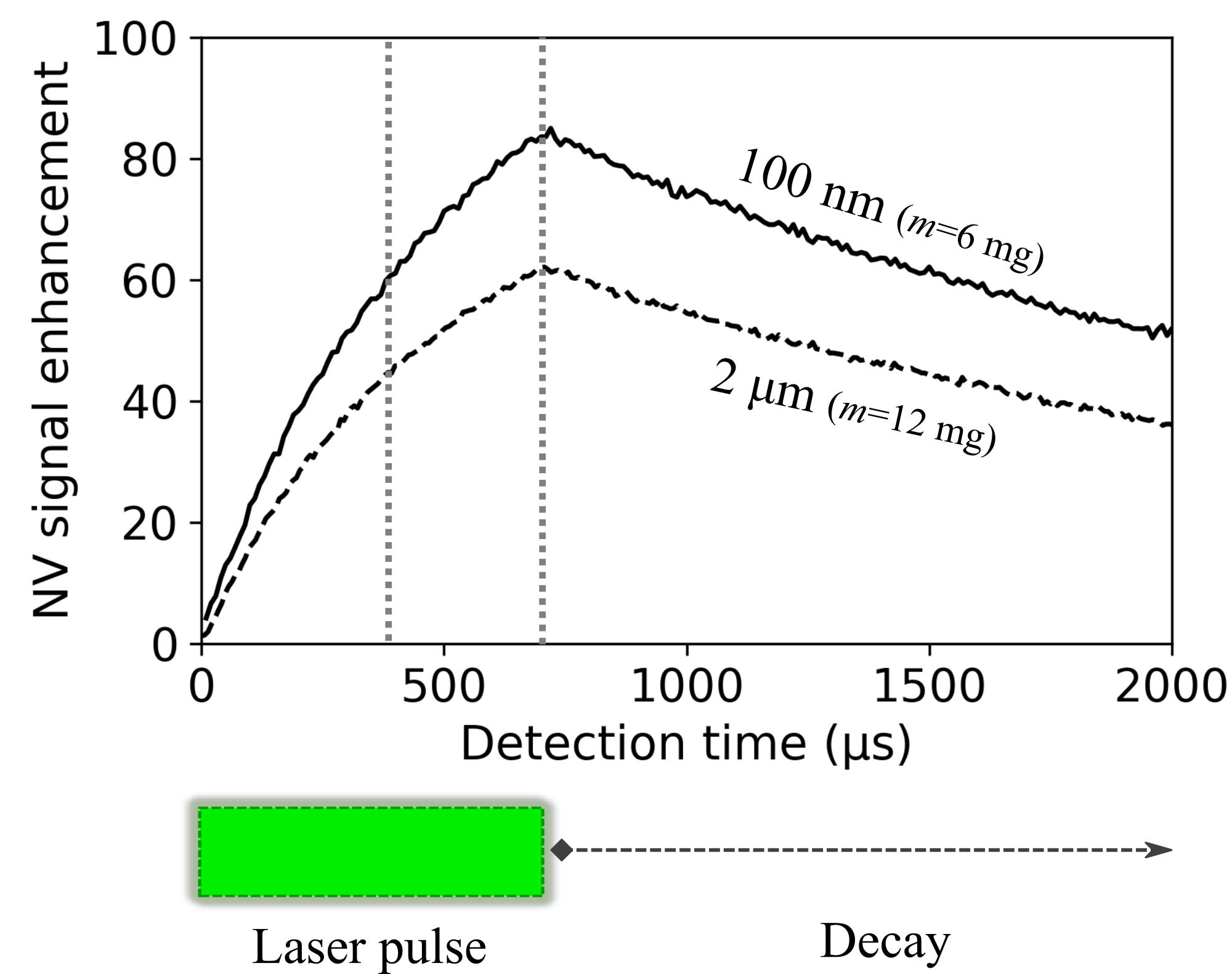}
\caption{Dynamics of the EPR-detected NV signal induced by a  \SI{700}{\micro\second} laser pulse for the  samples used in the experiments described in main text.
The magnetic field is $B=\SI{0.286}{T}$, and the microwave frequency $\nu=\SI{9.58}{\giga\hertz}$, corresponding to the low-field peak of the Pake doublet. The experimental trace gives the enhancement of NV signal, at different times. The dotted lines indicate the laser pulse duration used in the hyperpolarization experiments reported in main text (\SI{400}{\micro\second}), at which the enhancement  for the  \SI{2}{\micro\meter} and \SI{100}{\nano\meter} particles is 46 and 61 respectively, and the full  length (\SI{700}{\micro\second}) used in this experiment.}
\label{fig:SI_illumination_pulse}
\end{figure}

\clearpage
\subsection{Heating induced by illumination}
\label{SI:illumination:heating}
\begin{figure}[ht]
\centering
\includegraphics[width=0.8\linewidth]{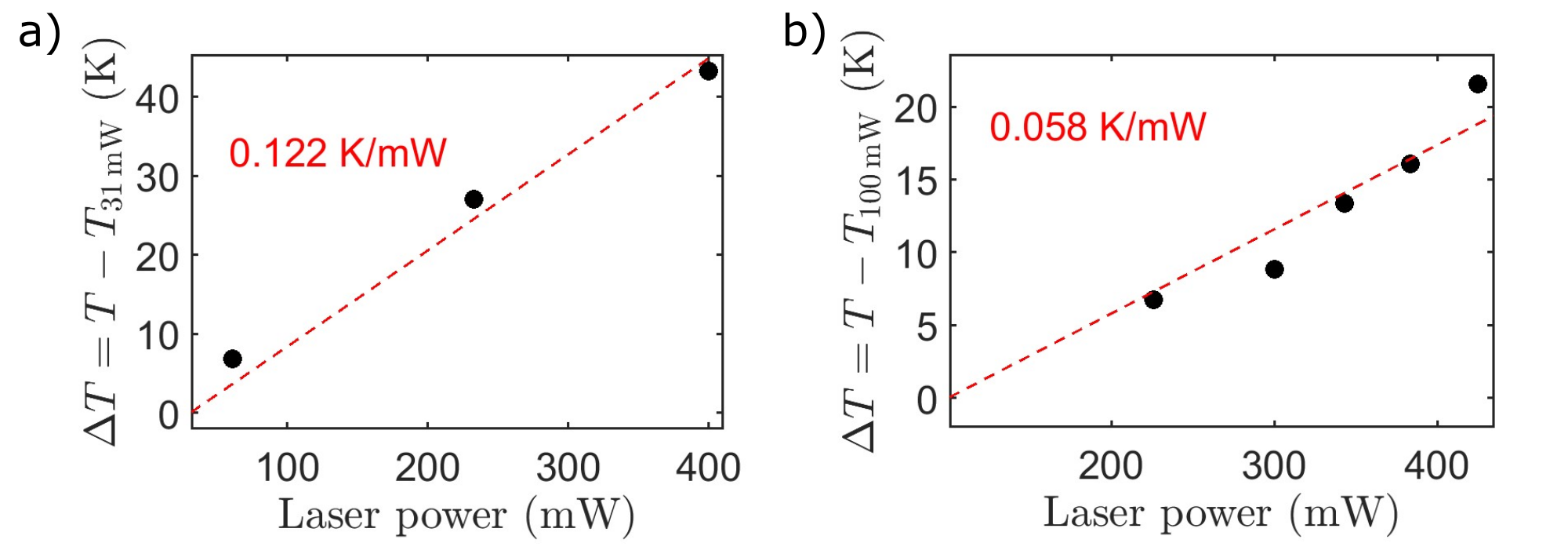}
\caption{Heating of the samples  used in the experiments described in main text (a. \SI{2}{\micro\meter} and b. \SI{100}{\nano\meter}) induced by continuous illumination at different laser power. Dashed lines are a linear fits, allowing to extract as heating rates \SI{0.122}{\kelvin\per\milli\watt} and \SI{0.058}{\kelvin\per\milli\watt} respectively.}
\label{fig:SI_heating}
\end{figure}

The local temperature increase of the diamond powder sample under illumination can be calculated by considering the temperature-induced change of NV zero-field splitting $D$ using the experimentally determined relation from Acosta et al. \cite{acosta2010}:
\begin{equation}
	\frac{dD}{dT}=-\SI{0.074}{\mega\hertz \per \kelvin}.
	\label{DoverT}	
\end{equation}
Combining Eq. \ref{DoverT} and \ref{HoverD}, following a change in temperature $\Delta T$, the low field NV EPR peak is expected to shift by magnetic field $\Delta B$ as follows:
\begin{equation}
	\frac{\Delta B}{\Delta T} =\frac{dB_{12}}{dD}\frac{dD}{dT}=\SI{0.0158}{G \per \kelvin}.
\end{equation}
In this work, for precise determination of temperature changes induced by illumination, we perform the fit as detailed in section \ref{SI:illumination:spectrum} of the  EPR spectrum around the low-field peak for different values of continuous laser power (Figure  \ref{fig:SI_signal_enhance_optic_polar}), which provides the value of $D$ at each laser power.  Comparing to that obtained at a chosen laser power (31 mW and 100 mW for \SI{2}{\micro\meter} and \SI{100}{\nano\meter}, respectively), we can extract the corresponding temperature change (Fig. \ref{fig:SI_heating}).

From the heating rates determined by the linear fits in Fig.~\ref{fig:SI_heating}, we can determine the temperature elevation of the  samples during the application of the NV initialization and polarization transfer cycle described in main text, Fig.~\ref{fig:hp_protocol}c. As an average laser power of \SI{420}{\milli\watt} was used, the temperature changes are +\SI{51}{\kelvin} and +\SI{24}{\kelvin} for the \SI{2}{\micro\meter} and \SI{100}{\nano\meter} respectively. 

Here, the comparatively high temperature increase for the \SI{2}{\micro\meter} sample  can be attributed to the two-times higher  mass employed and the higher NV concentration, both increasing  the heat deposition by photon absorption.

\clearpage
\subsection{Transient NV signal under pulsed illumination}
\label{SI:illumination:pulsed}

To investigate the NV initialization dynamics, the evolution of the NV signal during a \SI{700}{\micro\second} laser pulse was acquired with EPR  (Fig.~\ref{fig:SI_illumination_pulse}). The magnetic field and frequency correspond to the transition between the two lower energy spin states, $s_{1,90\degree}$ and $s_{2,90\degree}$ (depicted in Fig.~\ref{SI_energy_levels}).  

In the hyperpolarization experiments, shorter, \SI{400}{\micro\second} laser pulses were used. From Fig.~\ref{SI_energy_levels}, corresponding NV signal enhancements  for the  \SI{2}{\micro\meter} and \SI{100}{\nano\meter} particles, are $\epsilon_{\mathrm{NV}}=46$ and 61 respectively.

We now turn to the interpretation of the value of $\epsilon_{\mathrm{NV}}$, in terms of state occupation probabilities. The EPR signal scales with the population difference between the states involved in the transitions, and the number of centers contributing to the signal. We note here that by photoluminescence measurement, no conversion towards the neutrally charged NV$^0$ was observed in the samples in the used laser power range, therefore we consider the number of NV centers to be constant. In these conditions, 

\begin{equation}
	\epsilon_{\mathrm{NV}} = \frac{(\Delta\rho)_{\mathrm{NV}}}{(\Delta\rho)_{\mathrm{NV,dark}}},
    \label{eq:epsilon_NV}
\end{equation}

\noindent where  $(\Delta\rho)_{\mathrm{NV}} = \rho_{s_1}-\rho_{s_2}$, and $(\Delta\rho)_{\mathrm{NV,dark}} = \rho_{s_1,\mathrm{dark}}-\rho_{s_2,\mathrm{dark}}$, here $\rho_{s_{1,2}}$ are the occupation probabilities of states $s_{1,90\degree}$ and $s_{2,90\degree}$ (`$90\degree$' is omitted in the subscript for simplicity). The additional subscript `dark' indicates the case without illumination.

In the dark, the occupation probabilities are given by the Boltzmann statistics:
\begin{equation}
	\rho_{s_i,\mathrm{dark}} = e^{\frac{-u_i}{kT}}/\left(\sum\limits_{j=1}^3 e^{\frac{-u_j}{kT}}\right),
\end{equation}

\noindent where $u_i$ ($i=1,2,3$) are the state energies given in Eq.~\ref{eq:E_NV90}. Using $D=\SI{2869}{\mega\hertz}$,  $B=\SI{2860}{G}$, at $T=\SI{293}{\kelvin}$ one obtains: 
\begin{equation}
(\Delta\rho)_{\mathrm{NV,dark}} = \rho_{s_1,\mathrm{dark}}-\rho_{s_2,\mathrm{dark}} = 5.234\times 10^{-4}.
\label{eq:thermal_population_NV_90deg}
\end{equation}
%

From Eq.~\ref{eq:thermal_population_NV_90deg}, it is straightforward to estimate the population difference as $(\Delta\rho)_{\mathrm{NV}} = \epsilon_{\mathrm{NV}} (\Delta\rho)_{\mathrm{NV,dark}}$.   We obtain $(\Delta\rho)_{\mathrm{NV}} = 0.024$ and 0.032   for the \SI{2}{\micro\meter} and \SI{100}{\nano\meter} particles, respectively. We note that CW illumination brings similar populations in the steady-state, as comparable values of the  enhancement $\epsilon_{\mathrm{NV}}$ are observed at high laser power in Fig.~\ref{fig:SI_signal_enhance_optic_polar}. The determined  $(\Delta\rho)_{\mathrm{NV}}$ thus can be compared to the value $\approx 0.07$  reported in NV-ensemble studies of single crystal diamonds under continuous illumination~\cite{drake2015}. 



\clearpage
\pagebreak

\section{Build-up and decay of $^{13}$C polarization}
\label{SI:pol_buildup}

\subsection{\SI{2}{\micro\meter} and  \SI{100}{\nano\meter} samples, with sample rotation (25\degree/s)}

\begin{figure}[h!]
\centering
\includegraphics[width=0.8\linewidth]{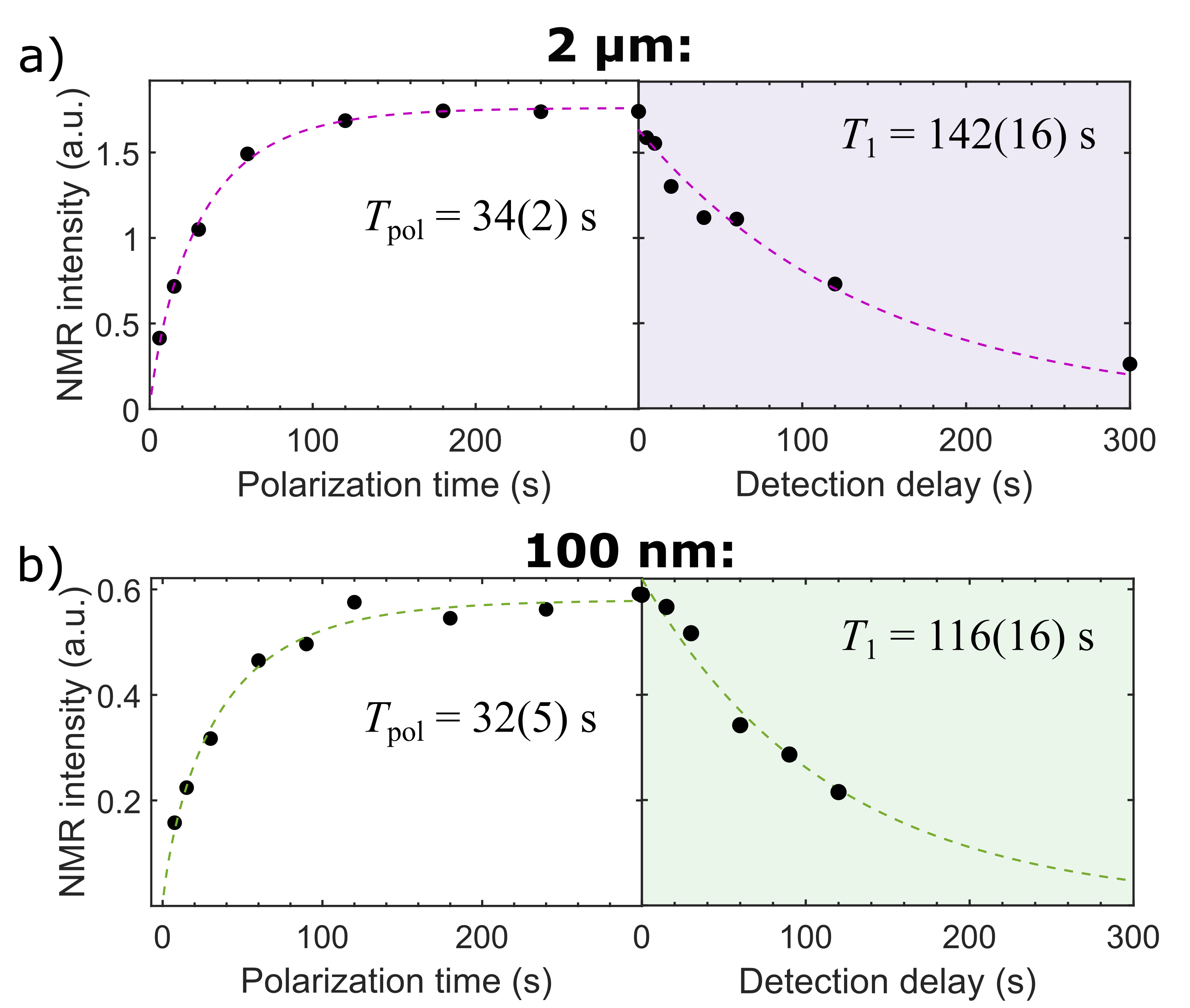}
\caption{ Build-up (left) and decay (right) of $^{13}$C polarization for a)  \SI{2}{\micro\meter} and b) \SI{100}{\nano\meter} diamond powders, respectively. The dashed lines are the fits of the data with stretched exponential function for the build-up time and single exponential for the decay of $^{13}$C polarization.}
\label{fig:buildup}
\end{figure}

 The build-up time, obtained from fitting the data in Figure \ref{fig:buildup} with a stretched exponential function $s(t)=A \left (1-\exp{-(\frac{t}{T_{\mathrm{pol}}})^\beta} \right )$, is $T_{\mathrm{pol}}=34\pm2$ s (with $\beta=0.85\pm0.06$) for the \SI{2}{\micro\meter} sample and $T_{\mathrm{pol}}=32\pm5$ s (with $\beta=0.88\pm0.12$) for the \SI{100}{\nano\meter} sample. The $^{13}$ spectra after buildup ( $t=\SI{300}{\second}$) are shown in main text, Fig.~\ref{fig:Hyper_signal}.
The decay of polarization at 1 T, estimated from the fitting the data in Figure \ref{fig:buildup} with a single exponential decay, is $T_1=142\pm16$ s and
$T_1=116\pm16$ for \SI{2}{\micro\meter} and \SI{100}{\nano\meter} diamond powders, respectively.

\clearpage
\pagebreak
\subsection{Buildup for the \SI{2}{\micro\meter} sample, without rotation}

\begin{figure}[h!]
\centering
\includegraphics[width=0.45\linewidth]{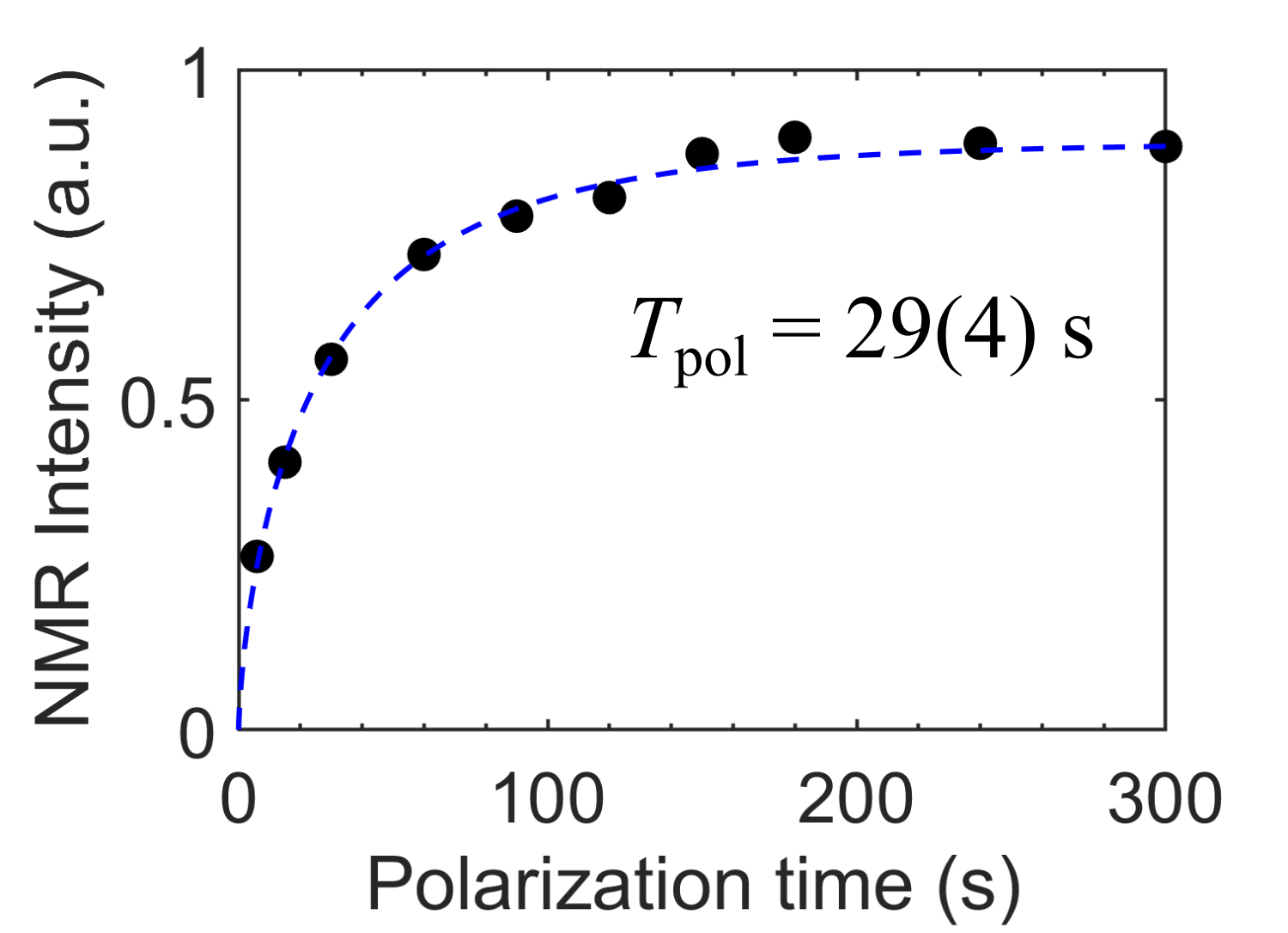}
\caption{ Build-up  of $^{13}$C polarization for the \SI{2}{\micro\meter} sample in the absence of sample rotation and fit with a stretched exponential function giving $T_{\mathrm{pol}}=29\pm4$ s and  $\beta=0.70\pm0.08$.}
\label{fig:SI_buildup_2um_norotation}
\end{figure}

The build-up time, obtained from fitting the data in Figure \ref{fig:SI_buildup_2um_norotation} with a stretched exponential function  is $T_{\mathrm{pol}}=29\pm4$ s (with $\beta=0.70\pm0.08$).

\section{Evaluation of the hyperpolarization enhancement factor and the  absolute polarization}
\label{SI:enhancement_thermal}

The signal after hyperpolarization $S_{\mathrm{^{13}C}}$, the enhancement factor $\epsilon_{\mathrm{^{13}C}}$ and the absolute polarization $p_{\mathrm{^{13}C}}$ (defined in main text) obey 

\begin{equation}
	\epsilon_{\mathrm{^{13}C}}=\frac{S_{\mathrm{^{13}C}}}{S_{\mathrm{^{13}C,therm}}}=\frac{p_{\mathrm{^{13}C}}}{p_{\mathrm{^{13}C,therm}}},
 \label{eq:epsilon_13C}
\end{equation}

\noindent where $S_{\mathrm{^{13}C,therm}}$, $p_{\mathrm{^{13}C,therm}}$ are the values of the thermal signal and the corresponding absolute nuclear polarization at thermal equilibrium.

The value of the absolute polarization at thermal equilibrium  can be calculated  as  $p_{\mathrm{^{13}\mathrm{C},therm}}=\frac{\hbar \gamma_{^{13}\mathrm{C}} B}{2 k_{\mathrm{B}} T}$  , where  $k_{\mathrm{B}}$ is the Boltzmann constant, $T$ is the temperature, $\gamma_{^{13}\mathrm{C}}$ is the nuclear gyromagnetic ratio (from the Boltzmann statistics as $k_{\mathrm{B}}T \gg \hbar \gamma_{^{13}\mathrm{C}} B$).  From  Eq.~\ref{eq:epsilon_13C} it can be seen that, from the hyperpolarized signal,  both  $\epsilon_{\mathrm{^{13}C}}$ and  $p_{\mathrm{^{13}C}}$ can be determined if the thermal signal  $S_{\mathrm{^{13}C,therm}}$  can be measured. However, the sensitivity of the setup for in situ NMR detection is low and, as a result, insufficient to detect the thermal $^{13}$C  signal from the diamond powder samples. To address this limitation, we employed a reference sample of \SI{40}{\micro\liter} water, in which the high thermal signal from the $^1$H nuclei can  easily be measured. 
 
For the \SI{2}{\micro\meter} particles sample, the hyperpolarized spectrum seen in main text, Fig.~\ref{fig:Hyper_signal}a, corresponds to a signal (observed value of the free induction decay at $t\sim0$)  of \SI{1.407}{\micro\volt} for a mass $m=\SI{12}{\milli\gram}$ of sample. 
For this sample first derive $p_{\mathrm{^{13}C}}$  from the reference $^1{\mathrm{H}}$ measurement   in the water sample.  
Then we deduce $\epsilon_{\mathrm{^{13}C}}$ at the polarization field (\SI{0.287}{\tesla}). 

Similar calculations apply for the \SI{100}{\nano\meter} sample for which the spectrum in main text, Figure~\ref{fig:Hyper_signal}b corresponds to a signal of \SI{0.441}{\micro\volt}  for a sample mass $m=\SI{6}{\milli\gram}$.   For each sample, both $\epsilon_{\mathrm{^{13}C}}$ and $p_{\mathrm{^{13}C}}$ are reported in main text. 

\paragraph{Determination of $p_{\mathrm{^{13}C}}$} By measuring the intensity of the $^1$H thermal signal in the water sample ($S_{^1\mathrm{H,therm}}$), one can estimate the intensity of the $^{13}$C thermal signal as follows:
\begin{equation}
              S_{\mathrm{^{13}C,therm}}= S_{\mathrm{^1\mathrm{H},therm}}\Gamma_{\mathrm{ref}}.
\end{equation}
We now focus on the derivation of $\Gamma_{\mathrm{ref}}$. The intensity of thermal signal $S_{\mathrm{nucl}}$ in a magnetic field $B$ at temperature $T$ can be estimated as~\cite{webb2012}
\begin{equation}
	S_{\mathrm{nucl}}=C\times(\gamma B)\times(\frac{\gamma^2\hbar^2BN_{\mathrm{nucl}}}{4k_{\mathrm{B}}T}).
	\label{eq:NMRsignal}
\end{equation}
where $C$ is a constant, determined by parameters of the detection channel, $\hbar$ is the Planck constant, $N_{\mathrm{nucl}}$ is the number of detected nuclear spins, and $\gamma$ is the nuclear gyromagnetic ratio. This expression is valid in the limit  $k_{\mathrm{B}}T>>\hbar\gamma B$.

The $^{13}$C signal was detected at \SI{1.004}{T}. Unlike conventionally done in NMR, the detection of signals for $^1$H  nuclear spins was implemented at the same frequency, adjusting the magnetic field for the $^1$H nucleus ($B=\SI{0.252}{\tesla}$). Therefore, it is convenient to replace the magnetic field $B$ in equation \ref{eq:NMRsignal}, as
\begin{equation}
	B=\frac{2\pi f}{\gamma},	
\end{equation}
where $f$ is the detection frequency.
Furthermore, given that the measurement were both performed at ambient temperature, the ratio of thermal signals for $^1$H in water and $^{13}$C nuclear spins in the respective samples can be expressed as
\begin{equation}
	\Gamma_{\mathrm{ref}}=\frac{S_{\mathrm{^{13}C,therm}}}{S_{\mathrm{^1H,therm}}}=\frac{\gamma_{\mathrm{^{13}C}} N_{\mathrm{^{13}C}}}{\gamma_{\mathrm{\mathrm{^1H}}} N_{\mathrm{^1H}}}.
	\label{eq:enhan_1}
\end{equation}
The number of hydrogen and carbon nuclear spins can be estimated as follows:
\begin{equation}
\begin{aligned}
	N_{\mathrm{^{13}C}}=\frac{m_{\mathrm{diam.}} N_{\mathrm{A}} C_{\mathrm{^{13}C}}}{M_{\mathrm{diam.}}},
        N_{^1\mathrm{H}}=\frac{n_{\mathrm{^1H/mol}}m_{\mathrm{water}} N_\mathrm{A} C_{\mathrm{^{1}H}} }{M_{\mathrm{water}}},
  \end{aligned}
\end{equation}
where $m_{\mathrm{diam.}}$ and $m_{\mathrm{water}}$ are the masses of diamond and water samples, respectively, $N_\mathrm{A}$ is the Avogadro constant, $C_{\mathrm{^{13}C}}$ and $C_{\mathrm{^{1}H}}$ are  the  isotopic abundances of $^{13}$C and $^{1}$H, respectively, $n_{\mathrm{^1H/mol}}$ is the number of hydrogen atoms per molecule ($n_{\mathrm{^1H/mol}}=2$), $M_{\mathrm{diam.}}$ and $M_{\mathrm{water}}$ are the molar masses of water molecules and carbon in diamond at natural abundance, respectively. 
From the parameters listed in Table~\ref{table:enhancement_det_gamma_ref} we determine $\Gamma_{\mathrm{ref}}=1/1652$ for the $\SI{2}{\micro\meter}$ sample described in main text.

\begin{table}[ht]
    \centering
    \begin{tabular}{|c|c|c|c|c|}
        \hline
        sample & mass  &  $C_{^{13}\mathrm{C}}$ or $ C_{^{1}\mathrm{H}}$  &  molar mass &  $\gamma_{^{13}\mathrm{C}}$ or $\gamma_{^{1}\mathrm{H}}$  \\ \hline
        \SI{2}{\micro\meter}& \SI{12}{\milli\gram} &   1.07\%  &  \SI{12.011}{\gram\per\mol} &   $(2\pi)$\SI{10.705}{\mega\hertz\per\tesla} 
    
        \\ 
         water reference & \SI{40}{\milli\gram} & 99.98\%    &   \SI{18.015}{\gram\per\mol}  &  $(2\pi)$\SI{42.576}{\mega\hertz\per\tesla}  \\ \hline 
    \end{tabular}
    \caption{Sample properties used for determining the parameter $\Gamma_{\mathrm{ref}}$, in Eq.~\ref{eq:enhan_1}, used for determination of the signal enhancement (illustrated here for the $\SI{2}{\micro\meter}$ sample).}
    \label{table:enhancement_det_gamma_ref}
\end{table}

Eq.~\ref{eq:enhan_1} and Eq.~\ref{eq:epsilon_13C} yield:

\begin{equation}
p_{^{13}\mathrm{C}} = \frac{p_{^{13}\mathrm{C,therm}} \times  S_{^{13}\mathrm{C}}}{S_{^{13}\mathrm{C,therm}}} =  \frac{p_{^{13}\mathrm{C,therm}} \times S_{^{13}\mathrm{C}}}{\Gamma_{\mathrm{ref}} \times  S_{^{1}\mathrm{H,therm}}}
\end{equation}


We calculate,  for the absolute thermal $^{13}$C polarization at the detection field of \SI{1.004}{\tesla} ($T=\SI{293}{\kelvin}$), $p_{^{13}\mathrm{C,thermal}}= \SI{8.80e-7}{}$. For the reference water sample, we measure  $S_{^{1}\mathrm{H,therm}}=\SI{5.420}{\micro\volt}$. 
Thus, for the \SI{2}{\micro\meter} sample: 

\begin{equation}
p_{^{13}\mathrm{C}} = \frac{ \SI{8.80e-7}{} \times  \SI{1.407}{\micro\volt}  }{\frac{1}{1652} \times \SI{5.420}{\micro\volt} } =  \SI{3.77e-4}{}
\label{eq:absolute_pol_2_micrometer}
\end{equation}

Similarly, for the \SI{100}{\nano\meter} sample, $p_{^{13}\mathrm{C}} = \SI{2.36e-4}{}$ is obtained. 

\paragraph{Determination of $\epsilon_{\mathrm{^{13}C}}$ at the  polarization field (\SI{0.287}{\tesla})}

For the absolute thermal $^{13}$C polarization at the \emph{polarization} field of \SI{0.287}{\tesla} ($T=\SI{293}{\kelvin}$), we calculate $p_{^{13}\mathrm{C,thermal}}= \SI{2.52e-7}{}$. For the \SI{2}{\micro\meter} sample, it  follows from Eq.~\ref{eq:absolute_pol_2_micrometer} that: $\epsilon_{^{13}\mathrm{C}} = \frac{p_{^{13}\mathrm{C}}}{p_{^{13}\mathrm{C,thermal}}} \approx 1500$. Similarly, for the \SI{100}{\nano\meter} sample, we obtain $\epsilon_{^{13}\mathrm{C}} \approx 940$.


\section{PulsePol parameter  tuning}

\subsection{Testing the robustness to detuning of the phase-offset PulsePol sequence}
\label{SI:pulsepol_phaseoffset}

The original PulsePol sequence (main text, Fig~\ref{fig:hp_protocol}a) was proposed by Schwartz \emph{et al.}~\cite{schwartz2018}. 
In our experiments, we however used the `phase-offset' sequence described in (main text, Fig.~\ref{fig:hp_protocol}b). 
Tratzmiller \emph{et al.} introduced a general set of sequences based on PulsePol, in which the phase definition contains a free parameter $\phi$~\cite{tratzmiller2021}. In this definition, the original PulsePol and the sequence presently labelled  `phase-offset'  are obtained by choosing  $\phi=\pi/2$ and $\phi=3\pi/4$, respectively. 

\begin{figure}[htp]
\centering
\includegraphics[width=0.7\linewidth]{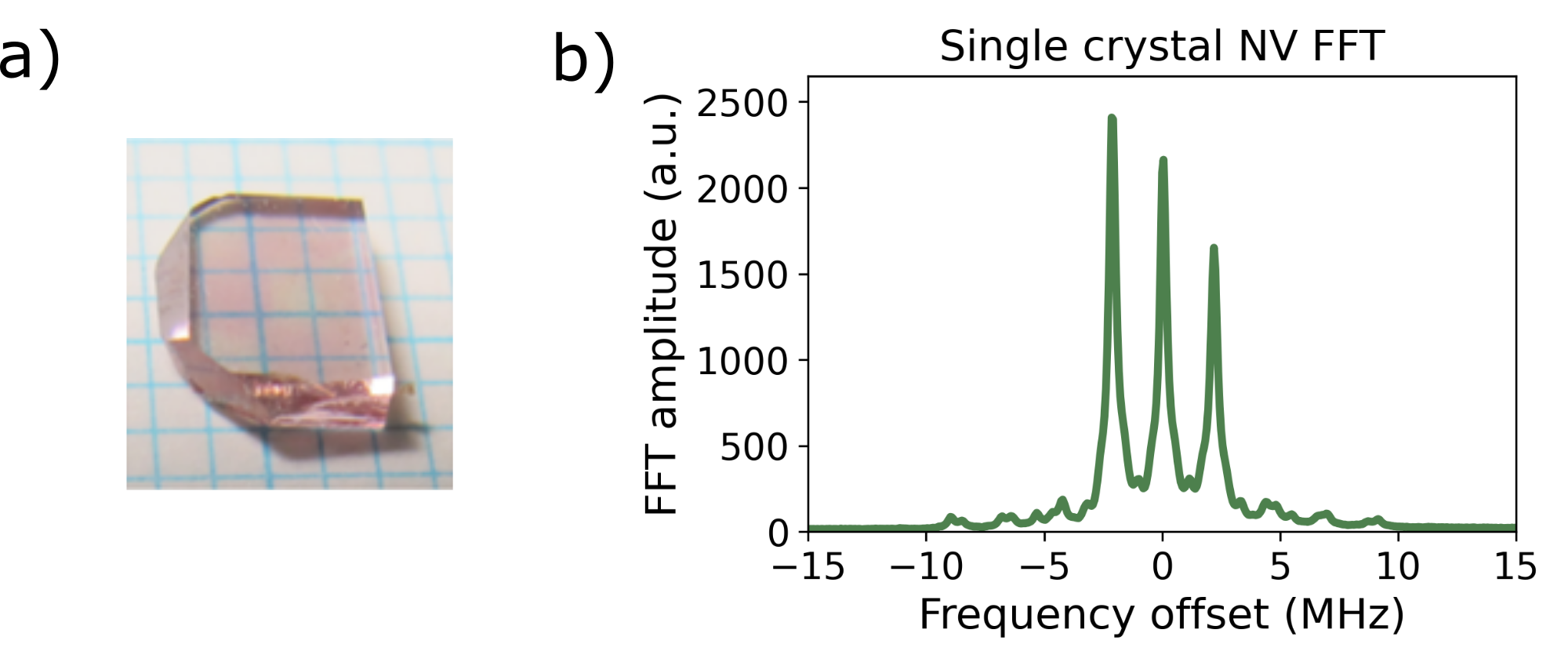}
\caption{a) Single crystal diamond used for the sequence robustness tests. b) Spectrum of NV around the resonator frequency $\nu=\SI{9.7}{\giga\hertz}$ at $B=\SI{2436}{G}$ aligned to the NV axis. The several lines correspond to the hyperfine structure for the $m_S = 0$ to +1 transition of NV. Coupling to $^{14}$N nuclear spin ($I=1$) explains the three-lines pattern, and  nearby $^{13}$C provide weak additional satellites. The spectrum is obtained by performing Fourier Transform of the Hahn echo signal, acquired in EPR using  pulses generated by the \SI{1}{\kilo\watt} TWT of Rabi frequency $\Omega_1=(2\pi) \SI{25}{\mega\hertz}$, under continous illumination at $P=\SI{100}{\milli\watt}$ laser power. }
\label{fig:SI_single_crystal_NV_FFT}
\end{figure}

In order to characterize the  robustness to spectral detuning, we employ a single-crystal diamond, represented in Fig.~\ref{fig:SI_single_crystal_NV_FFT}a, that was utilized in an earlier work~\cite{scheuer2016}. 
Addressing NV in single-crystal diamond instead of powder drastically reduces their  inhomogeneous broadening, so that the sequence robustness can be tested by varying  the pulse carrier frequency with respect to a group of NV transitions occupying a narrow spectral range.

\begin{figure}[h]
\centering
\includegraphics[width=0.55\linewidth]{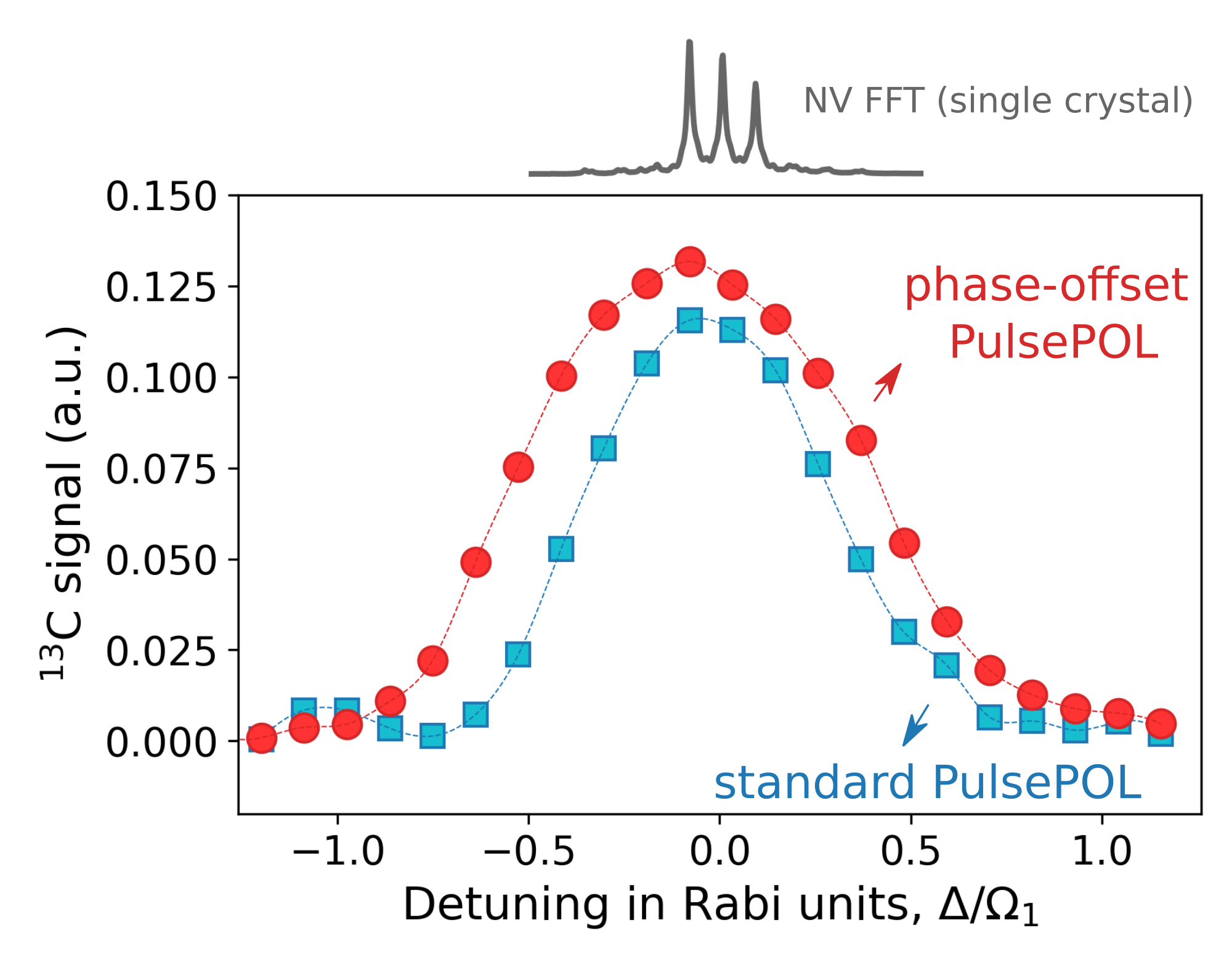}
\caption{Testing the robustness to detuning of the original PulsePol sequence and its variant, represented in main text Fig.~\ref{fig:hp_protocol}a and ~\ref{fig:hp_protocol}b, respectively. $^{13}$C hyperpolarized signal is acquired at different detunings $\Delta$ of the pulse carrier frequency with respect to the central line ($m_I=0$ for $^{14}$N) of the NV signal. The detuning is represented in units of the ordinary Rabi frequency $\nu_1=\Omega_1/2\pi$. The top curve shows the FFT spectrum from Fig.~\ref{fig:SI_single_crystal_NV_FFT}b  on the same scale. A high-power amplifier (TWT, \SI{1}{\kilo\watt})  was used in order to ensure a high $\nu_1=\SI{25}{\mega\hertz}$, driving the different hyperfine ($^{14}$N) transitions similarly. Resonance conditions $n=3$ and $n=3.5$ were used for PulsePol and phase-offset PulsePol respectively. The number of repetitions was $M=10$. Lines in the main plot are a guide to the eye. }
\label{fig:SI_phase_offset_detuning_robustness}
\end{figure}

\clearpage

The $^{13}$C signal was acquired after hyperpolarization with both the standard and phase-offset PulsePol, at different magnetic fields positively and negatively detuned from the resonance of the central   hyperfine line ($m_I=0$ for $^{14}$N) visible in Fig.~\ref{fig:SI_single_crystal_NV_FFT}b. A change in magnetic field $\delta B$   shifts the NV spectrum frequency for the central line ($\nu_{\mathrm{NV}}$) creating  for the pulse carrier frequency ($\nu_{\mathrm{MW}})$ a detuning  $\Delta = 2\pi (\nu_{\mathrm{MW}} - \nu_{\mathrm{NV}}) = -\gamma  \delta B $, where $\Delta$ is given in angular frequency unit, $\gamma =(2\pi)\SI{2.8032}{\mega\hertz\per G}$ is the gyromagnetic ratio of the NV center.  Figure~\ref{fig:SI_phase_offset_detuning_robustness} shows the hyperpolarized $^{13}$C signal acquired with both the standard and phase-offset PulsePol. It is well visible that the latter demonstrates  increased robustness to detuning, i.e. ensures the   polarization transfer to be efficient in a wider spectral range.

\subsection{Comparison of resonance indices $n=3.5$ and $n=4.5$}
\label{SI:pulsepol_4p5vs3p5}

\begin{figure}[ht]
\centering
\includegraphics[width=0.8\linewidth]{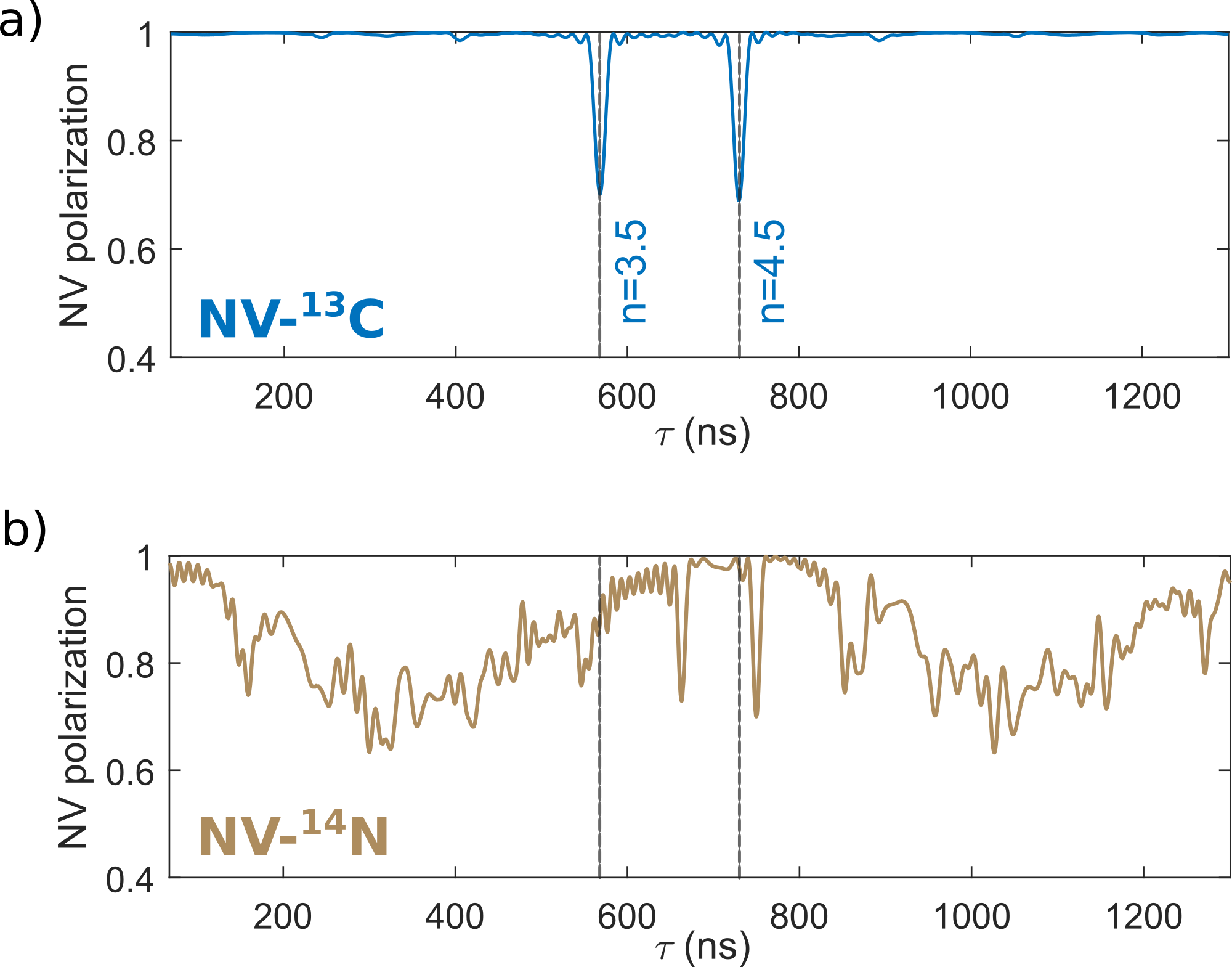}
\caption{Simulation of polarization dynamics  for NV oriented at 90\degree to the magnetic field, in the conditions of our experiment  ($B=\SI{0.287}{\tesla}$). The horizontal axis represents the $\tau$ delay in the sequence and the values on the vertical axis correspond to the remaining polarization of the NV,  after $M=20$ repetitions of the phase-offset PulsePol sequence. a) Case of a $^{13}$C spin coupled to NV with hyperfine tensor elements $A_{ZX}=(2\pi) \SI{80}{\kilo\hertz},A_{ZY}=0$, without interaction to $^{14}$N b) Case of NV-$^{14}$N interaction, where the $^{14}$N Hamiltonian is taken from Felton \emph{et al.}~\cite{felton2009},
in the absence of interaction to $^{13}$C. }
\label{fig:SI_PulsePol_sim_14N_13C}
\end{figure}

In Fig.~\ref{fig:SI_PulsePol_sim_14N_13C} is shown the simulated polarization loss of NV when applying the phase-offset PulsePol sequence for different $\tau$ intervals, considering in one case  the interaction to a $^{13}$C nuclear spin  (Fig.~\ref{fig:SI_PulsePol_sim_14N_13C}a) and in the other case  the interaction to the strongly coupled $^{14}$N nuclear spin of NV (Fig.~\ref{fig:SI_PulsePol_sim_14N_13C}b).

 The interaction to the $^{13}$C nuclear spin leads to the expected half-integer resonances.  The $n=3.5$ and $n=4.5$ resonances are the most intense in comparison to the other half-integer ones in the range  of represented $\tau$ values. Thus the  $n=3.5$ and $n=4.5$ resonance dips corresponds to a strong probability of transferring polarization to $^{13}$C. For the chosen number of repetition of the 6-pulse block ($M=20$ in the figure), the  $n=4.5$ resonance dip is slightly more pronounced than the $n=3.5$ one, however this advantage should disappear if we consider the same  interaction times (as  a given number of repetitions  is achieved in a time that is shorter by $1-3.5/4.5 \approx 22\% $ with the $n=3.5$ resonance). From these considerations, i.e., assuming similar interaction times, the $n=3.5$ resonance is expected to be advantageous.  

\begin{figure}[p]
\centering
\includegraphics[width=0.35\linewidth]{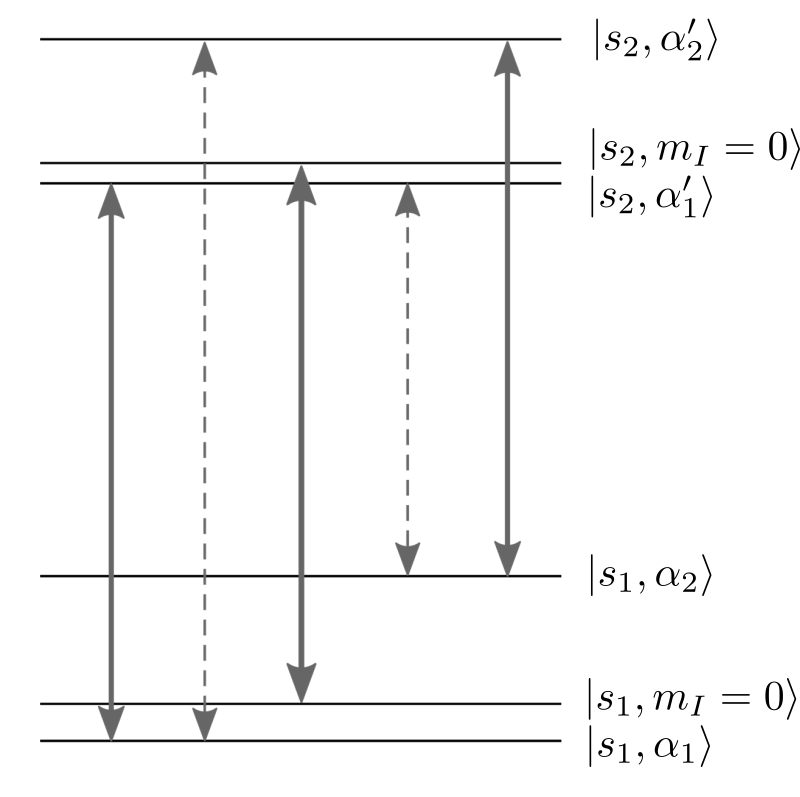}
\caption{Energy diagram and transitions induced by microwave on  the NV center spin  coupled to the $^{14}$N nuclear spin for a magnetic field perpendicular to the  $C_{3v}$  symmetry axis of NV. The diagram is restricted to the low energy spin states $\ket{s_1}$ and $\ket{s_2}$ of the NV center,  which were the one driven in our experiment.   In this configuration  the $\ket{m_I=0}$ state of $^{14}$N (choosing the axis of the applied magnetic field for quantization) is the only one that is unaffected by the hyperfine coupling to the NV. The other $^{14}$N spin states are $\ket{\alpha_1},\ket{\alpha_2}$ in the $\ket{s_1}$ manifold of NV and $\ket{\alpha'_1},\ket{\alpha'_2}$ in the $\ket{s_2}$ manifold. 
The arrows represent the  microwave transitions: solid for the allowed transitions, and dashed for the  `forbidden' transitions, respectively.
}
\label{fig:SI_14N_transition_diagram}
\end{figure}

However, the  presence of the $^{14}$N nuclear spin must be included and leads to more complex dynamics. This dynamics is specific to the case of  NV misaligned to the magnetic field, which leads to microwave transitions affecting simultaneously the state of the $^{14}$N and NV spin. These transitions, which can be predicted from the diagonalization of the Hamiltonian, are illustrated  in Fig.~\ref{fig:SI_14N_transition_diagram}. The simulation of PulsePol in Fig.~\ref{fig:SI_PulsePol_sim_14N_13C}b, which considers a NV exclusively coupled to its  $^{14}$N spin, shows that applying the  sequence leads to a polarization loss on the NV center due to this interaction. The polarization loss on NV shows complex dependence on the parameter $\tau$.

To prevent the interaction to $^{14}$N  from disturbing the dynamics of polarization transfer to $^{13}$C, we choose a $\tau$ value for which the simulation predicts the polarization loss on NV induced by the NV-$^{14}$N interaction to be negligible. It is visible from Fig.~\ref{fig:SI_PulsePol_sim_14N_13C}b that in the conditions of our experiment the $n=4.5$ resonance fullfills this criterion  better than the $n=3.5$ one, therefore the former is chosen. 

\begin{figure}[p]
\centering
\includegraphics[width=0.5\linewidth]{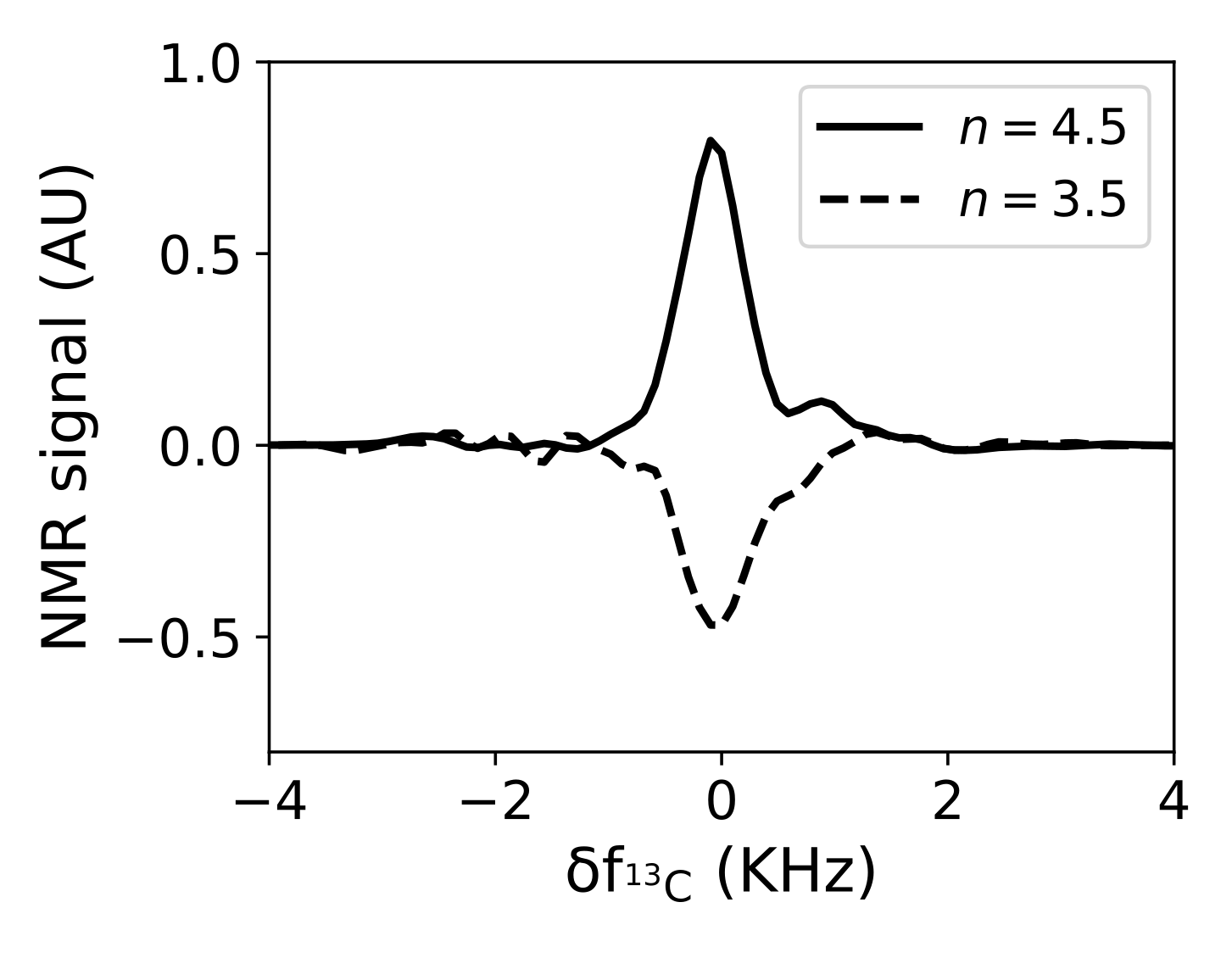}
\caption{Hyperpolarized signal obtained using the $n=3.5$ and $n=4.5$ resonances, after $M=20$ repetitions of the phase-offset PulsePol (\SI{2}{\micro\meter} sample). The change in sign of polarization is an expected characteristic of the sequence. 
The signal in the case $n=4.5$ is estimated to be 70\% higher in absolute intensity than in the case  $n=3.5$. }
\label{fig:SI_PulsePOL_var_n}
\end{figure}

To confirm the benefit of using the $n=4.5$ resonance, we performed polarization transfer at both the  $n=3.5$ and  $n=4.5$ resonances, which leads to the hyperpolarized signal seen in Figure~\ref{fig:SI_PulsePOL_var_n} (\SI{2}{\micro\meter} particles). Using the $n=4.5$ resonance yields a 70\% higher absolute signal than using the $n=3.5$ one. This confirms the advantage of choosing the $\tau$ value among available resonances so as to minimize the advert effects of the NV-$^{14}$N interaction.

As the evolution dynamics of the coupled NV and $^{14}$N  is affected by the magnetic field, our conclusions do not hold at any value of $B$. At other magnetic field values, it might be favorable to use yet other phases in PulsePol (by adjusting the parameter $\phi$ in the general sequence definition from ref.~\cite{tratzmiller2021}). Indeed, the resulting change in the resonance positions  might allow finding conditions where the $^{14}$N-related dynamics is refocused, in the same way as at our magnetic field, using the $\tau$ value corresponding to  $n=4.5$. However, the simulation of the NV-$^{14}$N interaction dynamics at different magnetic fields is outside the scope of this work.

\clearpage
\pagebreak

\clearpage
\subsection{The composite pulses}
\label{SI:pulsepol_composite}
Composite pulses consist of (quasi-)contiguous  pulses of different phase  and/or amplitude. This provides degrees of freedom that can be used for improving the robustness properties of sequences~\cite{shaka1987}. The composite pulses used in this work  require applying microwave only on a `on' or `off' basis with no amplitude variation during the `on' time. This makes their application of experimental convenience as one can use any microwave amplifier in its  saturation regime, without the need to correct for response non-linearities.

The composite pulses used in the present work for increasing the bandwidth of PulsePol are a variant of the symmetric phase-alternating composite pulses introduced in ref.~\cite{shaka1987}: instead of analytically choosing the pulse durations in order to cancel low-order contributions from an expansion of the pulse-induced rotation in the detuning, we choose a numerical optimization which instead aims to reach a minimum fidelity threshold of the intended $\pi$-pulse in a maximal bandwidth. This can be advantageous as the additional error correction intrinsic to PulsePol loosens the constraints on the individual pulses.  In order to increase the available degrees of freedom for the pulse, we allowed for non-zero waiting times between sub-pulses. A pulse description is defined via a set of durations $t_i = a_i \ \pi/\Omega$ written as $[a_n,\ldots, a_1, a_0]$ which corresponds to a composite pulse with subsections of duration $(t_n,\ldots,t_1,2t_0,t_1,t_2,\ldots,t_n)$ where the central $t_0$ durations correspond to positive-phase pulses, the odd durations ${t_{2k+1}}_{k\in\mathbb{N}}$ correspond to waiting times, and the even durations ${t_{2k}}_{k\in\mathbb{N}}$ have alternating positive and negative phase. In this notation, a non-composite rectangular $\pi$ pulse is defined as $[0.5]$

In Table \ref{tab:comp-pulses}, the composite pulses are named by the number of pulses to each side of the central pulse ('sidebands'), the overall duration and power of one composite pulse is measured in $\pi$ pulse multiples. As there are waiting times involved, these values do not coincide.
As robustness, we give the range of $\Delta/\Omega$ for which PulsePol gives strong polarisation transfer (approximate values). A visual representation of the pulses, as well as the theoretical inversion profile and detuning- and robustness properties of the pulses as part of PulsePol can be found in Fig.~\ref{fig:compPulses}.

\begin{table}[ht]
    \centering
    \begin{tabular}{ccccl}
        pulse & duration & power & robustness & definition \\ \hline
        normal & 1 & 1 & $<0.4$ & [0.5] \\
        2 & 3.7 & 2.7 & 0.7 & [0.187, 0.251, 0.408, 0.253, 0.721] \\ 
        3 & 5.1 & 3.6 & 0.75 & [0.13  0.231 0.386 0.261 0.527 0.236 0.771] \\
        5 & 8.1 & 5.6 & 1 & [0.167, 0.32,  0.427, 0.348,   \\
        & & & & 0.658,~0.263, 0.72, 0.292, 0.822]  \\
    \end{tabular}
    \caption{Overview over definitions and properties of the numerically optimized phase-alternating composite pulses of which the 2-sideband pulse was used in this work to increase the bandwidth of PulsePol. The values for robustness here refer to the range of detunings $\Delta$ in which polarisation transfer remains successful, such that value of $1$ corresponds to the bandwidth $\Delta\in[-1\Omega,+1\Omega]$ (cf.~Fig.~\ref{fig:compPulses}).} 
    \label{tab:comp-pulses}
\end{table}

\begin{figure}[bt]
\centering
\includegraphics[width=0.85\linewidth]{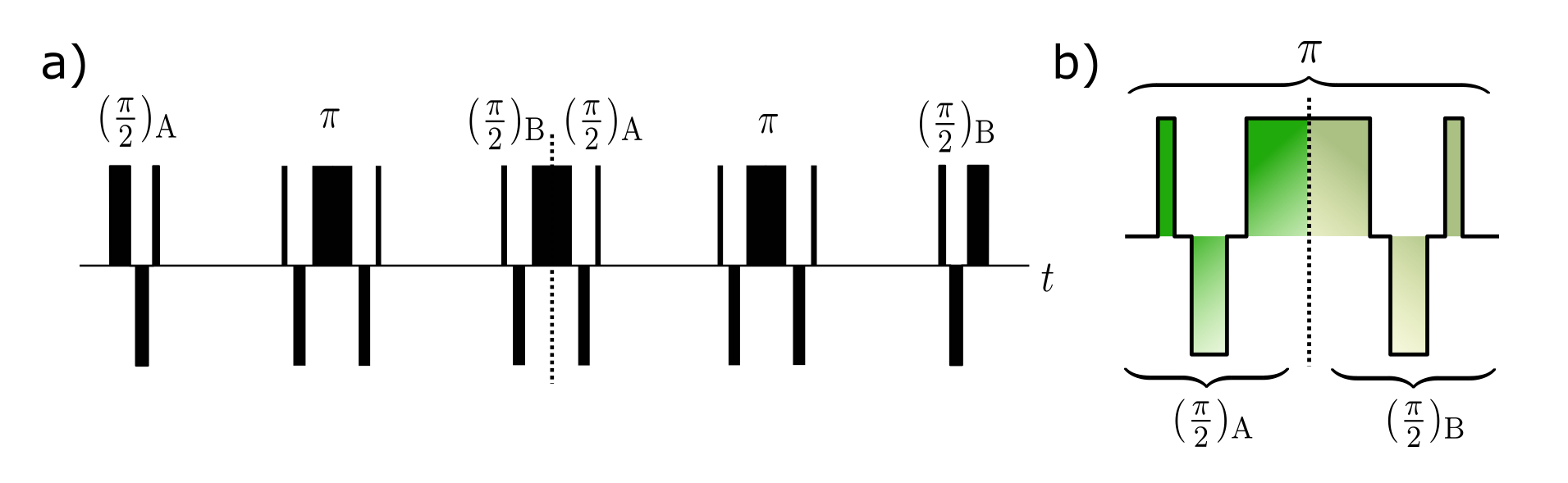}
\caption{a) Illustration of the PulsePol sequence, where composite pulses with two sidebands  replace rectangular pulses. b) Central pulse, consisting of two types of $\pi/2$ pulses, of symmetric shape.}
\label{brush}
\end{figure}

\begin{figure}[bt]
\centering
\includegraphics[width=1\linewidth]{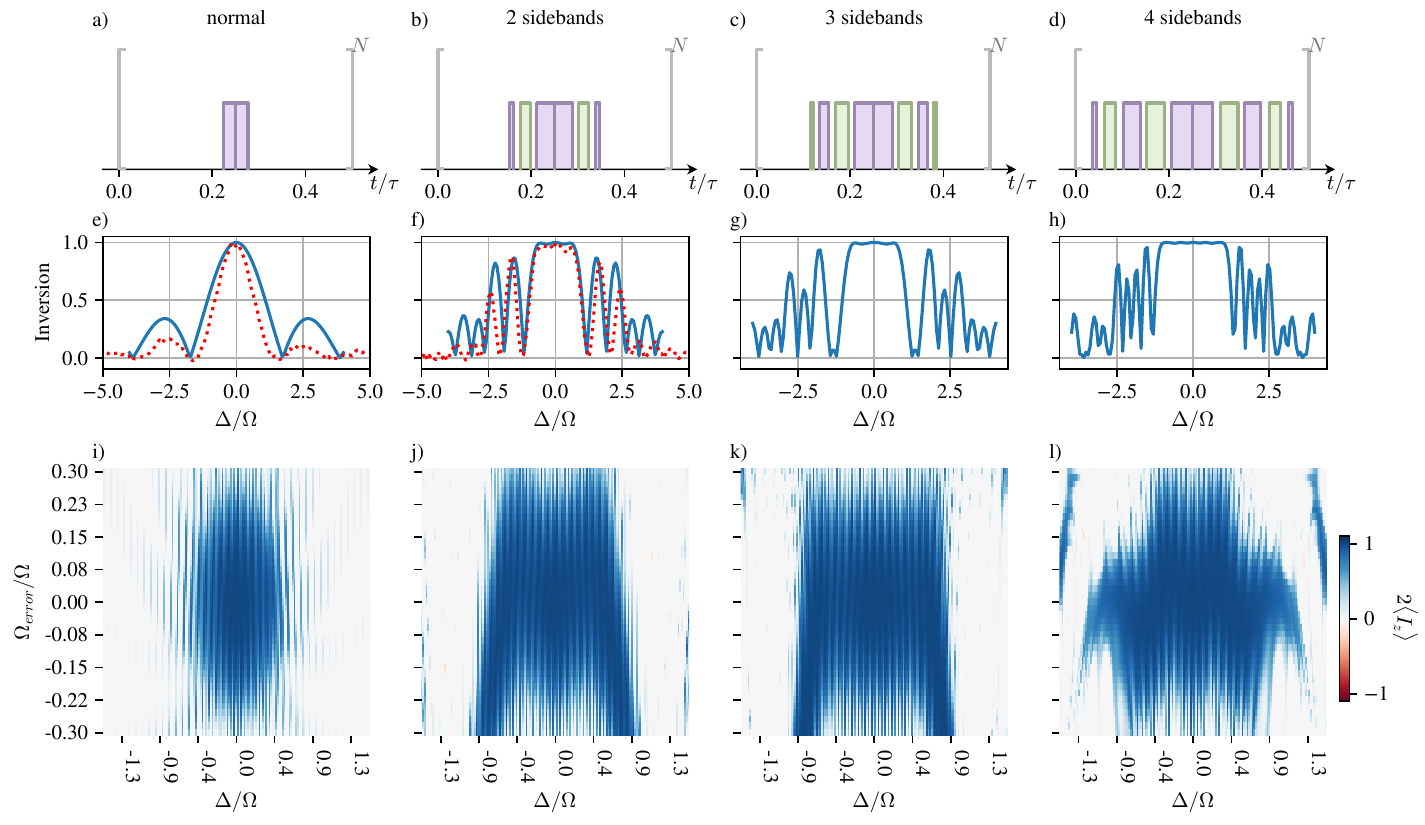}
\caption{Composite pulses (a--d), theoretical (solid line) and experimental (dotted line where available) inversion properties (e--h), both assuming the experimental Rabi frequency with $\pi/\Omega = 48\,$ns; and pulse-error robustness properties as part of PulsePol using an increased Rabi frequency of $\Omega=(2\pi)\ 30\,$MHz to allow for also using the longer composite pulses (i--l). Thus, (a,e,i) correspond to the standard rectangular pulse, (b,f,j) to the 2-sideband pulse, (c,g,k) to the 3-sideband pulse, and (d,h,l) to the 4-sideband composite pulse. The experimental profiles appearing as dotted lines in (e,f) were taken on NVs aligned to the magnetic field,  in the single crystal diamond also used for experiments described in section \ref{SI:pulsepol_phaseoffset}. }
\label{fig:compPulses}
\end{figure}

In our experiments, the two-sideband pulse was used, with Rabi $\Omega_1=(2\pi)\SI{10.5}{\mega\hertz}$. It can be seen from  Table \ref{tab:comp-pulses} that the width of the range of detunings for which PulsePol gives strong polarization transfer is $\Delta_{\mathrm{pol}} = 1.4 \Omega_1 = (2\pi)\SI{14.7}{\mega\hertz}$.

As derived in \cite{shaka1987}, the symmetry of the composite $\pi$-pulses implies that they are constant-phase pulses over their respective bandwidth, as well as that the left and right halves of the pulses are broadband $z$-to-$x$ and $x$-to-$z$ point-to-point rotations for the left and right half of the pulse, respectively. These properties make these composite pulses well-suited for the application in PulsePol, as the $(\pi/2)_X-(\pi)_Y-(\pi/2)_X$ combination can play out its full error correction for small deviations of the optimal rotation.
Furthermore, the oscillatory behavior of the sidebands of phase-alternating pulses makes it possible to fill a relatively large fraction of the waiting times of PulsePol without significantly degrading transfer properties.

\clearpage
\subsection{Other parameters adjustment}
\label{SI:pulsepol_otherparams}

\begin{figure}[ht]
\centering
\includegraphics[width=0.8\linewidth]{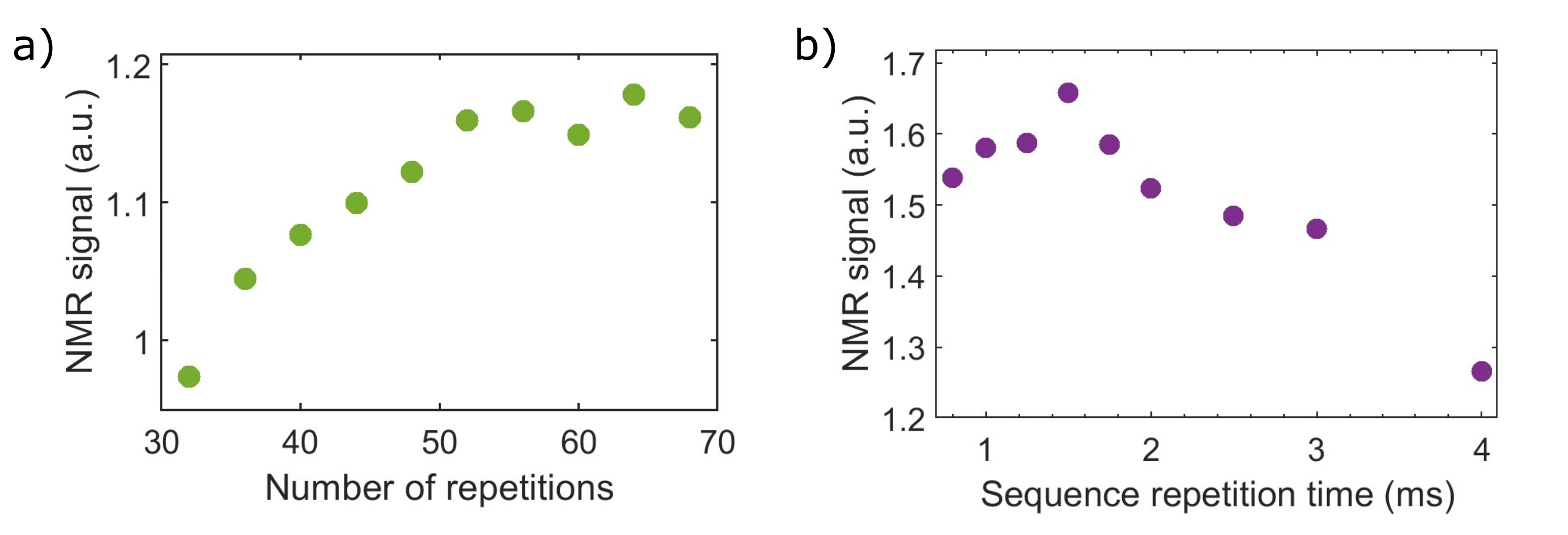}
\caption{Hyperpolarized $^{13}$C NMR signals for the \SI{2}{\micro\meter} sample  a)  as a function of the number of repetitions $M$  in the PulsePol sequence, using sequence repetition time $\tau_{\mathrm{SRT}}=\SI{1.5}{\milli\second}$ b)  as a function of the sequence  repetition time $\tau_{\mathrm{SRT}}$, using $M=68$.  To mitigate sample heating in b), the length of the laser initialization pulse was varied to ensure constant laser duty cycle, as $t_{\mathrm{laser}}=\frac{\SI{400}{\micro\second}}{\SI{1500}{\micro\second}}\times\tau_{\mathrm{SRT}}$. }
\end{figure}

\section{Potential   adaptations for further increase of the bandwidth $\Delta_{\mathrm{pol}}$}
\label{SI:increase_deltapol}

\paragraph{Using a higher power amplifier.} In our experiments, we used a \SI{40}{\watt} microwave amplifier for applying the polarization transfer sequences, and obtained $\Delta_{\mathrm{pol}} = (2 \pi) \SI{15}{\mega\hertz}$. 
Using a higher power amplifier  (\SI{500}{\watt}) would ensure    
 $\Delta_{\mathrm{pol}}\approx(2 \pi) \SI{50}{\mega\hertz}$ without change in the protocol. 
 \paragraph{Sequence multiplexing.} For further improvement, one could also use pulses fulfilling two criteria, that is not only performing efficient driving in a defined spectral range, but also leaving the spin polarization essentially unaffected outside this region. Although our approach based on composite pulses enabled us to meet the first criterion, it did not fulfill  the second. However, one could possibly resort to  envelope-modulated pulses~\cite{lurie1985,kobzar2012}. Despite their application being experimentally more involved  (as they requires good control over the non-linearities in the amplifier and the excitation channel overall) compared to our composite pulses, they would enable applying the polarization sequence  at nearby frequencies (this is in fact comparable to the idea of applying  parallel sweeps, i.e.  frequency combs in the ISE~\cite{ajoy2018pnas_combs}). 
 The frequency selectivity of the pulse would mitigate the advert effects of multiplexing, i.e. cross-talks. 
 
 We expect  that this approach could  extend further the proportion of addressed NVs, leading to $\Delta_{\mathrm{pol}}=(2 \pi) \SI{100}{\mega\hertz}$. The simulation shows that this corresponds to a 16\% fraction  of addressed NVs. To make implementation more straightforward and for ensuring faster polarization transfer, the frequency multiplexing and the use of the high power amplifier could be combined.  \\

\section{NV spin Hamiltonian}

In this section, we consider the NV spin Hamiltonian and derive properties related to experiments done in diamond powder, where all NV orientations are present, under a magnetic field.

First, we recall that the spin Hamiltonian for NV in an external magnetic field $\mathbf{B}$ is given by (neglecting the $^{14}$N and $^{13}$C hyperfine terms) 
\begin{equation}
     \hat{H}/h=D\hat{S}_z^2+\bar{\gamma} \mathbf{B}\cdot\hat{\mathbf{S}}
     +e_x (\hat{S}_y^2-\hat{S}_x^2)
     +e_y (\hat{S}_x\hat{S}_y+\hat{S}_y\hat{S}_x)
     ,
	\label{eq:Hamil_general}
\end{equation}

\noindent where $D$ is the zero-field splitting of the NV center, $z$ is the defect axis, $\bar{\gamma}=\gamma/ 2\pi$ ($\gamma$ is the gyromagnetic ratio of NV), $e_x$ and $e_y$ are terms related to strain and local charges~\cite{mittiga2018}.

\subsection{Spin Hamiltonian diagonalization for NV centers under   transverse $B$-field}
\label{SI:sec_hamiltonian_diag_90deg}

Hyperpolarization of nuclear spins was implemented by using NV centers oriented with 90° to the magnetic field. 
In this section, we will describe the diagonalization of the spin Hamiltonian of NV centers with 90° orientation to the magnetic field, which allows deriving analytically the position of the resonance peaks in the field-swept spectrum. The obtained results are used in section \ref{SI:illumination:heating}, to predict the shift in the magnetic field of the 90° resonance (low field) following a change in $D$, induced  by heating.

For diagonalization, we  follow steps that have already been described in the literature \cite{mabbs2013,yavkin2019}. For simplicity, we now assume $e_x, e_y \approx 0$ in Eq.~\ref{eq:Hamil_general}, as the variation of the EPR line position with the temperature, originates dominantly from the change in $D$. The Hamiltonian in Eq.~\ref{eq:Hamil_general} becomes 
\begin{equation}
     \hat{H}/h \approx D\hat{S}_z^2+\bar{\gamma} \mathbf{B}\cdot\hat{\mathbf{S}}
     .
	\label{eq:Hamil_x_original}
\end{equation}

Let’s now assume a magnetic field $B$ is applied along $x$ perpendicular to the axis of the NV. In this scenario, the Hamiltonian in Eq.~\ref{eq:Hamil_x_original} transforms into 

\begin{equation}
     \hat{H}/h=D\hat{S}_z^2+\bar{\gamma} B\hat{S}_x
     ,
	\label{eq:Hamil_x}
\end{equation}

\noindent where the operators $\hat{S}_z$ and $\hat{S}_x$ in Eq.~\ref{eq:Hamil_x} are the spin matrices:
\begin{equation}
	\hat{S}_z=
	\begin{pmatrix}
		1 & 0 & 0\\
		0 & 0 & 0\\
		0 & 0 & -1
	\end{pmatrix},
\:
	\hat{S}_x=\frac{1}{\sqrt{2}}
\begin{pmatrix}
	0 & 1 & 0\\
	1 & 0 & 1\\
	0 & 1 & 0
\end{pmatrix}.
	\label{eq:Pauli}
\end{equation}

In the basis of the zero-field eigenstates $\ket{-1}$, $\ket{0}$, $\ket{1}$, the 90° NV Hamiltonian can be expressed as
\begin{equation}
\hat{H}/h=
	\begin{pmatrix}
			D & \frac{\bar{\gamma} B}{\sqrt{2}} & 0\\
			\frac{\bar{\gamma} B}{\sqrt{2}} & 0& \frac{\bar{\gamma} B}{\sqrt{2}}\\
			0 & \frac{\bar{\gamma} B}{\sqrt{2}} & D
		\end{pmatrix}.
\end{equation}

In order to find energies, we must diagonalize NV Hamiltonian:
\begin{equation}
	\det(\hat{H}/h-u)=0,
\end{equation}
\begin{equation}
\begin{vmatrix}
\begin{pmatrix}
	D-u & \frac{\bar{\gamma} B}{\sqrt{2}} & 0\\
	\frac{\bar{\gamma} B}{\sqrt{2}} & -u& \frac{\bar{\gamma} B}{\sqrt{2}}\\
	0 & \frac{\bar{\gamma} B}{\sqrt{2}} & D-u
\end{pmatrix}
\end{vmatrix}
=0,
\end{equation}
\begin{equation}
	-u(D-u)^2-2(D-u)(\frac{\bar{\gamma} B}{\sqrt{2}})^2=0.
	\label{eq:H_diag}
\end{equation}
The solutions of equation \ref{eq:H_diag} are
\begin{equation}
	u_1=\frac{D-\sqrt{D^2+(2\bar{\gamma} B)^2}}{2},
	\:
	u_2=D,\\
	\:
	u_3=\frac{D+\sqrt{D^2+(2\bar{\gamma} B)^2}}{2}.	
	\label{eq:E_NV90}
\end{equation}
The corresponding dependency of energies versus  magnetic field is shown in Fig.~\ref{SI_energy_levels}b.  
EPR spectra are typically obtained by sweeping a magnetic field at a constant microwave frequency $\nu$. The resonance magnetic fields $B_{12}$ and $B_{23}$, corresponding to the transitions $u_2\rightarrow u_1$ and $u_3\rightarrow u_2$, are calculated as
\begin{equation}
B_{12}=\frac{1}{\bar{\gamma}}\sqrt{\nu(\nu-D)},
	\:
B_{23}=\frac{1}{\bar{\gamma}}\sqrt{\nu(\nu+D)}.
\end{equation}

The change of the low field $\SI{90}{\degree}$ NV resonance line position (for use in section  \ref{SI:illumination:heating}) can be written as a function of $D$ as
\begin{equation}
	\frac{dB_{12}}{dD}=\frac{-\nu}{2\bar{\gamma}\sqrt{\nu^2-D\nu}}=\SI{-0.213}{G\per\mega\hertz},
	\label{HoverD}
\end{equation} 
taking $\nu=\SI{9.6}{\giga\hertz}$,   $D=\SI{2869}{\mega\hertz}$, $\bar{\gamma}=\SI{2.8032}{\mega\hertz\per G}$.

The eigenstates corresponding to the energies in Eq.~\ref{eq:E_NV90} are:

\begin{align} 
\ket{s_{1,90\degree}}  &= \frac{1}{\sqrt{1+c^2}}\left(\ket{0} + c \frac{\ket{-1}+\ket{+1}}{\sqrt{2}}\right)   \nonumber \\ 
\ket{s_{2,90\degree}}  &= \frac{\ket{+1}-\ket{-1}}{\sqrt{2}}  \label{eq:states_90deg} \\ 
\ket{s_{3,90\degree}}  &= \frac{1}{\sqrt{1+c'^2}}\left(\ket{0} + c' \frac{\ket{-1}+\ket{+1}}{\sqrt{2}}\right),  \nonumber  
\end{align}

\noindent with $c = \frac{\frac{D}{2} - \sqrt{(\frac{D}{2})^2 + (\bar{\gamma}  B)^2 }}{\bar{\gamma}  B}$ and 
$c' = \frac{\frac{D}{2} + \sqrt{(\frac{D}{2})^2 + (\bar{\gamma} B)^2 }}{\bar{\gamma}  B}$.  At $B=\SI{0.287}{\tesla}$, one has     $c=-0.8375$ and $c'= 1.1941$.

\subsection{Fraction of NV centers in the bandwidth $\Delta_{\mathrm{pol}}$ ($\approx \SI{90}{\degree}$ to the magnetic field)}
\label{SI:sec_fraction_delta_pol}

We derive the condition for NV with an axis that is slightly misaligned from the perpendicular condition, i.e.  at an angle $90\degree+\delta$ from the magnetic field, to be present in the active bandwidth of the polarization sequence,  $\Delta_{\mathrm{pol}}$.

In the frame of the NV center $x,y,z$, the magnetic field can be written as $\mathbf{B} = (B\cos(\delta),0, B\sin(\delta))$. In fact, the effect of the tilt in the magnetic field can be treated considering the change in the magnetic  field vector $\mathbf{b}=(b_{\perp}, 0, b_{\parallel}) \approx B (- \delta^2/2, 0, \delta)$. We solve the Hamiltonian in the basis of the eigenstates for the exact, \SI{90}{\degree}  tilt of the magnetic field to the NV axis, which were determined previously in section~\ref{SI:sec_hamiltonian_diag_90deg}.

\begin{equation}
\begin{vmatrix}
\begin{pmatrix}
	u_1+\frac{2c }{1+c^2} \bar{\gamma}b_{\perp} - u' & \frac{c }{\sqrt{1+c^2}}\bar{\gamma}b_{\parallel} & \frac{c+c' }{\sqrt{(1+c^2)(1+c'^2)}} \bar{\gamma}b_{\perp}  \nonumber   \\
	\frac{c }{1+c^2}\bar{\gamma}b_{\parallel} & u_2 - u' &  \frac{c' }{1+c'^2}\bar{\gamma}b_{\parallel}\\
	\frac{c+c' }{\sqrt{(1+c^2)(1+c'^2)}} \bar{\gamma}b_{\perp} & \frac{c' }{1+c'^2}\bar{\gamma}b_{\parallel}   & u_3 +\frac{2c' }{1+c'^2} \bar{\gamma}b_{\perp}  - u'  \nonumber  
\end{pmatrix}
\end{vmatrix}
=0,
\end{equation}

\noindent where  $u_i, i=1,2,3$ are the energies in Eq.~\ref{eq:E_NV90} and $c,c'$ are the magnetic field-dependent terms defined in section~\ref{SI:sec_hamiltonian_diag_90deg}, $u'$ is the variable to be determined. 

Restricting to the terms up to the second order in $\delta$, the new energies read 
\begin{align}
u'_1 &= u_1   - \bar{\gamma} B \left[\frac{c}{1+c^2} +  \frac{c^2}{1+c^2}\frac{\bar{\gamma}B}{u_2-u_1}\right]  \delta^2  &= u_1+ \frac{D}{2}\left(1-\frac{D}{\sqrt{D^2+(2\bar{\gamma}B)^2}}\right)  \delta^2,   \\ 
u'_2 &= u_2 +  \bar{\gamma} B \left[  \frac{c^2}{1+c^2}\frac{\bar{\gamma}B}{u_2-u_1} -  \frac{c'^2}{1+c'^2}\frac{\bar{\gamma}B}{u_3-u_2}   \right]  \delta^2   &= u_2 - D \delta^2, \\ 
u'_3 &\approx u_3 +   \bar{\gamma} B \left[ +  \frac{c'^2}{1+c'^2}\frac{\bar{\gamma}B}{u_3-u_2} - \frac{c'}{1+c'^2} \right]  \delta^2  &= u_3 + \frac{D}{2}\left(1+\frac{D}{\sqrt{D^2+(2\bar{\gamma}B)^2}}\right)  \delta^2 . 
\label{eq:90deg_diag_secondorder}
\end{align}

 We now consider the transition $\ket{s_1}\rightarrow\ket{s_2}$ between the states of energy $u'_1$ and $u'_2$ at $B=\SI{0.287}{\tesla}$. The transition frequency $\nu_{12}' = u'_2-u'_1$, varies with the tilt $\delta$ so that 

\begin{equation}
\nu_{12}'\approx\nu_{12} -   (\SI{4052}{\mega\hertz}) \times   \delta^2,  
\end{equation}

\noindent where $\nu_{12} = u_2-u_1$. Now taking into account the width (here in ordinary frequency) $\Delta_{\mathrm{pol}}/2\pi = \SI{15}{\mega\hertz}$, it is possible to cover all the NV centers for which 
\begin{equation}
    |\delta| < \delta_{\mathrm{M}} = \sqrt{\frac{15}{4052}} = \SI{0.0608}{\radian} \hspace{20pt} [3.48\degree]. 
\end{equation}

The fraction of NV centers that is addressed is the covered portion of the solid angle, that is exactly $\delta_{\mathrm{M}}$. This suggests that 6.08 \% of NV centers are within the bandwidth $\Delta_{\mathrm{pol}}$. 

However, this calculation neglects the inhomogeneous broadening from sources other than the orientation-dependence of the transition frequency. Including the additional sources of inhomogeneous broadening, determined from the spectral fits (section \ref{SI:illumination:spectrum}), we find from a numerical simulation  that the fraction of NV centers is \textbf{4.15 \%} for the $\ket{s_1}\rightarrow\ket{s_2}$ transition. The other transition $\ket{s_2}\rightarrow\ket{s_3}$ contributes by a small amount,  \textbf{0.21 \%}. Thus we keep $\mathbf{\approx 4 \%}$ as the number of NV addressed by the polarization sequence, in the \SI{15}{\mega\hertz} bandwidth, and correspondingly $\delta_{\mathrm{M}}=\SI{0.04}{\radian}\sim \mathbf{2.5\degree}$. 

\subsection{NV transition dipole and Rabi inhomogeneity}

\begin{figure}[htp!]
\centering
\includegraphics[width=0.7\linewidth]{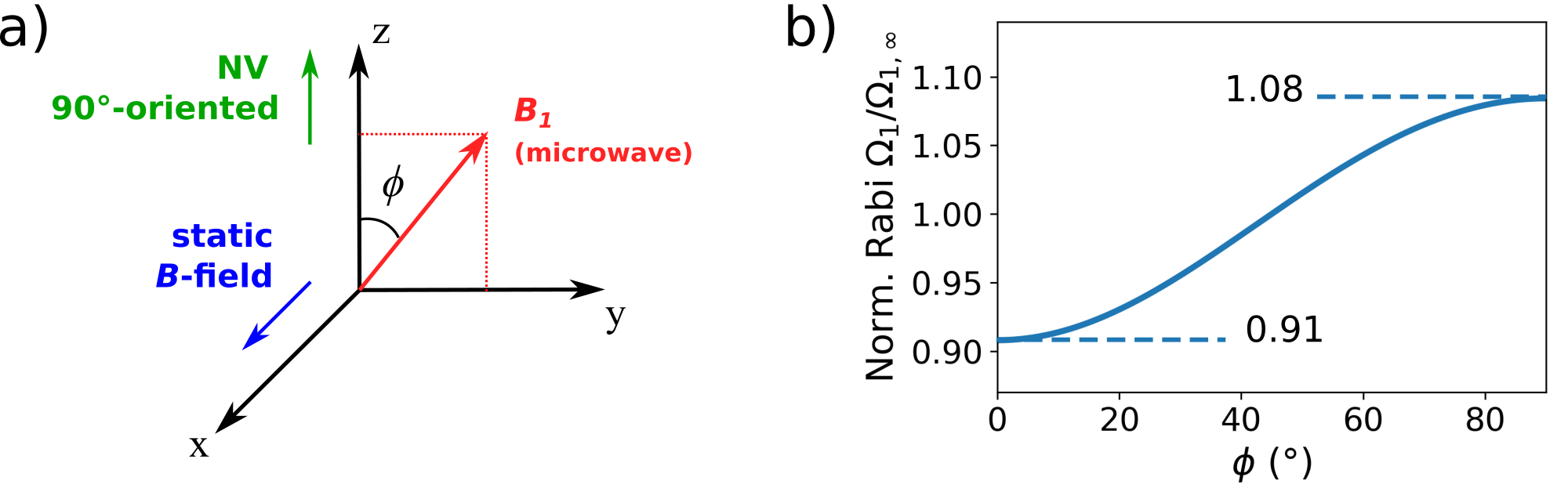}
\caption{a) Coordinate system of the NV frame,  in the case of a  perpendicular  magnetic field ($\parallel x$), and definition of the orientation of the  linearly polarized microwave field (in $yz$ plane). b)  Rabi vs $\phi$ angle for the employed $\ket{s_1}\rightarrow\ket{s_2}$ transition of NV at $B=\SI{0.287}{\tesla}$ after normalization to the Rabi frequency $\Omega_{1,\infty}$ obtained in the high-field regime ($\bar{\gamma} B\gg D$).}
\label{fig:SI_90deg_Rabi_var}
\end{figure}

The different orientations of the nanocrystals leads to a variation of   the magnetic transition dipole, even if the orientation of the NV center to the external magnetic field is fixed ($\approx 90\degree$ in our experiments). It is possible to derive the corresponding  inhomogeneity of the Rabi frequency for NV centers. For a linearly polarized microwave field and a transition between two states $\ket{s_i}$, and $\ket{s_j}$, the transition dipole can be expressed as

\begin{equation}
d_{ij} = \bra{s_i}\hat{S}_1 \ket{s_j}, 
\end{equation}

\noindent where $\hat{S}_1$ is the projection on the spin-operator in the direction of the linearly polarized $\mathbf{B_1}$-field.  In our configuration (EPR resonator), the $\mathbf{B_1}$-field is perpendicular to the static magnetic field $\mathbf{B}$. Using the coordinate system in Fig.~\ref{fig:SI_90deg_Rabi_var}a, it follows that  $\hat{S}_1 = \cos \phi \hat{S}_z + \sin \phi \hat{S}_y$.

Considering the 90\degree-oriented NV centers and the transitions between the states $\ket{s_1}$ and $\ket{s_2}$, using Eq.~\ref{eq:states_90deg} we obtain:
$d_{12} = (c\cos(\phi) + i\sin(\phi))/{\sqrt{1+c^2}}$, where $c$ is the magnetic-field dependent term defined in section~\ref{SI:sec_hamiltonian_diag_90deg}. As $c<1$, The magnitude of $d_{12}$ is the weakest when the microwave field points in the direction of the axis ($z$) of the NV center, and the strongest when it is directed along $y$ (as $c'>1$, the situation is however reverted for the other   transition). It is straightforward to derive  the Rabi frequency, as it obeys $\Omega_1 \propto |d_{12}|$. One can write

\begin{equation}
\frac{\Omega_1}{\Omega_{1,\infty}} =  \frac{\sqrt{c^2\cos^2(\phi) + \sin^2(\phi)}}{\sqrt{1+c^2}}  
\end{equation}

\noindent where $\Omega_{1,\infty}$ is the Rabi frequency assuming the Zeeman eigenstates (obtained for $B \gg D$:  $c\rightarrow 1$ and therefore $d_{12} = 1/\sqrt{2}$).  

At $B=\SI{0.287}{T}$, the values of $\Omega_1 /  \Omega_{1,\infty}$ vary in the range [0.91,1.08] (Fig.~\ref{fig:SI_90deg_Rabi_var}b). Roughly speaking, the  Rabi frequency has variations included within a $\pm 10\%$ range at this magnetic field.

\subsection{NV transition frequencies in diamond powder at different magnetic fields}
\label{SI:NV_magnetic_field_spectrum}

\begin{figure}[htp!]
\centering
\includegraphics[width=1\linewidth]{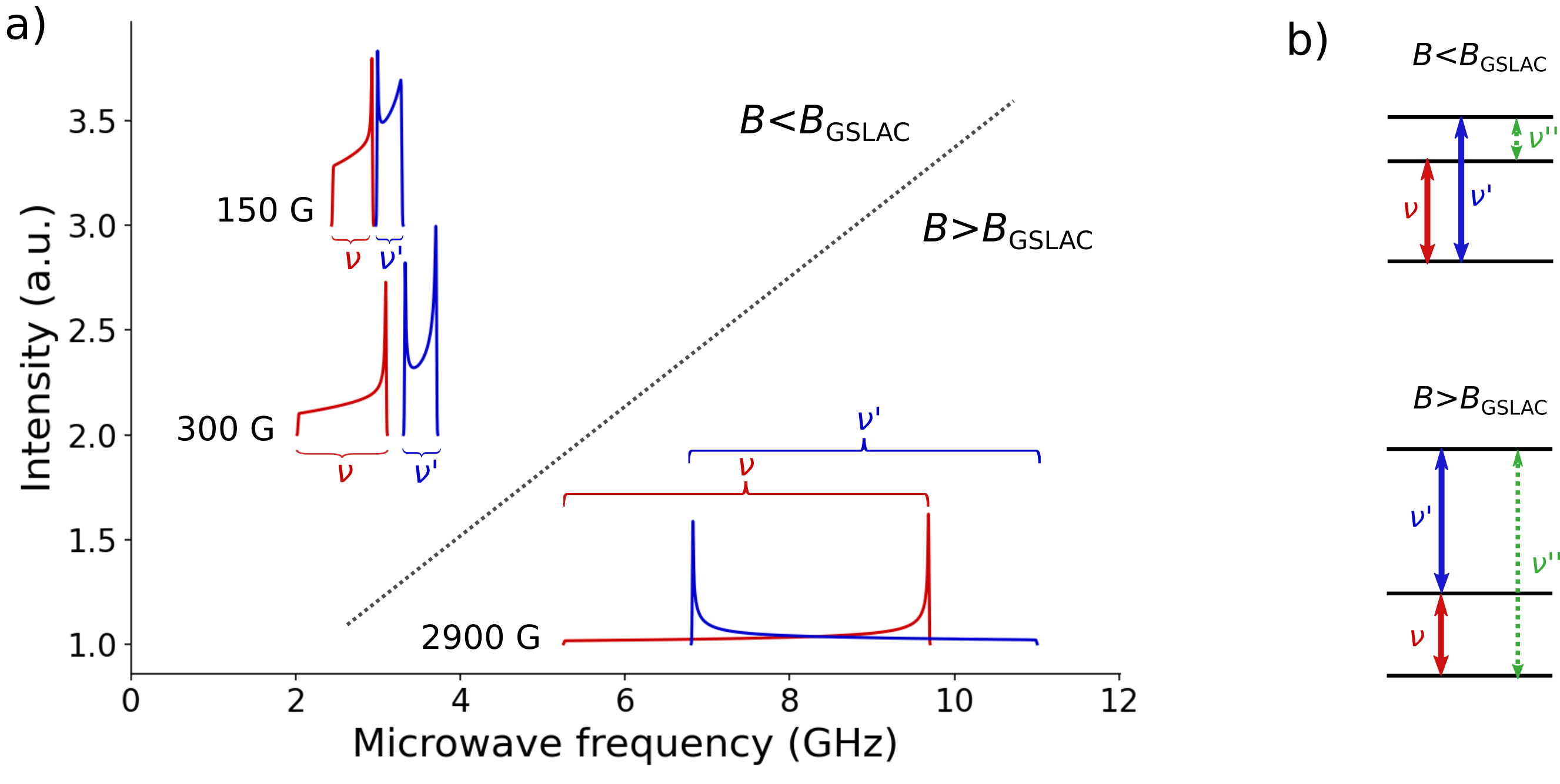}
\caption{a) Simulation of the distribution of the  transition frequencies of NV in diamond powder at magnetic fields $B=\SI{150}{G}, \SI{300}{G}$ and \SI{2900}{G}. The frequencies $\nu,\nu'$ correspond to the transitions as defined in b), where $B_{\mathrm{GSLAC}}$ is the magnetic field of the ground state level anticrossing ($B_{\mathrm{GSLAC}}\sim\SI{1024}{G}$).  The spectrum for the transition at $\nu''$ (dotted arrows) is not represented in a) (see text). }
\label{fig:SI_magnetic_field_sim}
\end{figure}

The distribution of transition frequencies in randomly oriented NV ensembles is shown in Figure~\ref{fig:SI_magnetic_field_sim}a for the main NV transitions. The simulation was performed with Easyspin. In contrast with the simulation described in section~\ref{SI:illumination:spectrum}, the setting \texttt{Temperature} was not defined (to give equal positive weight to both transitions) and the optional setting \texttt{Intensity} was set to \texttt{Off} (to remove the effect of varying transition dipoles  as a function of the orientation).
.

Besides the transitions represented in Figure~\ref{fig:SI_magnetic_field_sim}a, we indicate a third  transition   in Figure~\ref{fig:SI_magnetic_field_sim}b (at frequency $\nu''$). In the case of a magnetic field aligned to the NV axis, the transition is known as double quantum ($-1 \leftrightarrow +1$), and has vanishing  dipole. 
 In the case of a magnetic field tilted from the NV axis, this third transition can  be excited as a single quantum transition, owing to spin state mixing. At the considered magnetic fields, however, the corresponding dipole is typically lower than that of the other transitions, thus it was not represented in the spectrum in Figure~\ref{fig:SI_magnetic_field_sim}a

\bibliography{Hyper_article}